\documentclass[twocolumn]{aastex701}
\usepackage{amsmath}
\usepackage{subcaption}     
\usepackage{CJK}
\usepackage{bm}
\usepackage{lineno}
\usepackage{soul}
\usepackage{xcolor}
\soulregister\cite7
\soulregister\citet7
\soulregister\citealt7

\soulregister\ref7

\newcommand{\rev}[1]{\ifmmode\textcolor{blue}{#1}\else\textbf{\textcolor{blue}{#1}}\fi}
\newcommand{\del}[1]{\ifmmode\text{\textbf{\textcolor{blue}{\st{#1}}}}\else\textbf{\textcolor{blue}{\st{#1}}}\fi}

\begin{document}
\linenumbers

\begin{CJK*}{UTF8}{gbsn}

\correspondingauthor{Bingqiu Chen}
\email{bchen@ynu.edu.cn}

\author[0000-0003-2751-2172]{Lin Zhang(张琳)}
\affiliation{South-Western Institute for Astronomy Research,
Yunnan University,
Kunming, 650500, 
People's Republic of China}
\email{zhanglin@mail.ynu.edu.cn}

\author[0000-0003-2472-4903]{Bingqiu Chen(陈丙秋)}
\affiliation{South-Western Institute for Astronomy Research,
Yunnan University,
Kunming, 650500, 
People's Republic of China}
\email{bchen@ynu.edu.cn}

\author{Fei Qin(秦斐)}
\affiliation{Aix-Marseille University, CNRS/IN2P3, CPPM, Marseille 13288, France}
\email{qin@cppm.in2p3.fr}

\author[0000-0003-3144-1952]{Guangxing Li(李广兴)}
\affiliation{South-Western Institute for Astronomy Research,
Yunnan University,
Kunming, 650500, 
People's Republic of China}
\email{gxli@ynu.edu.cn}

\author[0000-0003-2471-2363]{Haibo Yuan(苑海波)}

\affiliation{Institute for Frontiers in Astronomy and Astrophysics, 
Beijing Normal University, 
Beijing 102206, 
People's Republic of China}

\affiliation{School of Physics and Astronomy, 
Beijing Normal University, 
No.19, Xinjiekouwai St, Haidian District, Beijing 100875, 
People's Republic of China}
\email{yuanhb@bnu.edu.cn}

\author[0000-0003-1218-8699]{Yi Ren(任逸)}
\affiliation{Department of Astronomy, College of Physics and Electronic Engineering, 
Qilu Normal University, 
Jinan 250200, 
People's Republic of China}
\affiliation{Shandong Key Laboratory of Space Environment and Exploration Technology, 
People's Republic of China}
\email{yiren@qlnu.edu.cn}

\title{A Face-on View of Interstellar Dust in the Galactic Plane}

\begin{abstract}
Interstellar dust is a fundamental component of the Milky Way, influencing star formation, galactic evolution, and observations across the electromagnetic spectrum. Using red clump stars selected from near- and mid-infrared photometry, together with stellar catalogs from previous studies, we construct dust density maps of the Galactic plane ({$|Z|<25$}\,pc) covering the full $360^\circ$ in longitude and reaching distances up to $7$\,kpc. By applying a U-Net convolutional neural network to invert the line-of-sight extinction distribution, we obtain dust density maps at resolutions of $10$, $50$, and $100$\,pc, which reveal detailed structures including spiral arms, inter-arm spurs, and giant cavities. The dust distribution in the Galactic plane exhibits a morphology closely resembling that of the so-called Phantom galaxy M74. The derived exponential scale length of the Galactic dust disk is $2.90$\,kpc, slightly larger than that of the stellar thin disk. Our publicly available dust maps provide a new benchmark for extinction correction, studies of Galactic structure, and the investigation of the interplay between star formation and the interstellar medium.
\end{abstract}

\keywords{\uat{Interstellar dust}{836} --- \uat{Interstellar medium}{847} --- \uat{Molecular clouds}{1072} --- \uat{Galaxy structure}{622} --- \uat{Red giant clump}{1370} }

\section{Introduction} 
\label{intro}

The study of interstellar dust is of great astrophysical importance, as dust is a key component of the Milky Way and plays a central role in the formation and evolution of stars and galaxies. Dust is widely distributed and can both hinder astronomical measurements and serve as an excellent tracer of Galactic structure and star-forming regions within dense molecular clouds. Traditional two-dimensional dust maps, however, lack distance information along the line of sight, and therefore cannot reveal how the dust distribution varies with distance or provide effective correction for extinction effects in the Milky Way. Constructing a three-dimensional (3D) dust map of the Galaxy is an essential step for accurately characterizing the dust distribution along different sightlines and for understanding the overall structure of the Milky Way and the spatial distribution of interstellar dust.

Before the release of high-precision trigonometric parallax data from the \textit{Gaia} mission, studies of the 3D dust distribution in the Milky Way relied mainly on two approaches: one was to infer the dust spatial distribution by comparison with Galactic models \citep{Drimmel2001, Marshall2006, Chen2013, Schultheis2014}, and the other was to estimate the variation of extinction with distance using photometric distances of stars \citep{Berry2012, Chen2014, Green2015}. These early works generally suffered from large distance uncertainties. The model-based approach introduced systematic biases, and distance estimates based on multi-band photometry were limited by uncertainties in stellar intrinsic luminosities and the effects of interstellar extinction, resulting in low distance resolution in the derived 3D dust maps and making it difficult to accurately trace the true dust distribution along the line of sight.

With the successive releases of trigonometric parallaxes for billions of stars from the \textit{Gaia} satellite \citep{Gaia2016,Gaia2018}, the precision of stellar distance measurements has been revolutionized, greatly advancing the study of the 3D dust distribution in the Milky Way. A number of recent works based on \textit{Gaia} data have combined trigonometric parallaxes with multi-band photometry to map reliable 3D dust structures in the Milky Way \citep[e.g.,][]{Chen2019_dust, Lallement2019, Green2019, Hottier2020, Vergely2022, Dharmawardena2024, Zucker2025, Gontcharov2025}. Additionally, the abundant low-resolution BP/RP spectra from \textit{Gaia} DR3 have enabled us to construct 3D extinction maps spanning distances of several kiloparsecs \citep{An2024,Edenhofer2024,Wang2025_dust}. Nevertheless, these works are still limited by the finite precision of \textit{Gaia} parallaxes, and the reliable distance range is typically confined to within about 4\,kpc. For regions of the Galactic disk farther from the Sun, the relative parallax error increases significantly, resulting in large uncertainties in the 3D reconstruction of the dust distribution.

In the study of the 3D dust distribution in the Milky Way, near-infrared (NIR) photometric surveys and Red Clump stars (RCs) play a crucial role. Optical observations are severely limited by extinction from interstellar dust and cannot penetrate regions of high column density, whereas the near-infrared band is less affected by dust extinction and can effectively reveal stars and structures hidden behind dense molecular clouds. Near-infrared surveys such as the Two Micron All Sky Survey (2MASS; \citealt{Skrutskie2006}), the VISTA Variables in the V\'ia L\'actea (VVV; \citealt{Minniti2010}), and the UKIRT Infrared Deep Sky Survey (UKIDSS; \citealt{Lawrence2007}) have provided key data for large-scale exploration of heavily extincted regions in the Galactic plane. Meanwhile, RCs are low-mass stars in the core helium burning phase. They have very small dispersions in absolute magnitude and intrinsic color and are particularly bright in the near-infrared, making them well recognized as excellent standard candles. Using this property, we can employ RCs as extinction tracers to accurately measure the interstellar extinction and distance of these stars by constructing the extinction-distance relation along the line of sight. This approach avoids the systematic errors inherent in model-dependent methods and enables the construction of high-precision, large-dynamic-range 3D dust distributions.

In this work, we aim to present a deep, guide-frame 3D dust map with high distance accuracy using near-infrared photometric data and RCs as tracers. We select RCs from intrinsic near-infrared color-magnitude diagrams (CMDs), employing photometric data from 2MASS, UKIDSS, VVV, the Galactic Legacy Infrared Mid-Plane Survey Extraordinaire (GLIMPSE; \citealt{Benjamin2003}), and the Wide-Field Infrared Survey Explorer (WISE; \citealt{Wright2010}). We then infer the extinction and distance for the selected RCs and derive a high-resolution dust distribution covering the entire Galactic plane.

\begin{figure*}
\centering
\includegraphics[width=0.95\textwidth,keepaspectratio]{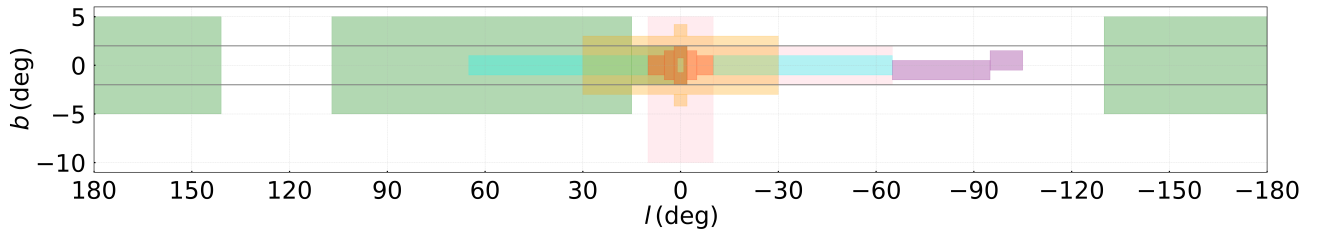}
\caption{Areal coverage of the VVV, GLIMPSE, and UKIDSS surveys in Galactic coordinates. The pink regions indicate the VVV bulge and VVV disk fields. The green regions show the UKIDSS GPS areas. The cyan, red, orange, and purple regions indicate the coverage of GLIMPSE I, GLIMPSE II, GLIMPSE 3D, and Vela-Carina, respectively. The horizontal lines mark the Galactic latitude boundaries of $b = \pm 2^\circ$}.
\label{fig:survey_region}
\end{figure*}

\section{Data and RC selection} \label{sec:data}

The overall strategy of this work is as follows. First, we combine near-infrared (2MASS, VVV, UKIDSS) and mid-infrared (MIR; ALLWISE, GLIMPSE) photometry and apply the Rayleigh-Jeans Color Excess (RJCE; \citealt{Majewski2011}) method to derive extinction values for individual stars. We then correct for extinction to obtain intrinsic colors and magnitudes, construct a reddening-corrected CMD, and finally select RC candidates based on their intrinsic color ranges in the diagram.

\subsection{Data}
2MASS used two 1.3\,m telescopes, one at Mount Hopkins in Arizona and one at Cerro Tololo in Chile, to observe the entire sky in the $J$, $H$, and $K_{\rm S}$ near-infrared bands. At a signal-to-noise ratio of $S/N=10$, the limiting magnitude in the $K_{\rm S}$ band is 14.3\,mag. For bright sources, the photometric uncertainty is typically less than 0.03\,mag. The VVV survey was conducted with a 4\,m telescope at the Paranal Observatory in Chile.  It mapped the Galactic bulge ($-10^{\circ}<l<10^{\circ}$ and $-10^{\circ}<b<5^{\circ}$) and the Galactic disk ($-65^{\circ}<l<-10^{\circ}$ and $-2^{\circ}<b<2^{\circ}$) in five bands: $Z$, $Y$, $J$, $H$, and $K_{\rm S}$, covering a total area of 520\,deg$^2$. The $K_{\rm S}$ band limiting magnitude reaches $\sim$18\,mag at $S/N \approx 3$, with saturation limits typically between 10 and 12\,mag. The UKIDSS Galactic Plane Survey (GPS; \citealt{Lucas2008}) used the 3.8\,m United Kingdom Infrared Telescope (UKIRT) to scan the Galactic plane in $J$, $H$, and $K_{\rm S}$ over regions of $-2^{\circ}<l<15^{\circ}$ with $-2^{\circ}<b<2^{\circ}$, $15^{\circ}<l<107^{\circ}$ with $-5^{\circ}<b<5^{\circ}$, and $141^{\circ}<l<230^{\circ}$ with $-5^{\circ}<b<5^{\circ}$. The photometric depths in $J$, $H$, and $K_{\rm S}$ are 19.9, 19.0, and 18.8\,mag, respectively, with a typical saturation limit of $\sim$11.5\,mag. WISE is an all-sky mid-infrared survey using a 40\,cm telescope \citep{Wright2010} in four bands: W1 (3.4\,$\mu$m), W2 (4.6\,$\mu$m), W3 (12\,$\mu$m), and W4 (22\,$\mu$m). The ALLWISE project combines data from WISE and NEOWISE \citep{Mainzer2011}. The ALLWISE source catalog \citep{Cutri2013} contains over 747 million objects and includes $J$, $H$, and $K_{\rm S}$ photometry from 2MASS. The GLIMPSE project observed the Galactic plane with the Spitzer Space Telescope \citep{Werner2004} in four bands centered at approximately 3.6, 4.5, 5.8, and 8.0\,$\mu$m. The survey comprises multiple sub-surveys. In this work, we focus on GLIMPSE I, GLIMPSE II, GLIMPSE 3D \citep{Churchwell2009}, and Vela-Carina \citep{Majewski2007, Zasowski2009}. GLIMPSE I covers the Galactic plane for $|l|=10^{\circ}$--65$^{\circ}$ and $|b|\leq1^{\circ}$, with a total area of 220\,deg$^2$. GLIMPSE II surveys the Galactic center region within $|l|\leq10^{\circ}$, with latitude coverage varying with longitude: for $|l|=10^{\circ}$ to $5^{\circ}$, $5^{\circ}$ to $2^{\circ}$, and $2^{\circ}$ to $0^{\circ}$, the latitude ranges are $\pm1^{\circ}$, $\pm1.5^{\circ}$, and $\pm2^{\circ}$, respectively. GLIMPSE 3D images $|b|\leq3^{\circ}$ and $|l|\leq30^{\circ}$, reaching $|b|\leq4.2^{\circ}$ at the Galactic center. Vela-Carina images $255^{\circ}\leq l \leq 295^{\circ}$ with latitude ranges of $-0.5^{\circ}\leq b \leq 1.5^{\circ}$ for $l=255^{\circ}$ to 265$^{\circ}$ and $-1.5^{\circ}\leq b \leq 0.5^{\circ}$ for $l=265^{\circ}$ to 295$^{\circ}$. Fig.~\ref{fig:survey_region} illustrates the survey areas of VVV, UKIDSS, and GLIMPSE in Galactic coordinates.

In our work, we focus on the Galactic plane and therefore select sources within $-2^{\circ} \leq b \leq 2^{\circ}$ from the three near-infrared surveys (2MASS, VVV, and UKIDSS). To construct an initial sample with high data quality, we select sources from all three near-infrared surveys with photometric errors $<0.1$\,mag in $J$, $H$, and $K_{\rm S}$ and with $J-K_{\rm S}>0.5$\,mag. We then apply $K_{\rm S}$-band magnitude cuts: $K_{\rm S}>10$\,mag for 2MASS, and $11<K_{\rm S}<18$\,mag for the deeper VVV and UKIDSS surveys. The bright-end cuts for VVV and UKIDSS are set by instrumental saturation, whereas the 2MASS $K_{\rm S}$ saturation limit is far brighter, so the $K_{\rm S}>10$\,mag cut is motivated by sample purity rather than saturation. For RCs, $K_{\rm S}<10$\,mag corresponds to distances within $\sim$2\,kpc, a regime already well sampled by precise \textit{Gaia} parallaxes \citep{Zhang2025} and not the one our RC method aims to extend; this bright end is instead dominated by upper-RGB and AGB giants that mimic RC colors but bias distances if misclassified, so the cut removes such contamination at negligible cost to genuine RCs. No bright-end cut is applied to the mid-infrared bands, which serve only to estimate extinction via RJCE. The $K_{\rm S}>10$\,mag cut already keeps the cross-matched mid-infrared photometry unsaturated.

Because the photometric systems of UKIDSS and VVV differ from that of 2MASS, we adopt 2MASS as the photometric reference. We convert VVV and UKIDSS magnitudes to the 2MASS system using the transformation equations from \citet{Soto2013} and \citet{Wegg2015}, respectively. The three near-infrared surveys have partially overlapping sky coverage. For sources that appear in multiple catalogs, we retain each measurement as an independent entry in our sample after photometric conversion, rather than averaging or selecting a single catalog. This approach maximizes the available stellar sample. It is appropriate here because our subsequent analysis relies on ensemble statistics of extinction along each sightline rather than on individual stellar properties. We then cross-match the near-infrared sample with the GLIMPSE I, GLIMPSE II, GLIMPSE 3D, Vela-Carina, and ALLWISE catalogs to obtain sources with common near-infrared and mid-infrared photometry. For GLIMPSE data, sources are retained if the photometric errors in all four mid-infrared bands are less than 0.5\,mag. For ALLWISE, we remove sources with signal-to-noise ratio less than 30 in the W2 band.

\begin{figure}
\centering
\includegraphics[width=0.45\textwidth,keepaspectratio]{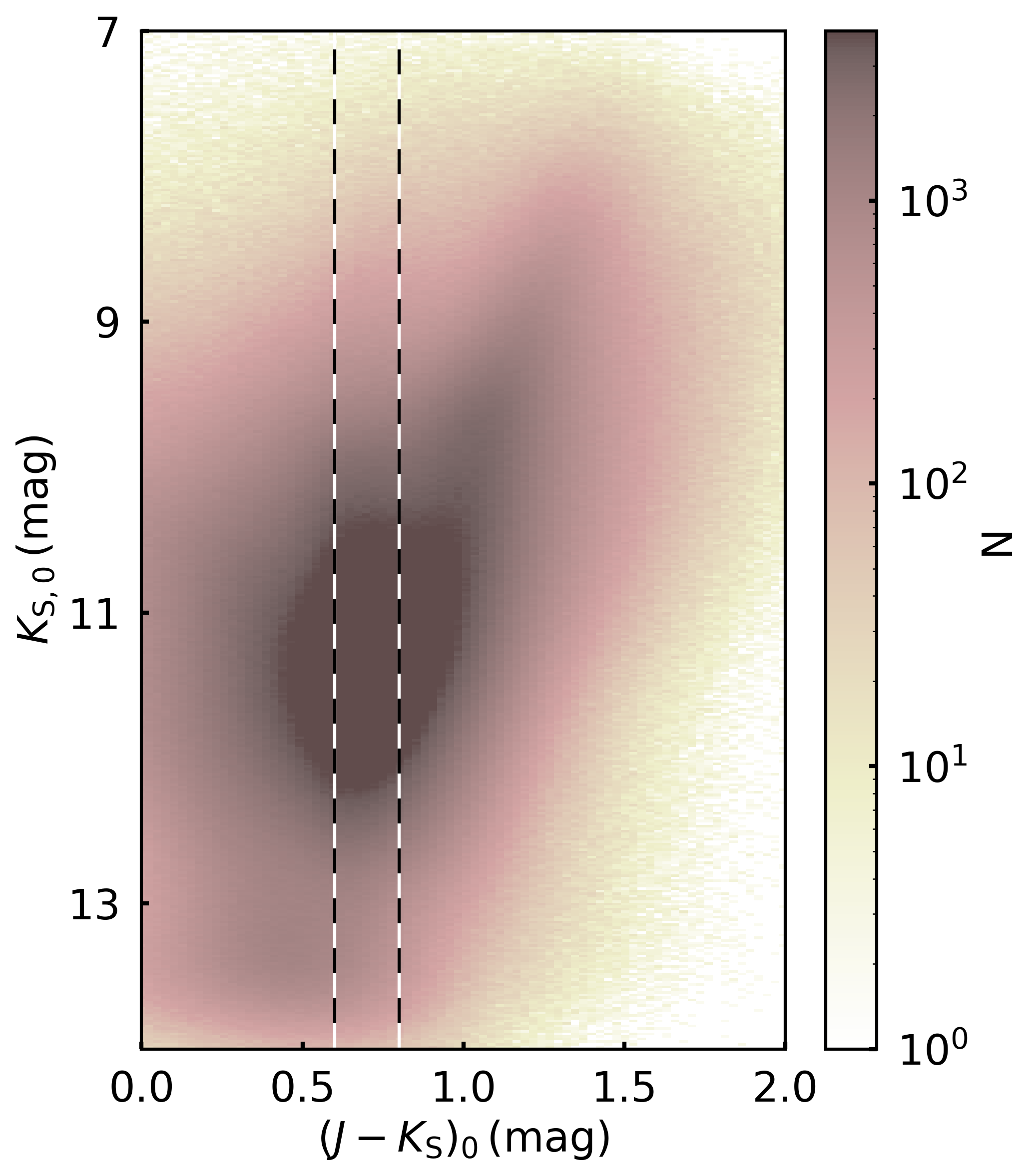}
\caption{Reddening-corrected CMD for all sources in the sample. The vertical dashed lines mark the RC color selection boundaries.}
\label{initial_sample_CMD}
\end{figure}

\begin{figure}
\centering
\includegraphics[width=0.45\textwidth,keepaspectratio]{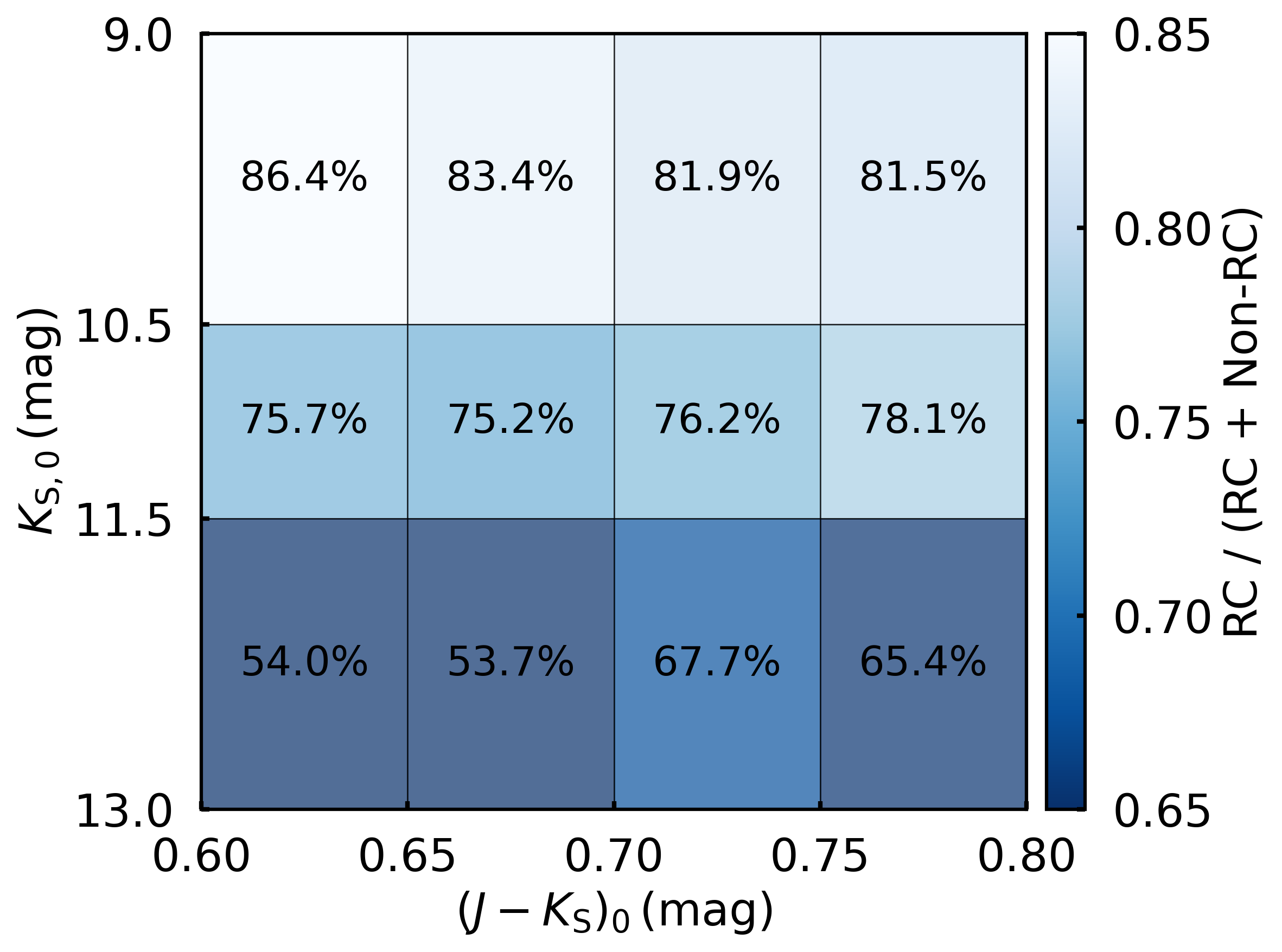}
\caption{Distribution of the selection precision of our RC candidates in the CMD.}
\label{fig:precision}
\end{figure}

 \subsection{RC selection} \label{sec:rcselect}

The RJCE method estimates stellar extinction using the color difference between near-infrared bands (e.g., $H$) and mid-infrared bands (e.g., $[4.5\mu]$ or W2). The principle is that stellar radiation in the infrared approximately follows the Rayleigh-Jeans law, making this color difference sensitive to extinction while being relatively insensitive to the intrinsic stellar parameters.

Following \citet{Majewski2011}, we adopt the extinction coefficients from \citet{Indebetouw2005}. We measure extinction for GLIMPSE sources using the relation $A_{K\rm_S}=0.918\times(H-[4.5\mu]-0.08)$. Since \citet{Indebetouw2005} does not provide an extinction coefficient for the $W2$ band, we cross-match the GLIMPSE sources with the ALLWISE sources and, from the color-color diagram of the common sources, derive $A_{K\rm_S}=0.897\times(H-W2-0.08)$ for ALLWISE sources. We note that a more recent extinction curve was presented by \citet{Gordon2023}, which provides a unified $R(V)$-dependent extinction relationship from the far-ultraviolet (FUV) to the MIR at spectroscopic resolution. When constrained to NIR bands only ($J$ and $K_{\rm S}$), the \citet{Gordon2023} curve yields an extinction coefficient of 0.660, consistent with the 0.667 value from \citet{Indebetouw2005} adopted here. However, when the 4.5\,$\mu$m band is involved (as in the RJCE method), the \citet{Gordon2023} curve gives 0.781, which differs from the \citet{Indebetouw2005} value. The discrepancy arises because the 4--5\,$\mu$m region falls in a gap between the NIR and MIR spectroscopic data used to constrain the \citet{Gordon2023} curve, and the authors themselves note that the NIR slopes in this range may be affected by residual atmospheric features. Given the consistency of the two curves in the purely NIR regime and to maintain continuity with the original RJCE calibration of \citet{Majewski2011}, we adopt the \citet{Indebetouw2005} extinction curve throughout this work.

Fig.~\ref{initial_sample_CMD} presents the reddening-corrected CMD of $(J-K_{\rm S})_0$ versus $K_{\rm S,0}$ for these stars, where a distinct clustering of RC stars is visible. Following \citet{Majewski2011}, we select RC candidates with $0.6 \leq (J-K_{\rm S})_0 \leq 0.8$, yielding 6,279,460 RC candidates. We adopt a single de-reddened color rather than additional near-infrared colors because dwarfs and cool giants (RC, RGB, and AGB) form an essentially one-dimensional sequence in the near-infrared, with $(J-H)_0$, $(H-K_{\rm S})_0$, and $(J-K_{\rm S})_0$ tightly correlated; a second color therefore adds little independent information for separating RCs from contaminates. We use $(J-K_{\rm S})_0$ as it has the longest wavelength baseline in the near-infrared, is the most sensitive to temperature, and is the standard color adopted for RC selection in the literature \citep{Majewski2011}.

Because the intrinsic color and absolute magnitude of RCs are nearly constant, we adopt $(J-K_{\rm S})_{0}= 0.7$\,mag \citep{Grocholski2002} and $M_{\rm {K_S}} =-1.61$\,mag \citep{Alves2000,Hawkins2017,Ruiz-Dern2018}. The distance and extinction for these stars are then derived as
\begin{equation}
\begin{aligned}
A_{K\rm_S} &= 0.667\times(J-K_{\rm S}-0.7), \\
D &= 10^{0.2\times(K_{\rm S}-(-1.61)-A_{K_{\rm S}})+1},\
\label{equ:boundary}
\end{aligned}
\end{equation}
where the extinction coefficient 0.667 is taken from \citet{Indebetouw2005}. We also provide uncertainties in extinction and distance for all RC candidates, combining systematic and statistical contributions. For distance, we adopt an uncertainty of 0.05\,mag in both the intrinsic color $(J-K_{\rm S})_{0}$ and the absolute magnitude $M_{K_{\rm S}}$ of RCs \citep{Salaris2002,Chen2017,Plevne2020,Yu2025RAA}. In addition, we adopt a typical uncertainty of 5\% for the infrared extinction coefficients \citep{Alonso2017}. Overall, this yields a typical distance uncertainty of $\sim$7\%. We have also compared the two extinction estimates, namely those derived from Eq.~(1) and from the RJCE method, for the selected RC stars, and find excellent agreement between them. In this work, we adopt the RC-based estimate from Eq.~(1), because the [4.5\,$\mu$m] photometry used in the RJCE method typically has larger uncertainties than the $K_{\rm S}$ band, which leads to relatively larger errors in the derived extinction. Since Eq.~(1) is calibrated specifically using the intrinsic $(J-K_{\rm S})$ color of RC stars, it is the natural choice for an extinction tracer based on RC stars.

RC candidates within the selection boundaries may include contaminants such as red giants or dwarfs due to measurement errors, uncertainties in interstellar extinction, and the partial overlap of RC stars with red giants and dwarfs on the CMD. Because giants and dwarfs have different luminosities from RCs, misclassifying a giant or dwarf as an RC biases the inferred distance. Specifically, if a red giant (dwarf) is misclassified as an RC, its inferred RC distance ($D_{\rm RC}$) is underestimated (overestimated). We therefore use \textit{Gaia} DR3, which provides reliable parallaxes for billions of stars, to remove a fraction of these contaminants. We cross-match our selected RC candidates with \textit{Gaia} DR3 and retain only common sources with $\sigma_{\varpi}/\varpi < 0.2$. All Gaia parallaxes in this work are corrected for the zero-point offset following \citet{Lindegren2021}. For most candidates, $D_{\rm RC}$ and $D_{\rm Gaia}$ are in good agreement, while a small fraction of sources exhibit large discrepancies. We remove sources with $D_{\rm Gaia} < D_{\rm RC}/2$ or $D_{\rm Gaia} > 2 D_{\rm RC}$. This yields 5,932,862 RC candidates. The absolute magnitude of RCs is well separated from those of the main contaminants: upper-RGB and AGB giants are intrinsically brighter, while dwarfs that mimic RC colors are several magnitudes fainter. A misclassified contaminant therefore has an inferred RC distance that is biased by an amount far exceeding the RC intrinsic scatter. The simple factor-of-two distance window gives essentially the same selection as a more sophisticated per-star treatment that explicitly propagates the individual uncertainties and the intrinsic scatter.

To quantify the selection precision of the RC candidates, we adopt the sample from \citet{Ting2018}, which provides a spectroscopically confirmed RC sample with a contamination level of only 3\% based on APOGEE spectroscopy. Following \citet{Ting2018}, we regard stars with $\Delta P <100$ and $\Delta \nu >5$ as non-RCs and treat the pure RC stars as reliable RCs. Cross-matching our initial sample with this reference yields 8,752 common sources, from which we obtain 4,150 positive (RC) and 1,553 negative (non-RC) samples after applying the above criteria. We then estimate the selection precision across the RC selection region as a function of color and magnitude. We divide color $(J-K_{\rm S})_{0}$ from 0.6\,mag to 0.8\,mag uniformly into four magnitude bins with a width of 0.05\,mag. Similarly, we divide  $K_{\rm S,0}$ from 9\,mag to 13\,mag into three uneven bins: 9--10.5\,mag, 10.5--11.5\,mag, and 11.5--13\,mag. Fig.~\ref{fig:precision} shows the variation in RC selection precision across these bins. As the magnitude becomes fainter, the selection precision decreases due to larger uncertainties and contamination from other stellar populations. Within the RC selection boundaries in the CMD, the average RC selection precision is approximately 72.8\%. Because the reference sample of \citet{Ting2018} does not include dwarfs, this precision of 72.8\% does not account for possible dwarf contamination. We note, however, that late-type dwarfs with RC colors are intrinsically faint and are predominantly nearby, high-parallax sources. If misclassified as RC stars, their Gaia distances would be much smaller than the distances inferred from the RC standard-candle relation; they are therefore largely removed by the $D_{\rm Gaia}$--$D_{\rm RC}$ consistency cut applied above.

\begin{figure*}
    \centering
    \includegraphics[width=0.32\textwidth,keepaspectratio]{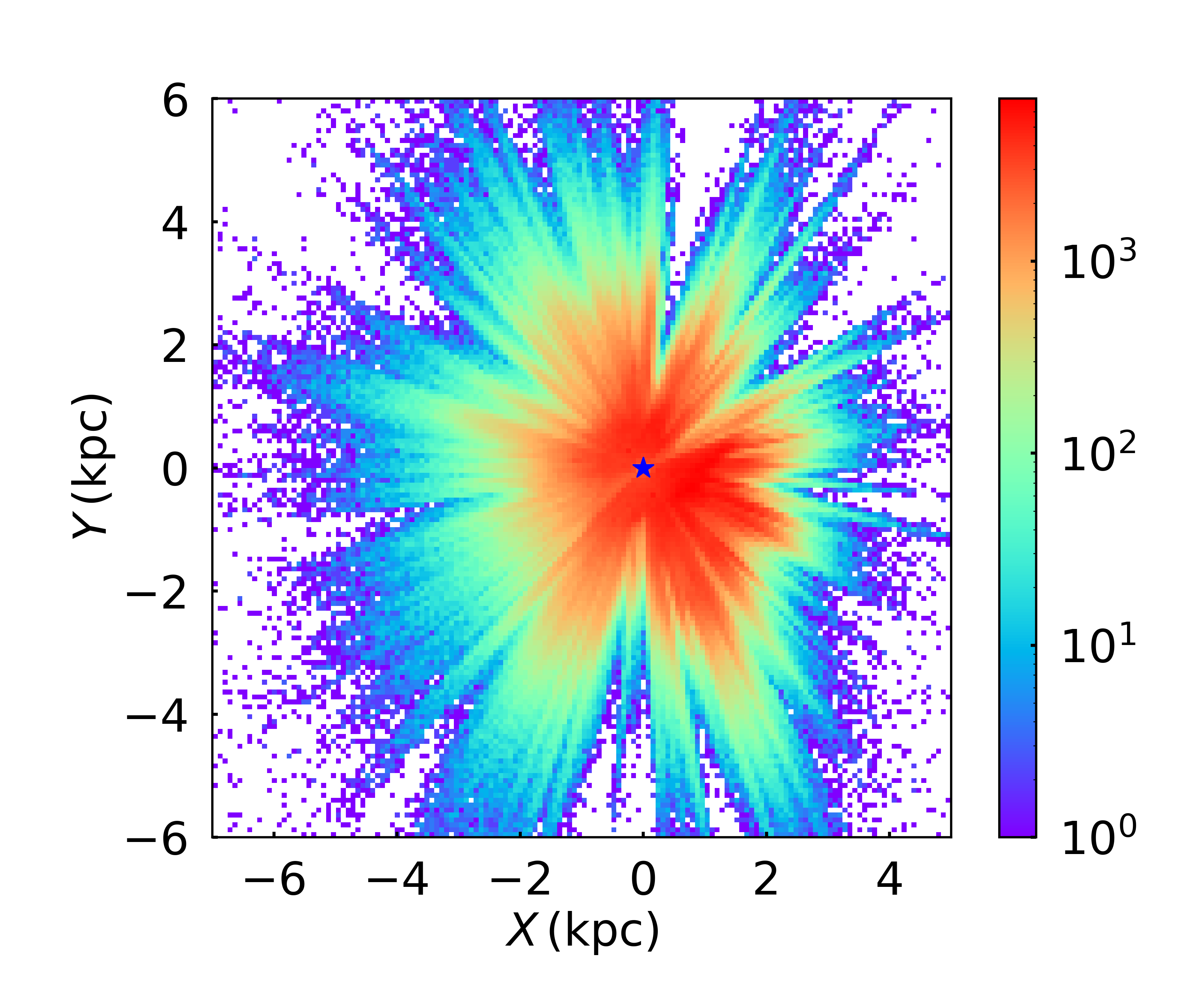}   
    \includegraphics[width=0.32\textwidth,keepaspectratio]{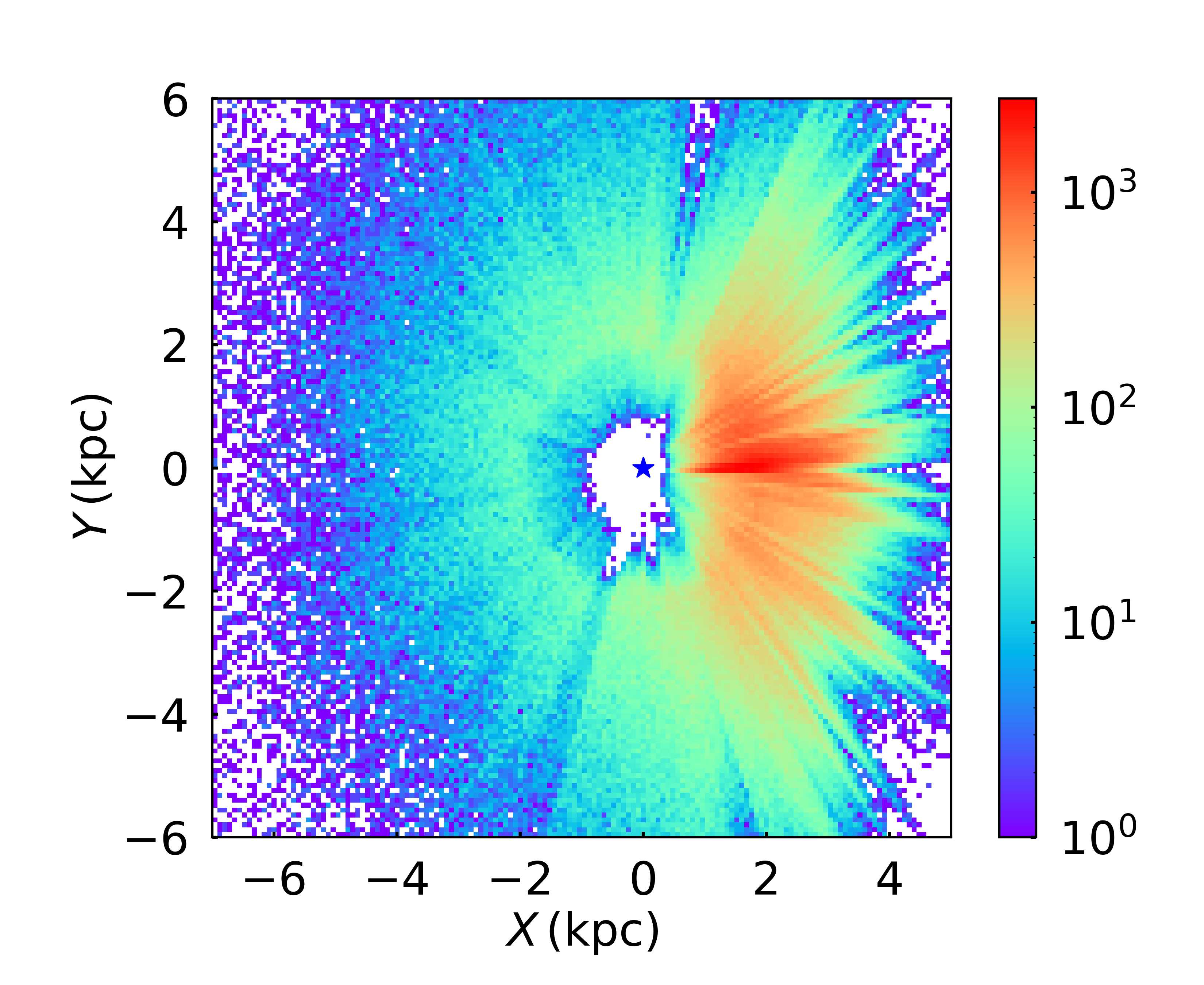}    
    \includegraphics[width=0.32\textwidth,keepaspectratio]{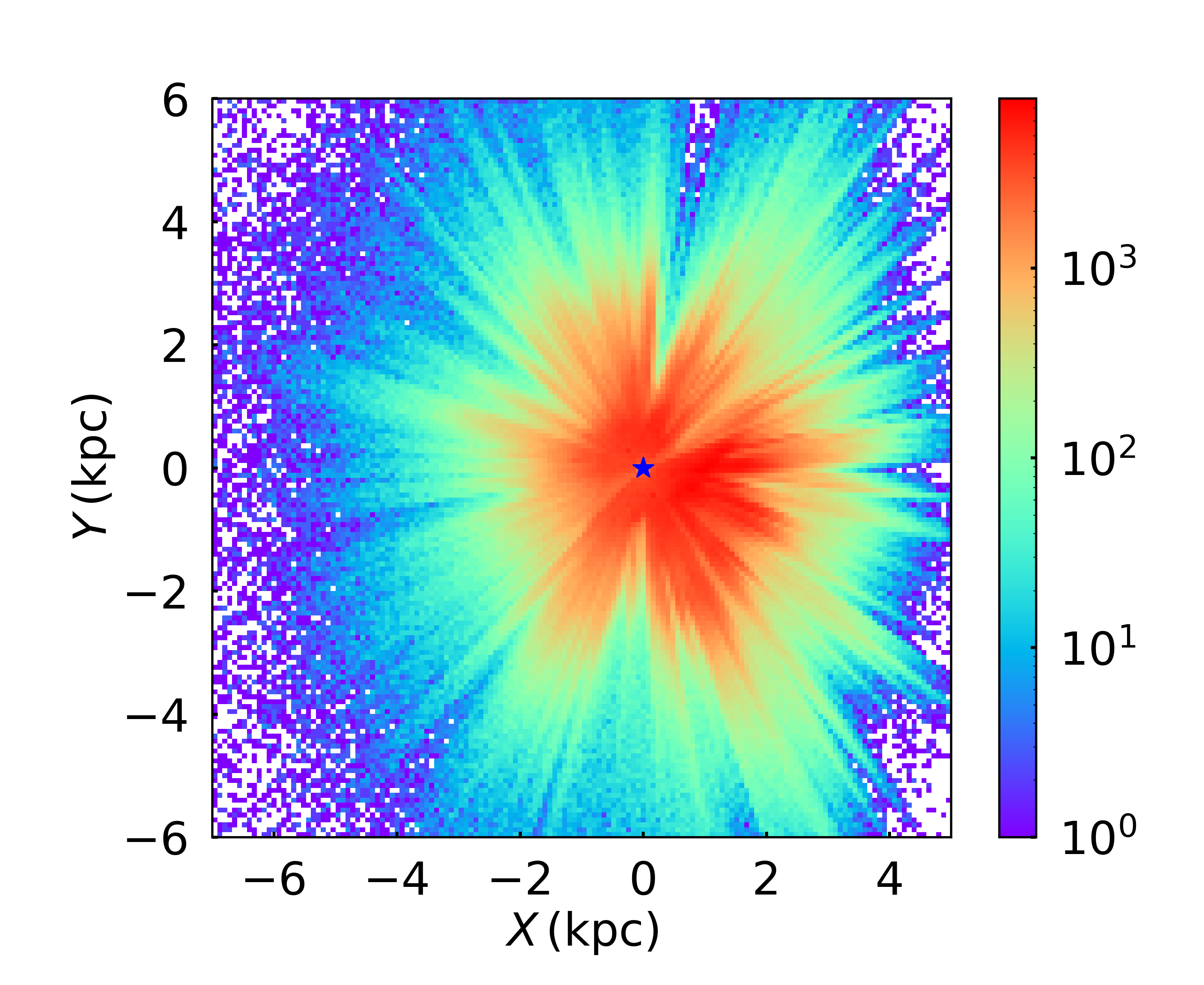}
    \caption{Spatial density distribution of stellar samples in Galactic plane. Left panel: the \textit{Gaia} BP/RP stellar sample from \citet{Zhang2025} in the Galactic plane. Middle panel: our RC sample combined with that of \citet{Lucey2020} in the Galactic plane. Right panel: the final adopted Galactic plane stellar sample, combining the \textit{Gaia} BP/RP stars from the left panel and the RCs from the middle panel to achieve a complete coverage from the solar neighborhood to distant regions. The Sun is located at ($X=0$ kpc, $Y=0$ kpc), marked by a blue star, and the Galactic center is located to the right of the Sun.}
  \label{fig:RC_XY}
\end{figure*}

\begin{figure}
    \centering
\includegraphics[width=0.45\textwidth,keepaspectratio]{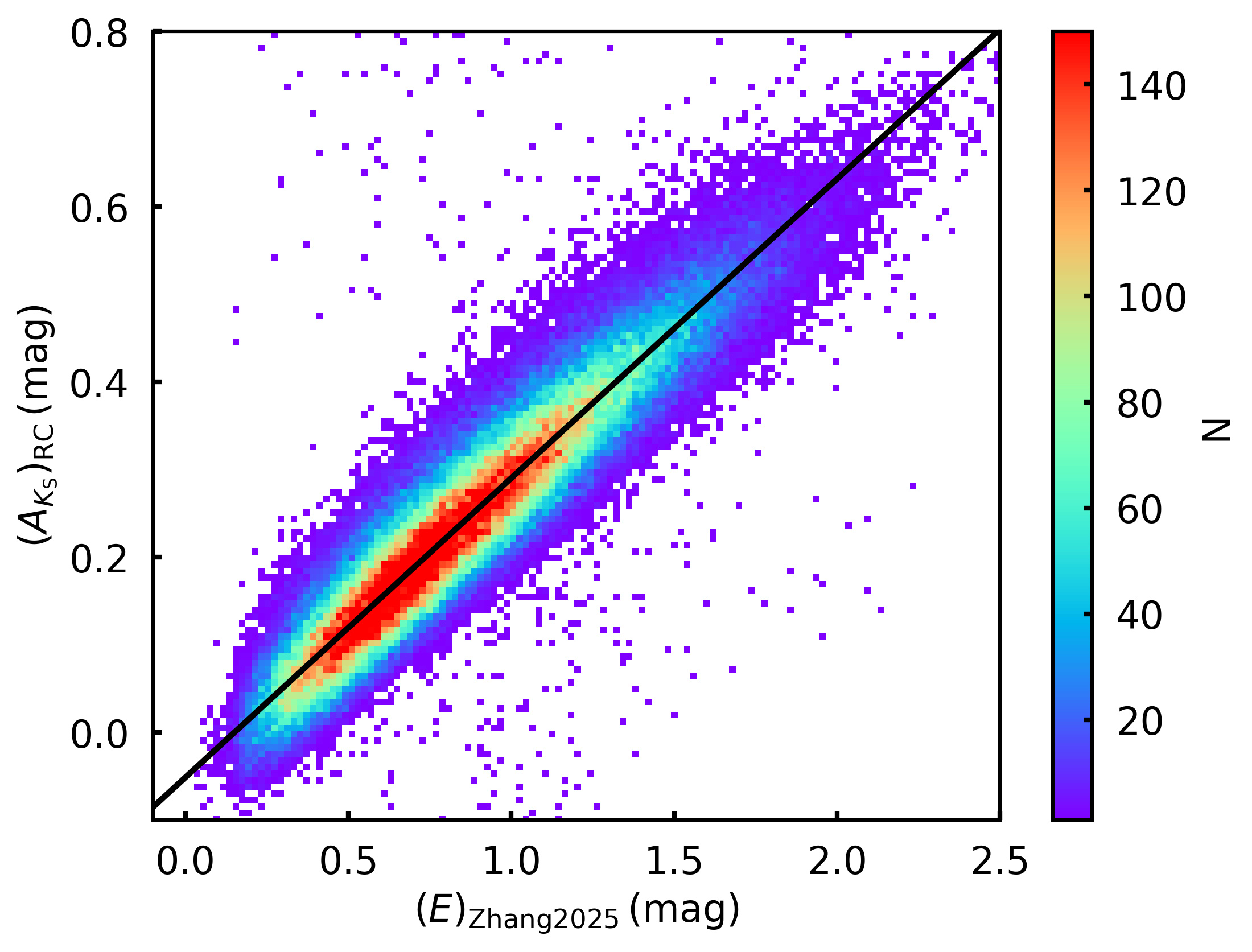}
  \caption{Comparison of $A_{K\rm_S}$ derived from RC candidates with the reddening $E$ from \citet{Zhang2025}. The black line shows the linear fit.}
  \label{fig:magnitude_correct}
\end{figure}

\section{Methodology for Deriving the Dust Distribution}

\subsection{Galactic Plane Stellar Sample} \label{sec:Extinction sample}

Because this work focuses on the Galactic plane ({$|Z|<25$}\,pc), we further restrict the RC candidates to those with a vertical distance $|Z|<25$\,pc from the Galactic plane for subsequent analysis. This constraint reduces the RC sample to 1,367,400 candidates. As a complement, we introduce two literature catalogs: the RC catalog from \citet{Lucey2020}, which contains 2.6 million RCs derived from 2MASS, AllWISE, \textit{Gaia}, and Pan-STARRS using inferred $\Delta P$, $\Delta \nu$, $T_{\rm eff}$, and $\log g$; and the stellar catalog from \citet{Zhang2025}, which uses a forward model to simultaneously fit stellar atmospheric parameters, parallax, extinction, and the extinction curve shape from Gaia BP/RP spectra augmented by 2MASS and WISE photometry. The model parallax is jointly constrained by both the Gaia DR3 parallax and the observed flux (which scales as $\varpi^2$). We adopt the model parallaxes and extinctions from this catalog, converting parallaxes to distances as $d = 1/\varpi$.

To construct a Galactic plane stellar sample with high reliability, we apply quality cuts to the \textit{Gaia} BP/RP stars from \citet{Zhang2025} to ensure reliable extinction measurements. Based on the basic reliability cuts recommended in \citet{Zhang2023} and the additional quality cuts described therein, we adopt the following criteria to remove outliers: \texttt{quality\_flags $>$ 8}, \texttt{teff\_confidence $<$ 0.2}, $E > 10$, $\sigma_{E} > 0.1$, and $\sigma_{\varpi}/\varpi > 0.2$.Here, $E$ is the extinction amplitude parameter in the forward model of \citet{Zhang2023,Zhang2025}. Their model predicts the observed flux as $f_\mathrm{pred} \propto \exp[-E \cdot R(\lambda)]$, where $R(\lambda)$ is the extinction curve learned from the data. $E$ is proportional to the total line-of-sight extinction, roughly corresponding to $E(B-V)$ in scale, but is determined jointly with stellar parameters ($T_{\rm eff}$, $\log g$, [Fe/H]), parallax, and the extinction curve shape from the BP/RP spectrum and NIR photometry rather than being derived from a color excess.For the RC candidates from \citet{Lucey2020}, we compute their extinction and distance using Eq.~(1). We further compare the inferred RC distance with the distance derived from \textit{Gaia} parallaxes to remove giants and dwarfs. Finally, only stars with $|Z|<25$\,pc are retained for further analysis. Our final sample contains over 6.36 million \textit{Gaia} BP/RP stars and 1.37 million RCs. The spatial distributions of the \textit{Gaia} BP/RP stars, the RC sample, and our final sample are shown in the left, middle, and right panels of Fig.~\ref{fig:RC_XY}, respectively. As shown in the figure, toward the Galactic center, the \citet{Zhang2025} sample reaches reliable distances of $\sim$3\,kpc, whereas the RC sample reaches $\sim$5\,kpc. Likewise, toward the anti-center direction, the \citet{Zhang2025} sample reaches reliable distances of $\sim$4--5\,kpc, while the RC sample extends to $\sim$7\,kpc. Overall, by using RCs as extinction tracers, we extend the reliable distance range by approximately 2--3\,kpc compared to that based on the \citet{Zhang2025} sample.

For RC stars, Equation~(1) yields the extinction in the $K_{\rm S}$ band ($A_{K\rm_S}$). For the \textit{Gaia} BP/RP stars, we convert the reddening $E$ given by \citet{Zhang2025} to $A_{K\rm_S}$. Cross-matching the 1.37 million RC stars with the 6.36 million Gaia BP/RP stars yields common sources. Fig.~\ref{fig:magnitude_correct} compares the extinction values $A_{K\rm_S}$ obtained for these common sources as RC candidates with the reddening values $E$ derived from \textit{Gaia} BP/RP spectra. A clear linear relation is seen between the two quantities, with the slope representing the extinction coefficient that relates the $K_{\rm S}$ band to the reddening $E$. A linear fit yields the scaling relation $(A_{K\rm_S})_{\rm RC} = 0.34\times(E)_{\rm Zhang2025}-0.05$. The negative intercept has two causes. The primary one is that \citet{Zhang2025} constrains $E\ge0$, which systematically inflates $E$ at the low-reddening end, whereas our $A_{K\rm_S}$ allows negative values when $(J-K_{\rm S}) < (J-K_{\rm S})_0$, which we retain. This asymmetry shifts the zero-extinction cloud to $E>0$, $A_{K\rm_S}\approx0$, producing the negative intercept. A secondary contribution may come from a small systematic offset in our adopted intrinsic RC color $(J-K_{\rm S})_0 = 0.7$ relative to the population mean used by \citet{Zhang2025}. In Fig.~\ref{fig:magnitude_correct}, the negative $A_{K\rm_S}$ points is visible. It is permitted for individual stars. However, when constructing extinction maps, any pixel with a negative median is assigned a floor value of $10^{-6}$\,mag. We apply the scaling relation to all stars in the \citet{Zhang2025} sample to obtain their extinction $A_{K\rm_S}$. Using the distance and extinction information of the final sample, we derive the integrated line-of-sight (LOS) extinction $A_{K\rm_S}$ distribution across the Galactic plane. Inverting these extinction distributions then yields the corresponding dust density distributions in the plane.

\subsection{Dust Density Inversion} \label{sec:U-Net}

\begin{figure*}
    \centering
    \includegraphics[width=0.93\textwidth,keepaspectratio]{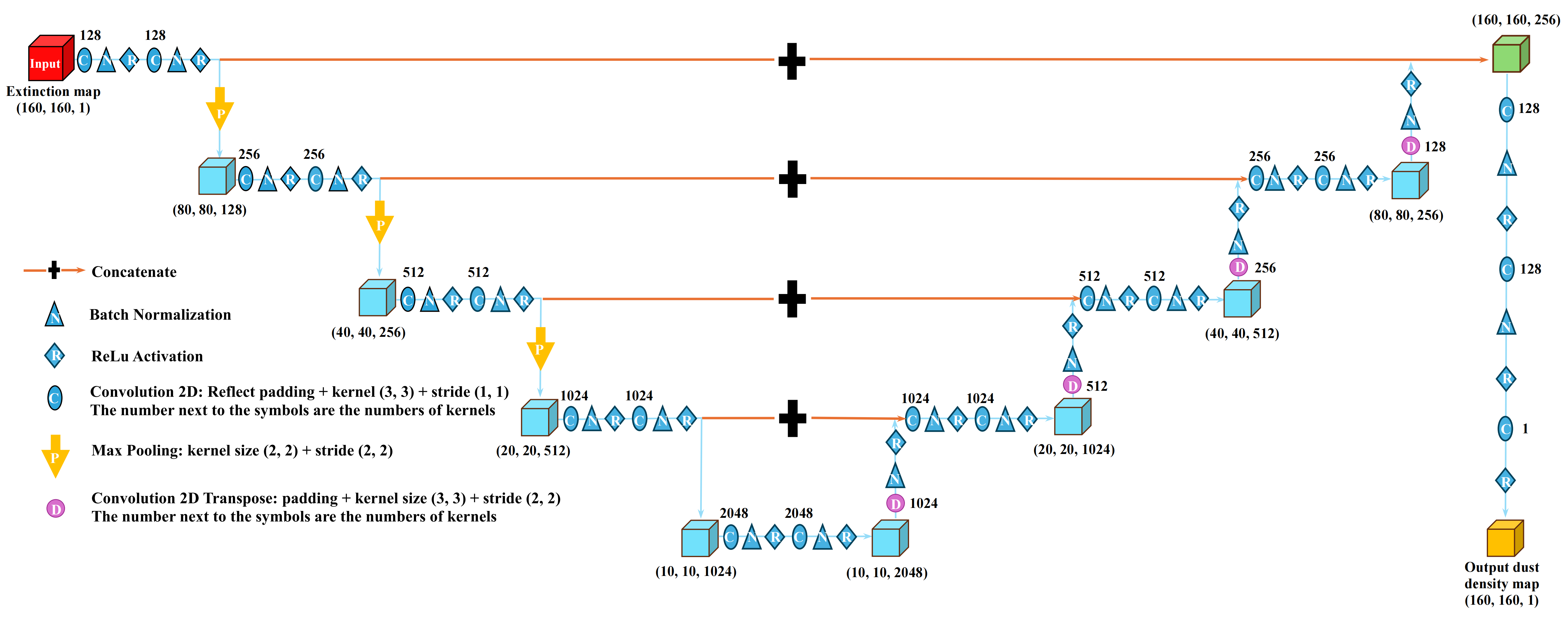}
  \caption{Schematic of the U-Net architecture, shown with the MR configuration as an example.}
  \label{fig:U-net}
\end{figure*}

In this work, we adopt a convolutional neural network (CNN) approach similar to that of \citet{Chen2024} to invert the integrated LOS extinction distribution into a dust density distribution. Unlike \citet{Chen2024}, we focus only on the dust density distribution in the region $|Z|<25$\,pc in the Galactic plane, which allows us to ignore the $Z$ direction and reduce the 3D dust density inversion problem in $XYZ$ space to a two-dimensional problem in $XY$. We therefore adopt the U-Net framework. U-Net is a CNN architecture designed for efficient and accurate image segmentation. Originally introduced by \citet{Ronneberger2015}, U-Net is specifically developed for analyzing two-dimensional maps. The network architecture consists of an encoding stage and a decoding stage. In the encoding stage, the machine increases the number of feature channels while reducing the map size per channel to extract features from the input map. In the decoding stage, it decreases the number of feature channels while increasing the map size per channel to reconstruct the output map. To prevent the loss of small-scale features, U-Net concatenates the outputs from the decoding stage with the inputs from the encoding stage. The U-Net architecture used in this work follows \citet{Qin2023}, as illustrated in Fig.~\ref{fig:U-net}.

\begin{figure*}
    \centering
\includegraphics[width=\textwidth,keepaspectratio]{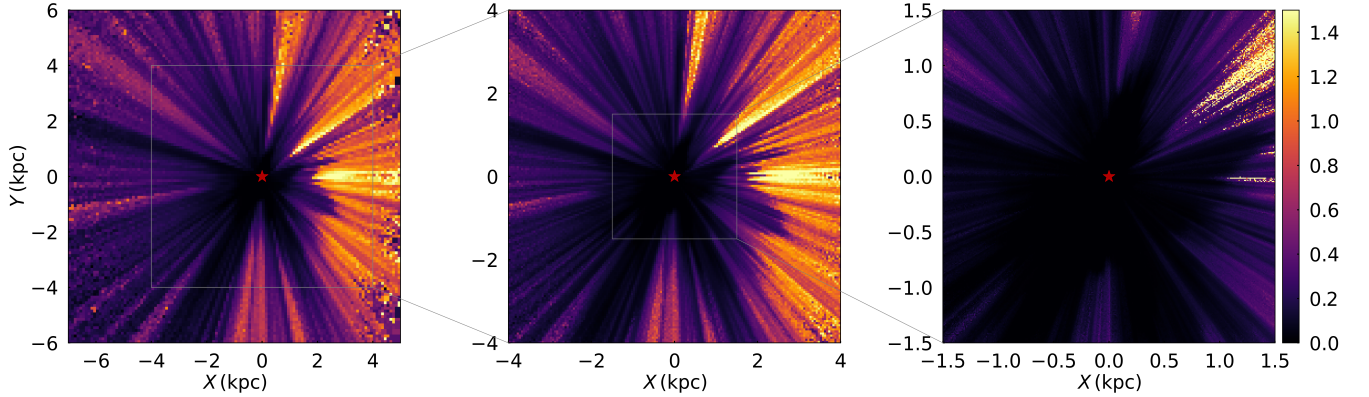}
  \caption{LOS extinction maps of the Galactic plane in the LR (left panel), MR (middle panel), and SR (right panel) fields. The Sun is located at the center ($X=0$\,kpc, $Y=0$\,kpc). The Galactic center lies to the right. From left to right, the panels show the median extinction values of the individual pixels with distance resolutions of 100, 50, and 10\,pc, respectively.}
  \label{fig:ext_maps}
\end{figure*}

Considering the trade-off between computational cost and resolution, we divide the Galactic plane into three regions for dust density reconstruction: a small region (SR) defined as $-1.5 \leq X \leq 1.5$\,kpc and $-1.5 \leq Y \leq 1.5$\,kpc; a medium region (MR) defined as $-4 \leq X \leq 4$\,kpc and $-4 \leq Y \leq 4$\,kpc; and a large region (LR) defined as $-7 \leq X \leq 5$\,kpc and $-6 \leq Y \leq 6$\,kpc. The distance resolutions for SR, MR, and LR are set to 10\,pc, 50\,pc, and 100\,pc, respectively. We then compute the median extinction per bin to construct the LOS extinction maps. Fig.~\ref{fig:ext_maps} shows the resulting LOS extinction maps for the three regions. For pixels with insufficient stellar coverage, especially in distant subfields, we fill the corresponding pixels in the extinction maps using values from maps with substantially coarser binning (up to 800\,pc for LR). 

Training the U-Net model requires a large number of training samples. In this work, we adopt an approach similar to that of \citet{Chen2024}. Training samples are constructed by generating simulated dust density distributions and integrating along the line of sight to obtain extinction distributions. We simulate the dust distribution assuming a logarithmic density profile that follows a Gaussian process as a function of position. For SR, MR, and LR, we generate 10,000 simulated dust density maps each. We then integrate each dust density map along the LOS to obtain the corresponding integrated LOS extinction map. 

Following \citet{Chen2024}, we normalize the integrated extinction and the dust density to bring them to similar orders of magnitude before training. For each region, we randomly select 1,000 simulated maps for testing. The remaining 9,000 simulated maps are split into two subsets: 7,200 maps are used for training the U-Net model, and the remaining 1,800 maps are used for validation.

\begin{figure*}
    \centering
    \includegraphics[height=0.93\textwidth,keepaspectratio]{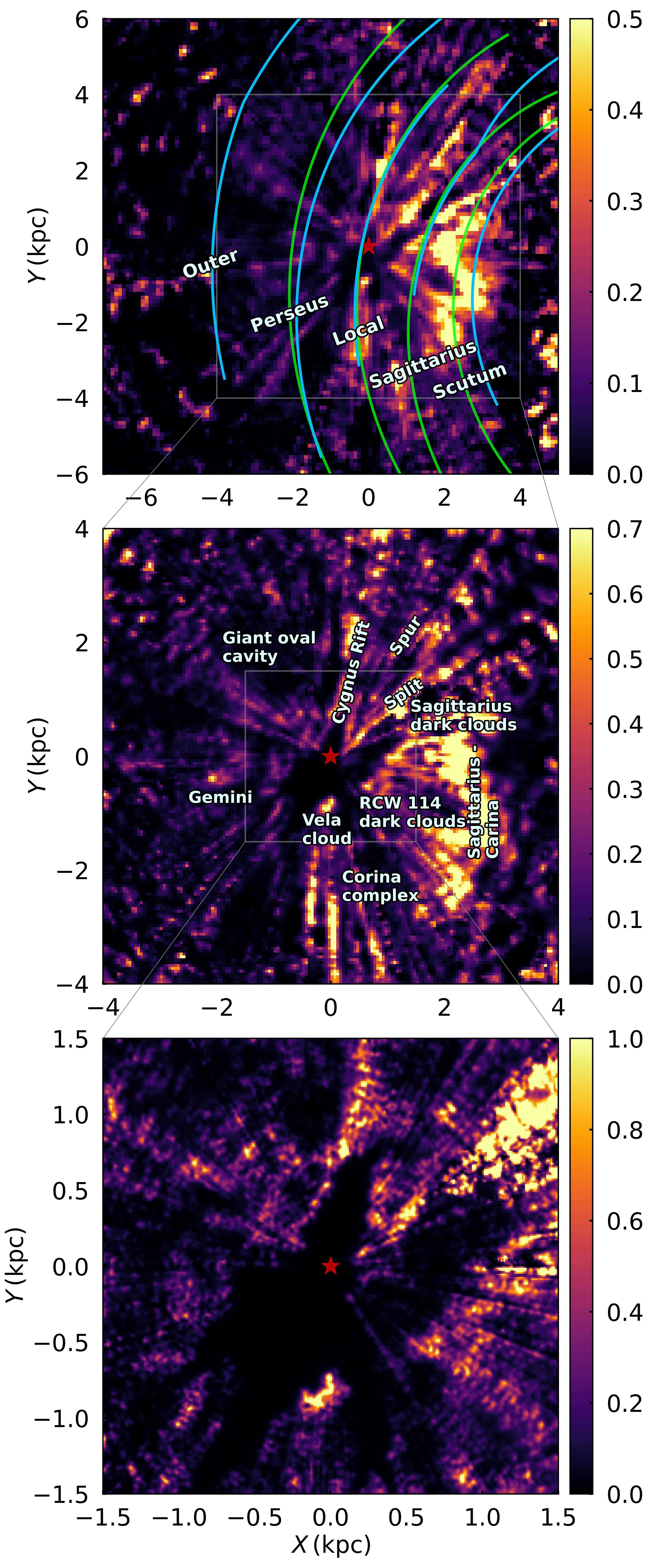}
  \caption{Dust density distribution in the Galactic plane ({$|Z|<25$}\,pc) from the U-Net model. The color scales represent the differential extinction $\delta A_{K\rm_S}$. The Sun is located at the center ($X=0$\,kpc, $Y=0$\,kpc), marked by a red star symbol. The Galactic center lies to the right of the figure. From top to bottom, the panels show the dust distribution in LR, MR, and SR, with distance resolutions of 100\,pc, 50\,pc, and 10\,pc, respectively. In the top panel, the green and  blue curves delineate the best-fit spiral arm models from \citet{Chen2019_OB} and \citet{Reid2019}, respectively, with arms labeled from left to right: Outer Arm, Perseus Arm, Local Arm, Sagittarius Arm, and Scutum Arm. In the middle panel, we mark known large-scale interstellar dust structures from the literature. } 
  \label{fig:dust_maps}
\end{figure*}

\section{Results and Discussion} \label{sec:Results}

\subsection{Dust Distribution in the Galactic Plane} \label{sec:dust distribution}

\begin{figure*}[t!]
    \centering
    \includegraphics[width=0.32\textwidth,keepaspectratio]{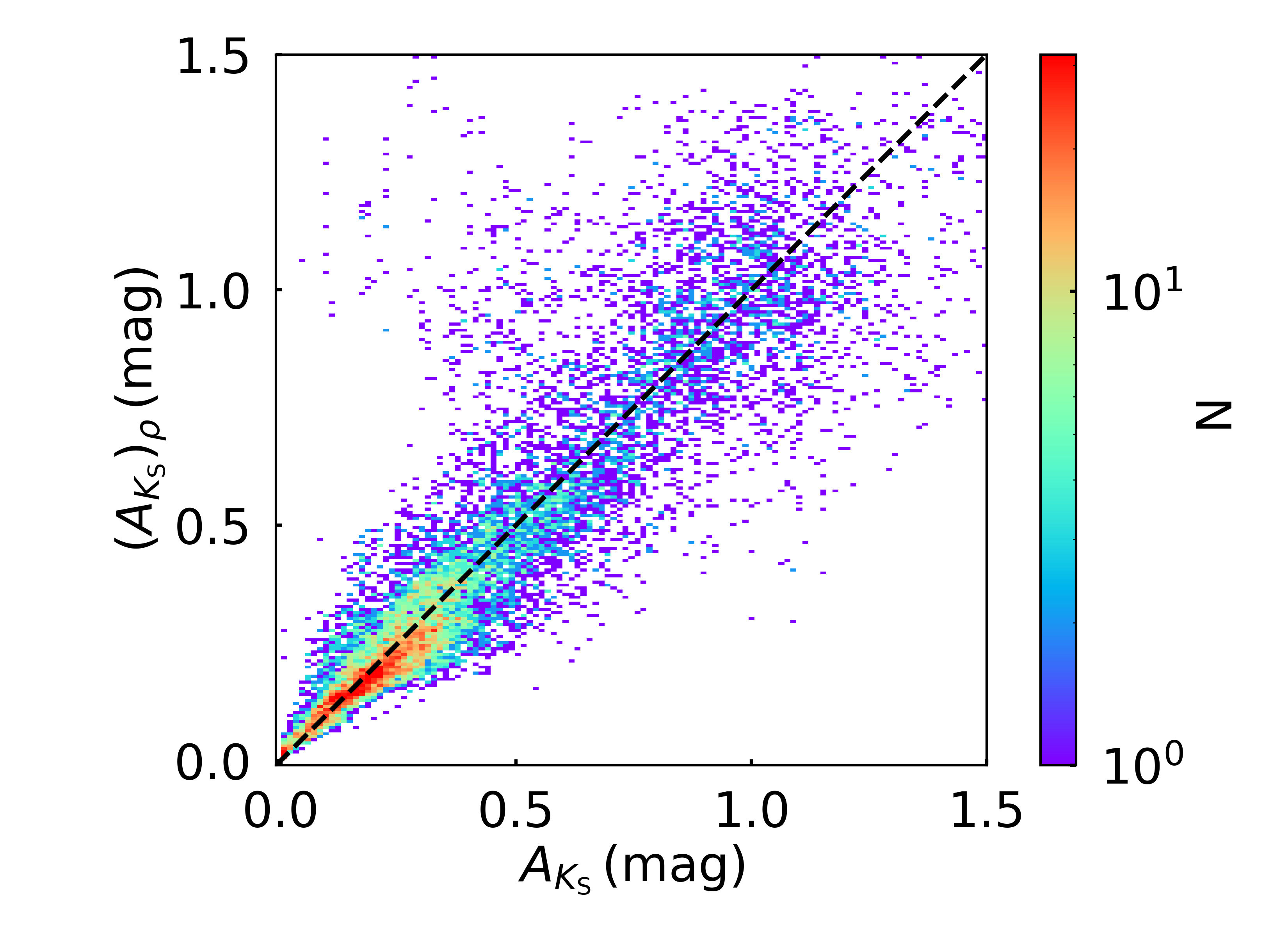}
    \includegraphics[width=0.32\textwidth,keepaspectratio]{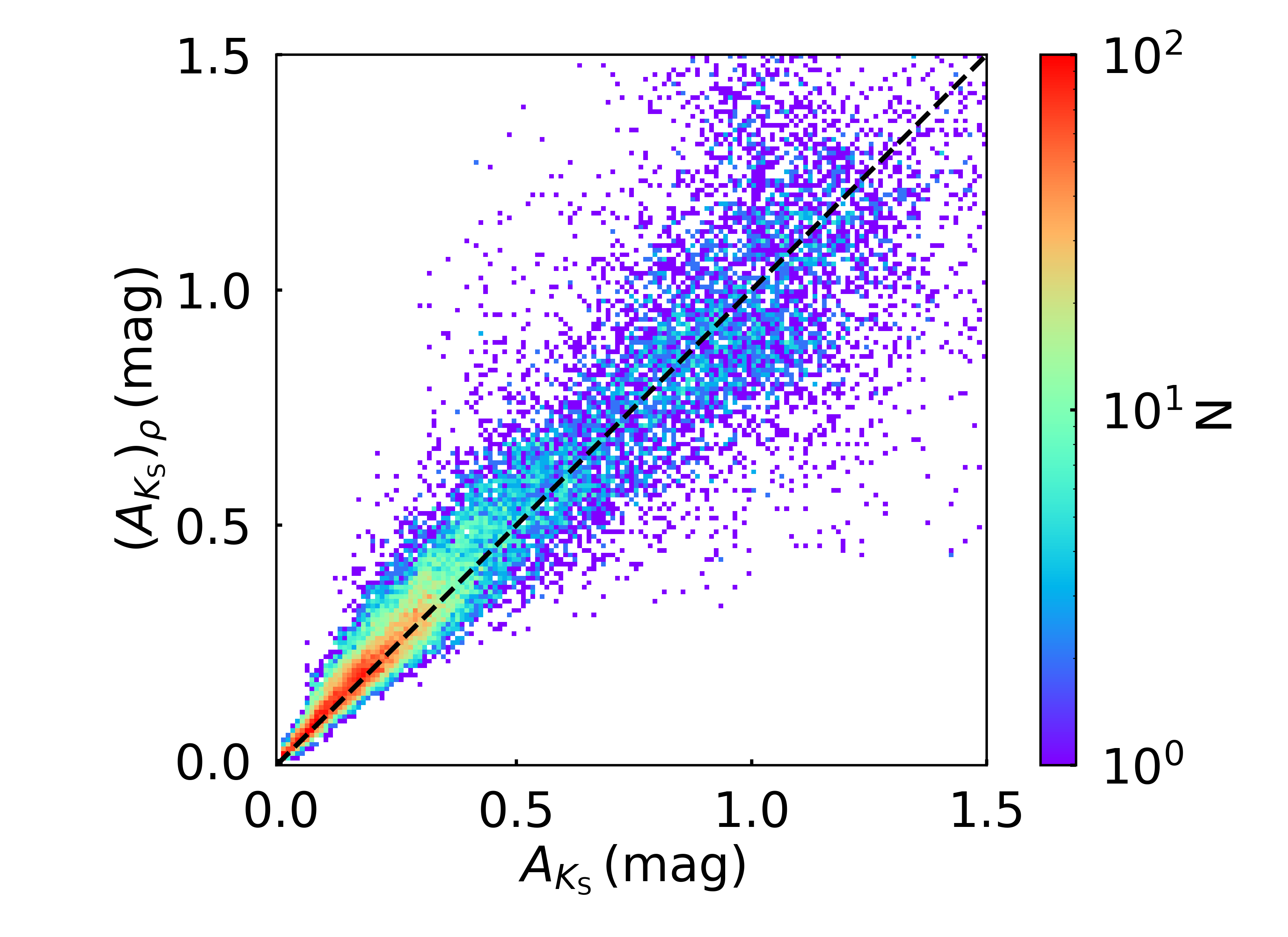}
    \includegraphics[width=0.32\textwidth,keepaspectratio]{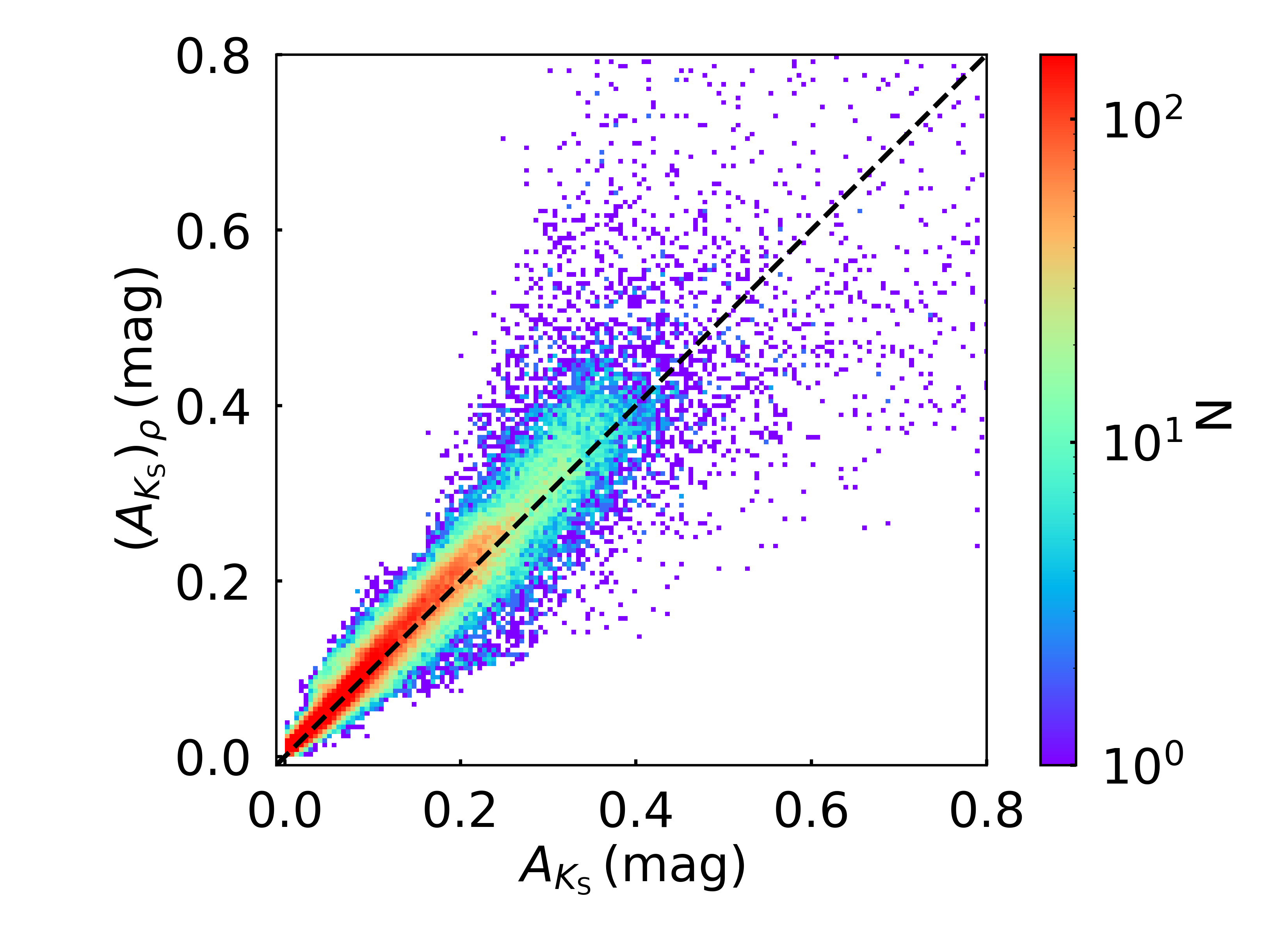}
  \caption{Comparison of $A_{K\rm_S}$ from sample stars with $({A_{K\rm_S}})_{\rho}$, the LOS  extinction integrated from the derived dust density maps, for LR (left), MR (middle), and SR (right). The black dashed line indicates the 1:1 relation.}
  \label{fig:true_data_pred_integral_com}
\end{figure*}

\begin{figure*}[t!]
    \centering
    \includegraphics[width=0.32\textwidth,keepaspectratio]{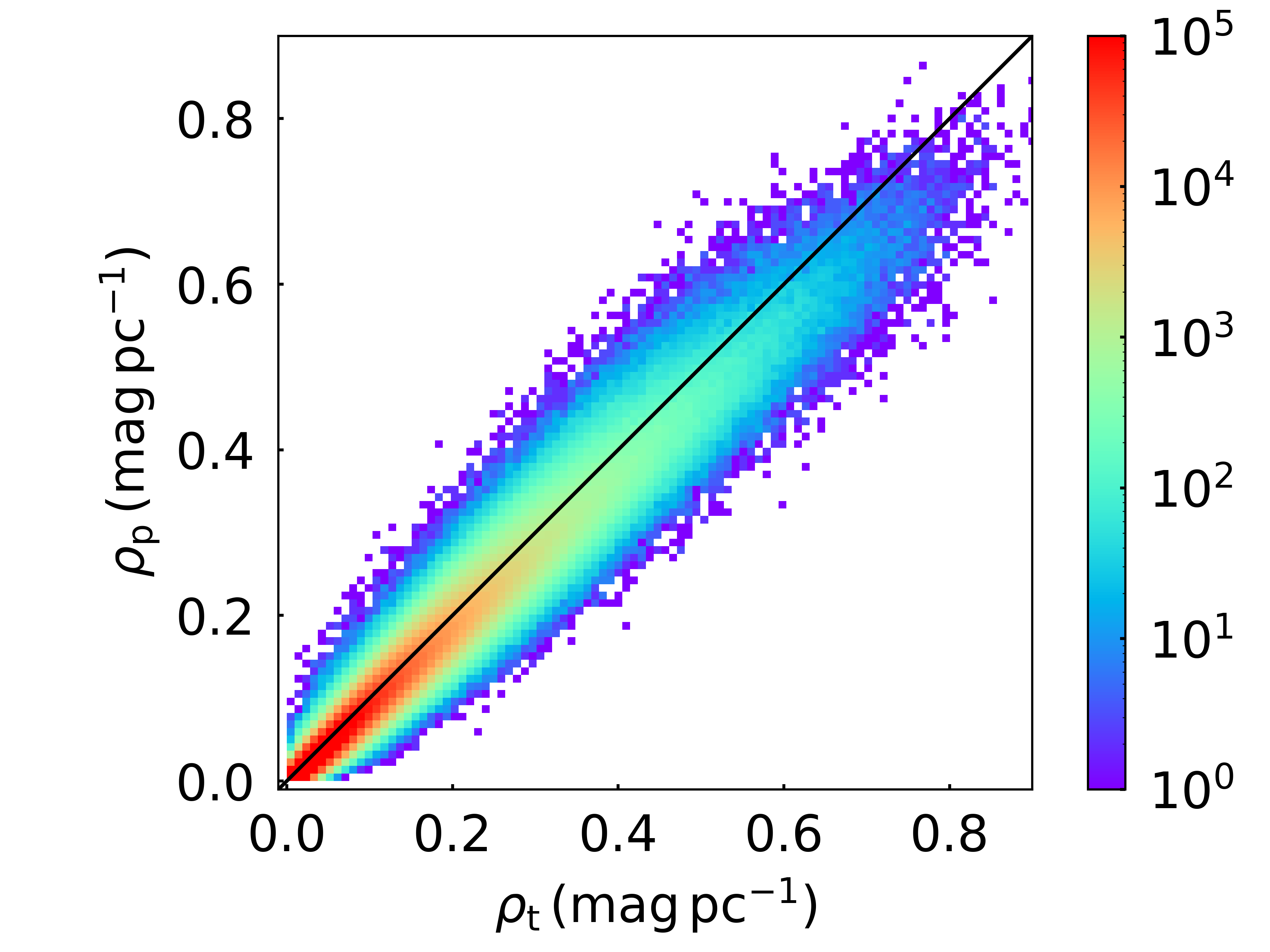}
    \includegraphics[width=0.32\textwidth,keepaspectratio]{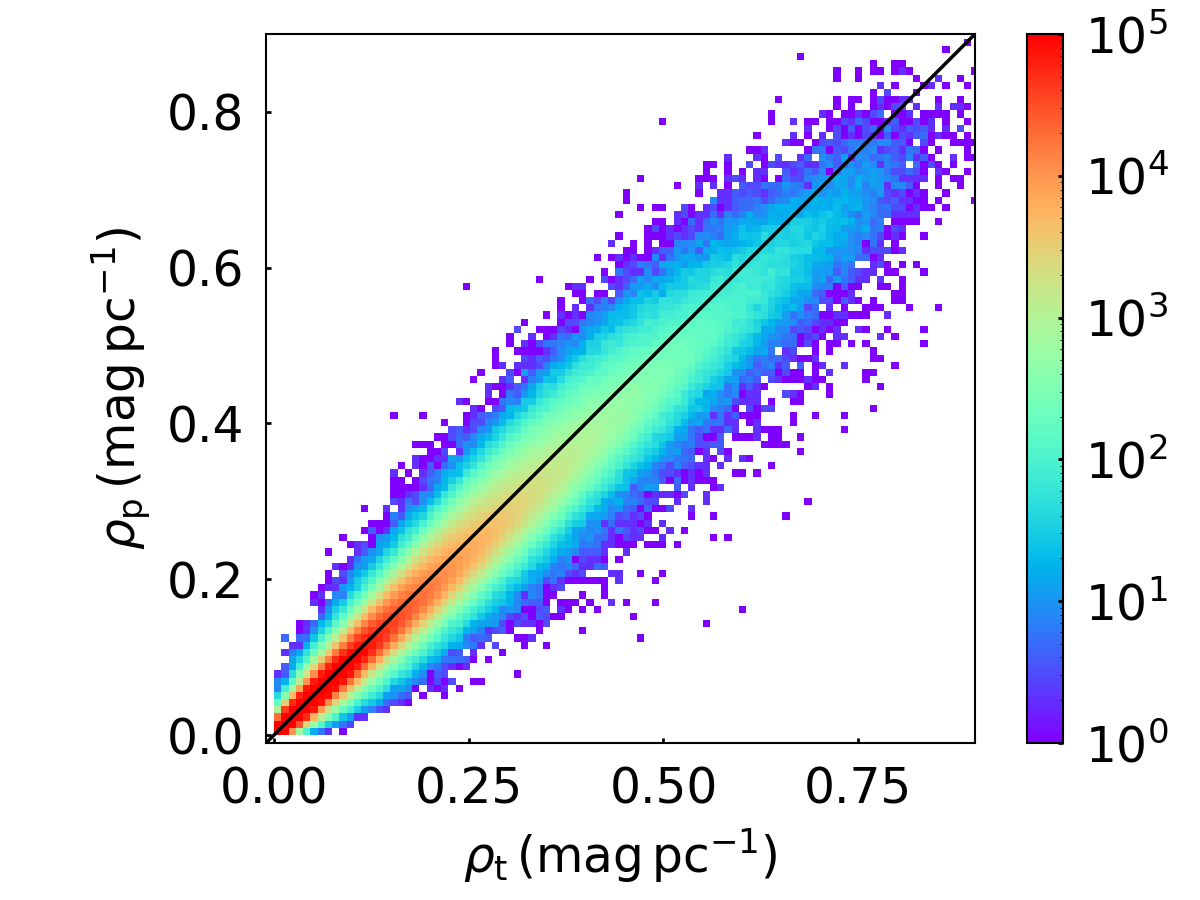}
    \includegraphics[width=0.32\textwidth,keepaspectratio]{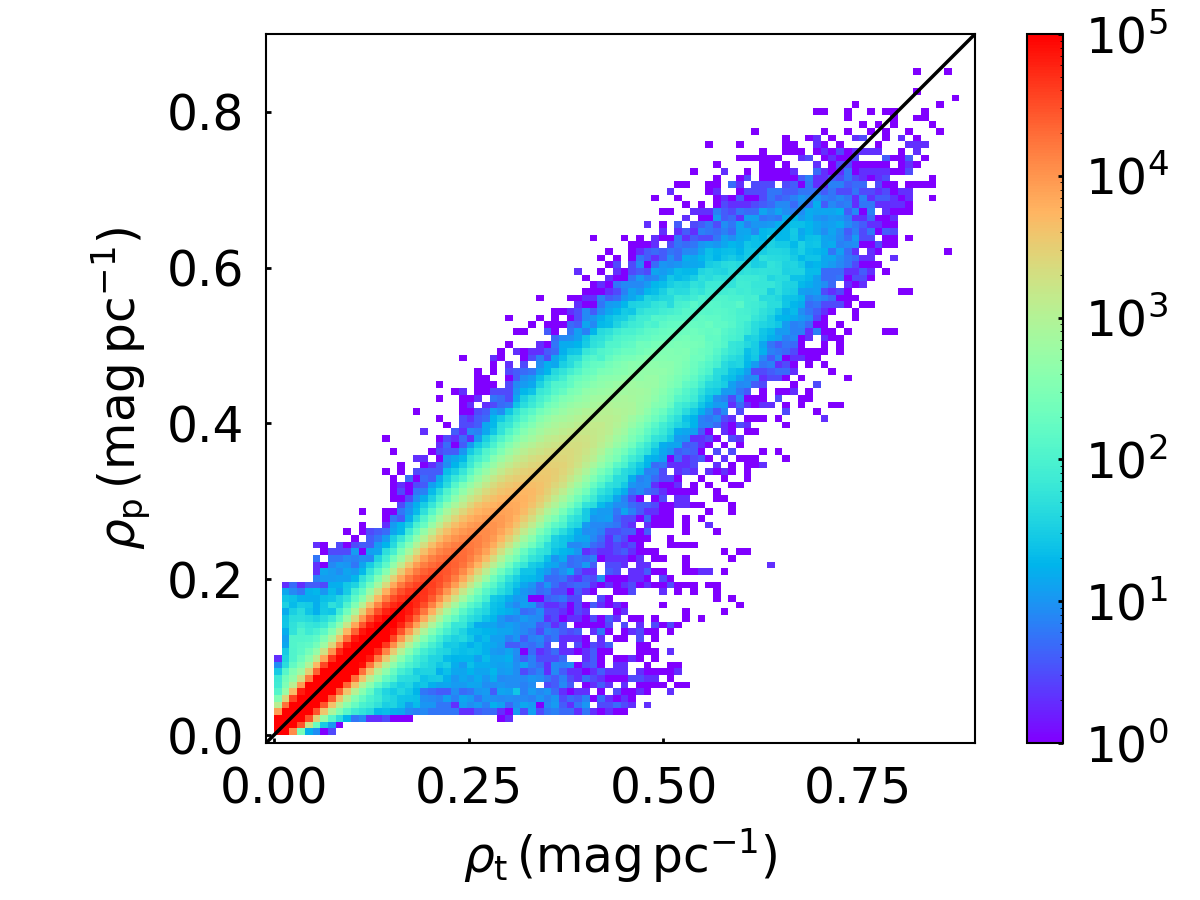}
  \caption{Comparison of true and predicted dust densities for the 1,800 validation maps in the LR (left), MR (middle), and SR (right) regions. The black line marks the 1:1 relation.}
  \label{fig:validation_dataset}
\end{figure*}

We apply the trained U-Net model to our extinction maps to obtain dust density maps of the Galactic plane. The derived dust density maps are publicly available at \url{https://doi.org/10.12149/101857}. Fig.~\ref{fig:dust_maps} presents the dust density maps for the LR, MR and SR fields from top to bottom. Integrating the resulting dust density maps along the LOS yields integrated extinction maps. Fig.~\ref{fig:true_data_pred_integral_com} compares the extinction derived from sample stars with the LOS extinction integrated from the generated dust density maps. For all three regions, the comparison between $A_{K\rm_S}$ and $(A_{K\rm_S})_{\rho}$ shows good agreement, confirming the reliability of the derived dust density maps. Fig.~\ref{fig:validation_dataset} shows the comparison between the true (input) and the U-Net-predicted dust densities for the 1,800 validation maps in the LR, MR, and SR regions. The predicted densities align well with the true values in all three regions, with no evidence of bias.

\begin{figure*}[t!]
    \centering
    \includegraphics[width=0.32\textwidth,keepaspectratio]{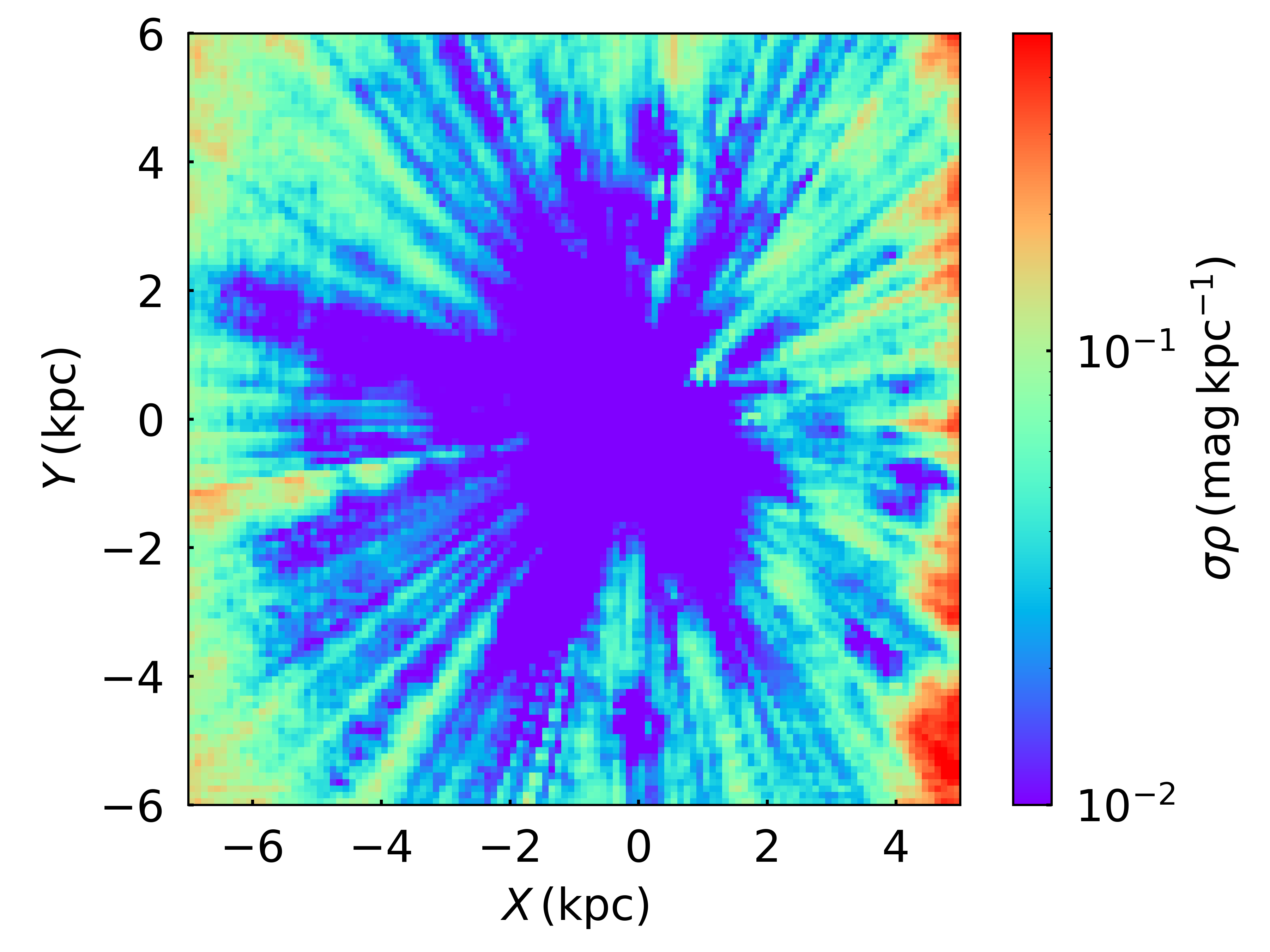}
    \includegraphics[width=0.32\textwidth,keepaspectratio]{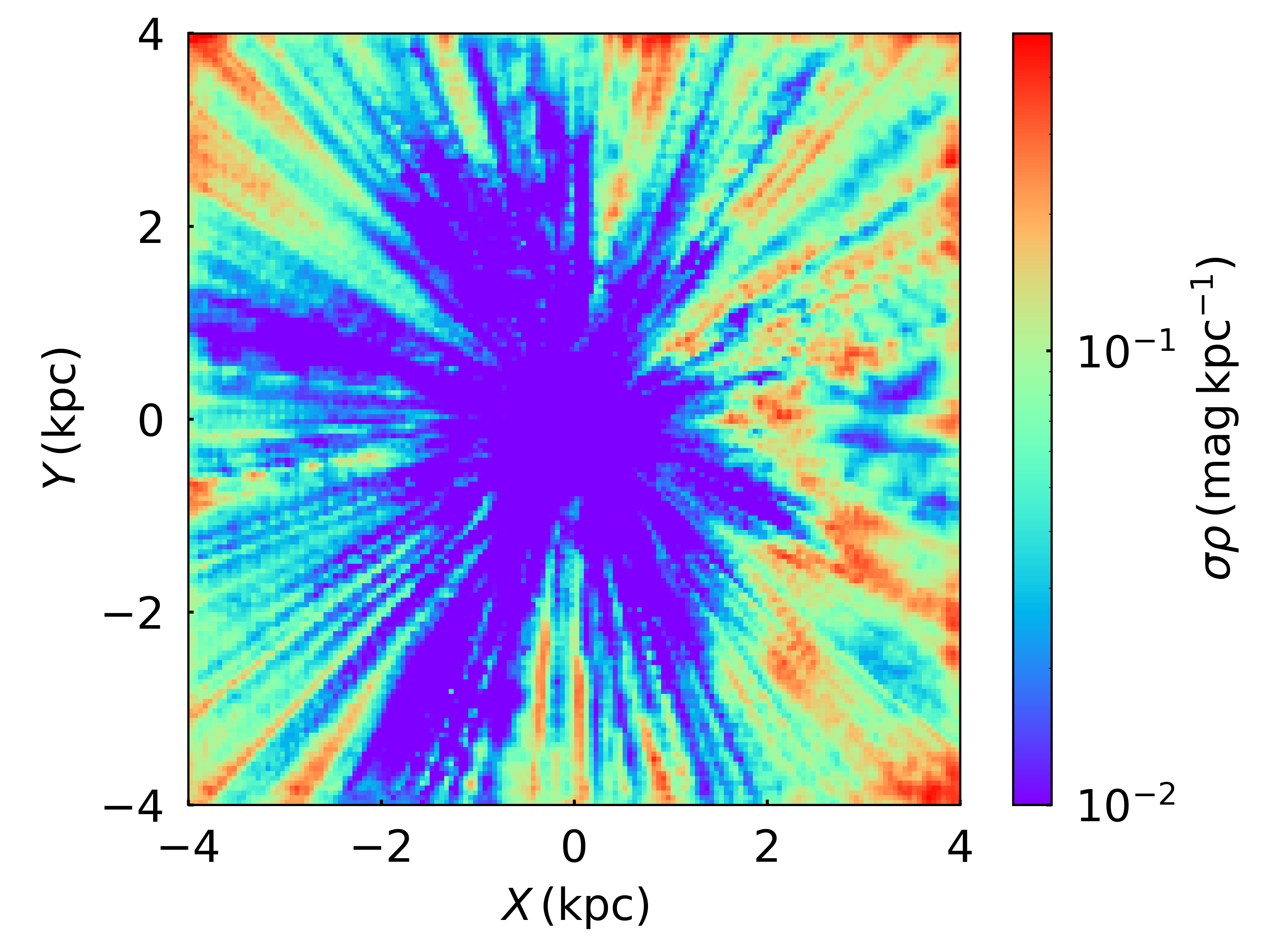}
    \includegraphics[width=0.32\textwidth,keepaspectratio]{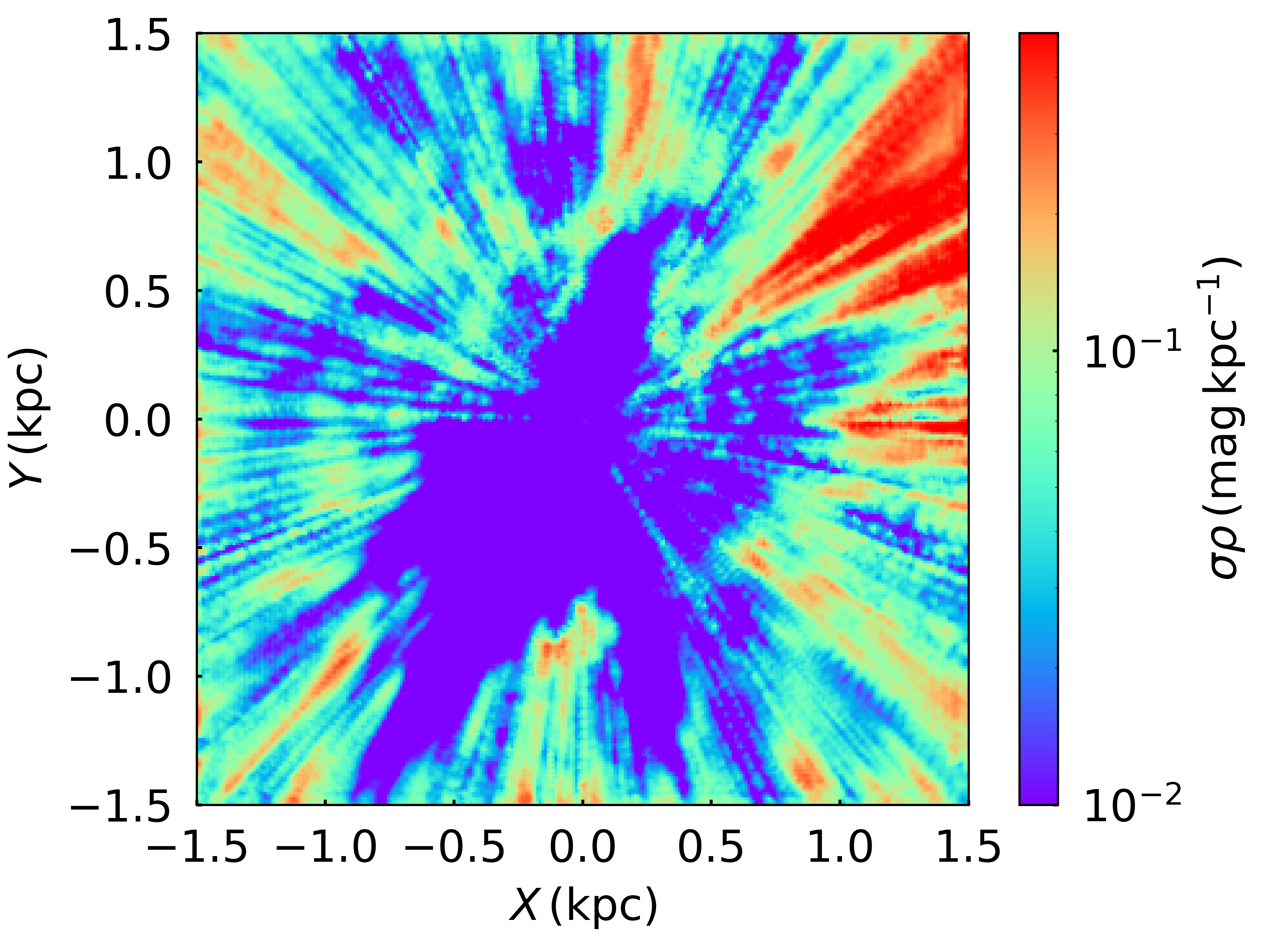}
  \caption{Uncertainties of the derived dust density maps for the LR (left), MR (middle) and SR (right) fields, respectively.}
  \label{fig:density_error}
\end{figure*}

The U-Net model does not provide uncertainty estimates. To obtain the errors of the generated dust density maps, we adopt the same method as \citet{Chen2024}. For each region, we apply the following bootstrap procedure. Each star has a measured extinction $A_{K\rm_S}$ and distance $d$, with associated uncertainties $\sigma_{A}$ and $\sigma_{d}$. For each of 300 realizations, we draw for every star a new extinction value from $\mathcal{N}(A_{K\rm_S}, \sigma_{A})$ and a new distance from $\mathcal{N}(d, \sigma_{d})$, then reconstruct the LOS extinction map from these perturbed values. We then apply the trained U-Net model to these 300 extinction maps to obtain the corresponding dust density maps. Finally, we adopt the pixel-wise dispersion among the 300 dust density maps as the uncertainty of our derived dust density maps. The resulting density error distributions are shown in Fig.~\ref{fig:density_error}.

Our dust maps cover the Galactic plane over the full $360^{\circ}$ in longitude and extend to approximately 7\,kpc, as shown in the top panel of Fig.~\ref{fig:dust_maps}. This reaches a greater depth than most existing dust maps. The distant probes originate mainly from RC candidates, whose distances are estimated independently of \textit{Gaia} parallaxes and rely on the properties of RC stars, allowing us to cover a larger region of the Galaxy. In addition, we use the near-infrared photometric data from VVV and UKIDSS, which have deeper detection limits and enable us to penetrate the Galactic plane more effectively. In the dust density map of the LR field, we observe large-scale structures in the dust density distribution. Prominent features such as spiral arms and bubble-like cavities are clearly visible. We note that, in addition to the well-known Local Bubble and Giant Oval Cavity \citep{Zucker2022, Vergely2022, Chen2025}, cavities spanning different spatial scales from a few hundred parsecs to several kiloparsecs are widely present in the Galaxy. This suggests that our Milky Way closely resembles the morphology of the so-called Phantom galaxy M74 as recently observed with JWST \citep{Barnes2023}. In the outermost parts of the LR map, our dust density map shows numerous clumpy dust structures, which are likely not real. The limited number of sample stars in these outer regions, combined with larger distance and extinction errors, leads to substantial uncertainties in the derived dust density maps.

The middle panel of Fig.~\ref{fig:dust_maps} shows the MR dust map. The resolution of this map is twice that of the top panel, allowing us to detect finer structures. We observe many elongated structures extending outward from the Sun. These structures are primarily caused by the ``Finger of God'' effect due to distance uncertainties. In this map, we see prominent cavities and kiloparsec-scale dust cloud structures. We mark several representative dust structures in the figure, including the Giant Oval Cavity, the Vela Cloud, the Carina Complex, and the Split \citep{Hottier2020,Vergely2022}. The most prominent feature is the Sagittarius-Carina Complex, which extends from the first quadrant to the fourth quadrant (from $l=30^\circ$ to $l=300^\circ$). Within this complex, we find an elliptical cavity previously reported by \citet{Hottier2020}. Molecular clouds located in front of this cavity and close to the Sun show lower extinction, while at a distance of about 2.5\,kpc from the Sun, we observe a compact region of dense dust. We also identify many giant cavities on scales of several kiloparsecs. The Giant Oval Cavity extends from $l=100^{\circ}$ to $l=150^{\circ}$, representing a large, long-lived, slowly expanding superbubble crossing the Perseus Arm \citep{Vergely2022,Chen2025}. In addition, we can clearly identify inter-arm spurs \citep{Dobbs2006, Lallement2019, Green2019}. The Split in the first quadrant extends from near 0\,kpc to beyond 2\,kpc.

The bottom panel of Fig.~\ref{fig:dust_maps} shows the highest-resolution dust map produced in this work, covering a $3\times3$\,kpc region. The distance resolution of 10\,pc allows us to distinguish structural features more clearly and identify nearby structures. The Local Bubble cavity at the center is clearly visible, and its boundaries can be easily traced in the dust map. In addition to the Local Bubble, many other cavities are present around the Sun. A structure composed of four such cavities, referred to as the butterfly by \citet{Hottier2020}, is also clearly identified in our dust map. Furthermore, we observe spurs extending outward from the center and connecting to the Sagittarius-Carina Complex in the first quadrant. In this highest-resolution map, the ``Finger of God'' effect is significantly reduced due to the improved distance accuracy.

\subsection{Spiral Arms}

In the top panel of Fig.~\ref{fig:dust_maps}, we overlay the spiral arm models from \citet{Reid2019}, based on maser sources, and from \citet{Chen2019_OB}, based on OB-type stars. Numerous dust cloud structures are clearly identifiable in our dust map and can be associated with the spiral arm models. 

The Outer Arm is likely traced by scattered dust clouds at a distance of about 4\,kpc from the Sun. These dust clouds enclose several superbubble structures. In the LR dust map, a few clumpy dust features are visible in the third quadrant (e.g., at $X\approx-4$\,kpc, $Y\approx-3$ to $-5$\,kpc), located beyond the Perseus Arm and at Galactocentric distances consistent with the extrapolation of the Outer Arm seen in the second quadrant. These structures are too weak and sparse to be identified as a definitive arm extension, but their positions and morphology are suggestive of an Outer Arm continuation into this quadrant.

The Perseus Arm appears to align with several discrete dust clouds. For example, the Gemini cloud \citep{Carpenter1995} is associated with the Perseus Arm. The Outer and Perseus Arms are associated with less compact clouds, whereas the Local, Sagittarius, and Scutum Arms correspond to regions of higher extinction. Multiple clouds associated with the Local Arm are clearly distinguishable and prominent, including the Cygnus Rift and the Vela Cloud. The Sagittarius Arm is associated with regions of dust overdensity, such as the Sagittarius Dark Cloud (approximately at $X\approx1.5$\,kpc, $Y\approx0.3$\,kpc) and the RCW 114 Dark Cloud (approximately at $X\approx1.0$\,kpc, $Y\approx-0.3$\,kpc) \citep{Wang2025_dust}. We also observe spurs between the Sagittarius and Local Arms, which serve as bridges connecting them. At a distance of approximately 2.5\,kpc to the right of the Sun, a large number of dense molecular clouds appear to be associated with the Scutum Arm. We cannot clearly distinguish the extinction region between the Sagittarius and Scutum Arms, as the high-extinction regions in this area all seem to be associated with the Sagittarius-Carina Complex. Furthermore, the Local, Sagittarius, and Scutum Arms extend into the fourth quadrant and are traced by several molecular clouds.

\subsection{Comparison with Previous Results} 
\label{sec:comepare to other map}

\begin{figure*}
    \centering
    \includegraphics[width=0.49\textwidth,keepaspectratio]{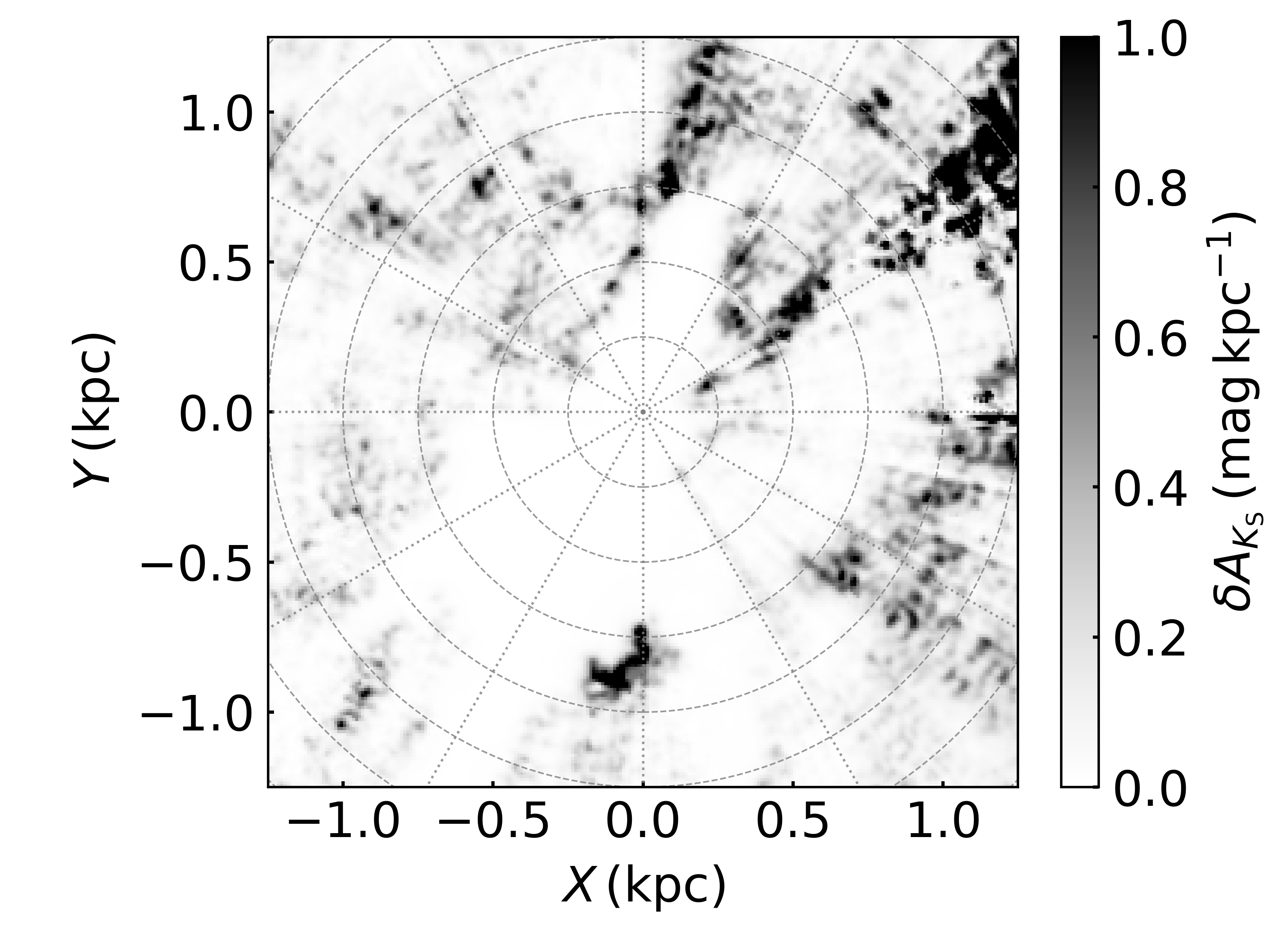}
    \includegraphics[width=0.49\textwidth,keepaspectratio]{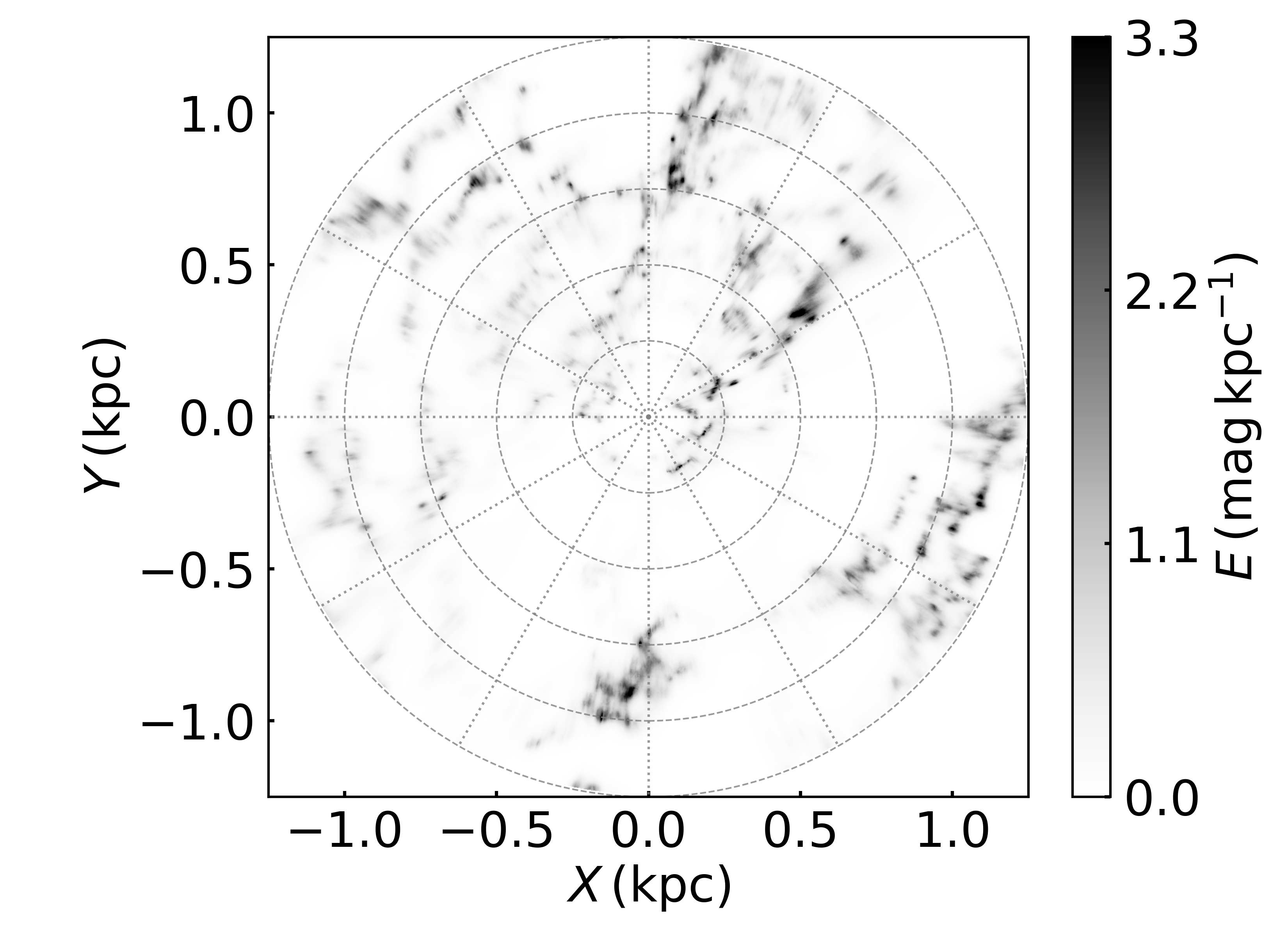}
  \caption{Comparison of our dust map in the SR region (left panel) with the dust map from \citet{Edenhofer2024} (right panel). The $X$ and $Y$ ranges are from $-1.25$\,kpc to $1.25$\,kpc. A polar grid is overlaid on the figure, with radial ticks at 0.25\,kpc intervals and Galactic longitude lines at $30^\circ$ intervals. The Sun is located at the center of the map ($X=0$\,kpc, $Y=0$\,kpc).}
  \label{fig:10pc}
  \vspace{0.3cm}
\end{figure*}

\begin{figure*}
    \centering
    \includegraphics[width=0.49\textwidth,keepaspectratio]{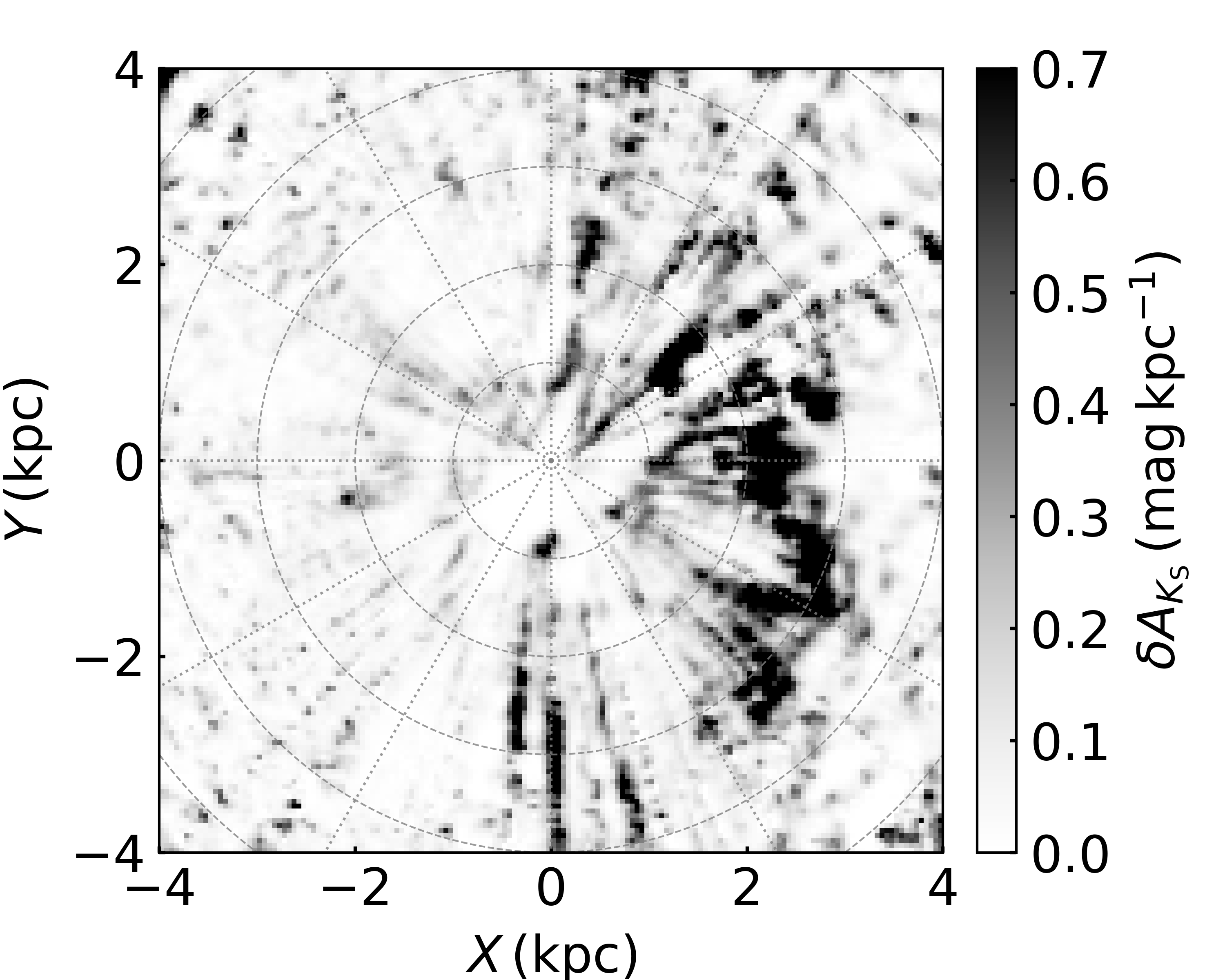}
    \includegraphics[width=0.49\textwidth,keepaspectratio]{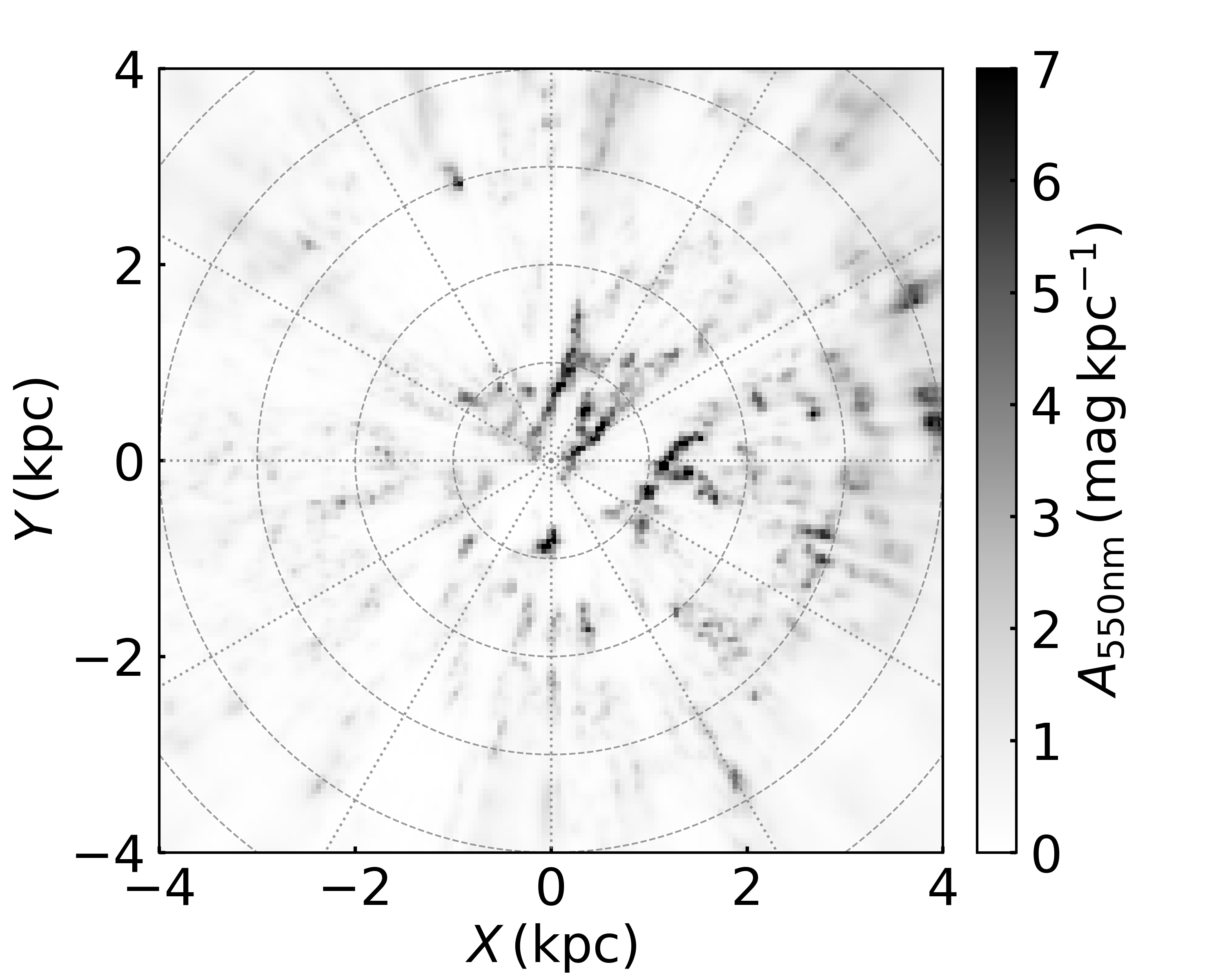}
  \caption{Comparison of our dust density distribution in the MR region (left panel) with the results from \citet{Vergely2022} (right panel). A polar grid is overlaid on the figure, with radial ticks at 1\,kpc intervals and Galactic longitude lines at $30^\circ$ intervals.}
  \label{fig:50pc}
  \vspace{0.3cm}
  \vspace{0.3cm}
\end{figure*}

\begin{figure*}
    \centering
    \includegraphics[width=0.32\textwidth,keepaspectratio]{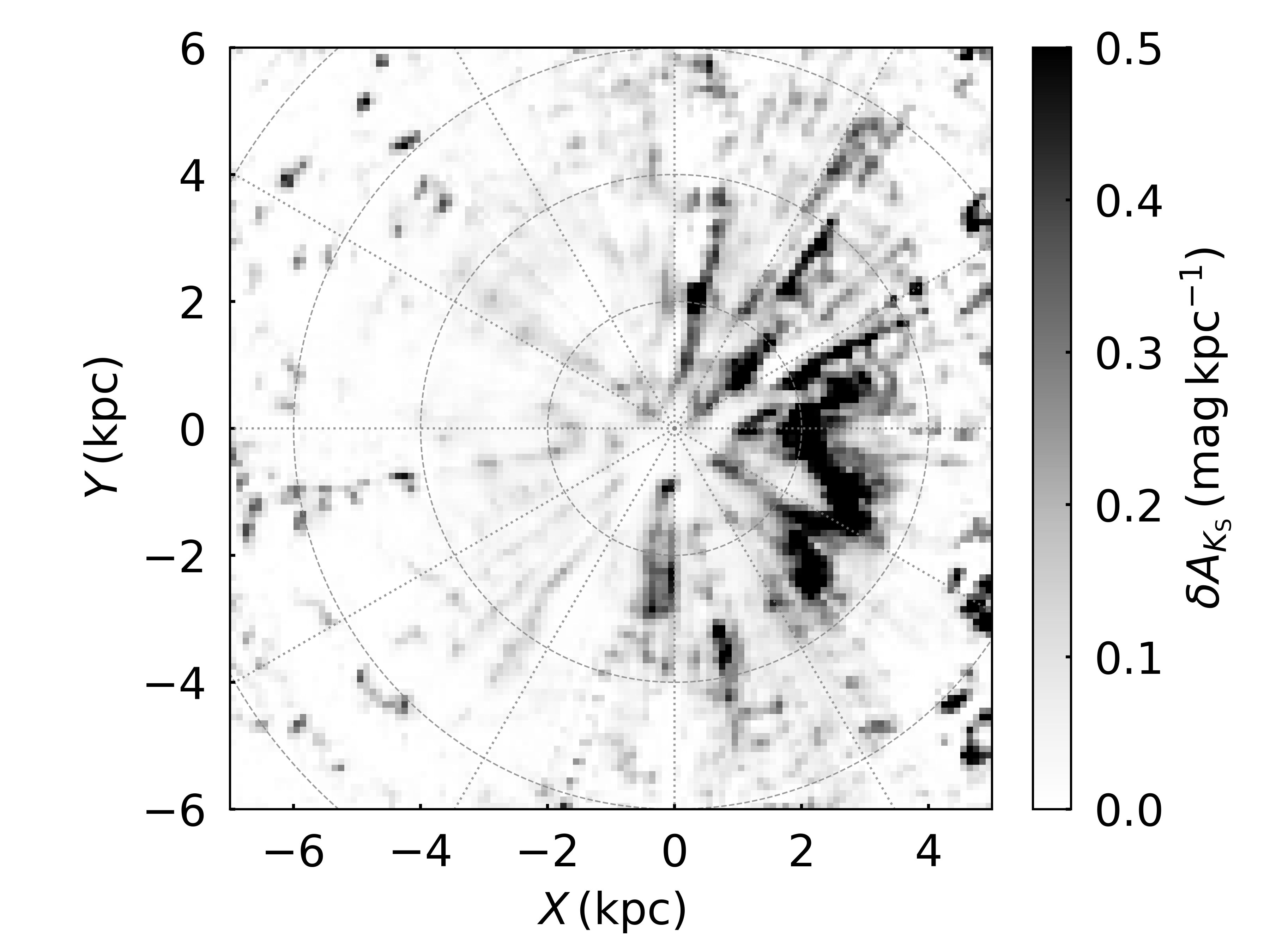}
    \includegraphics[width=0.32\textwidth,keepaspectratio]{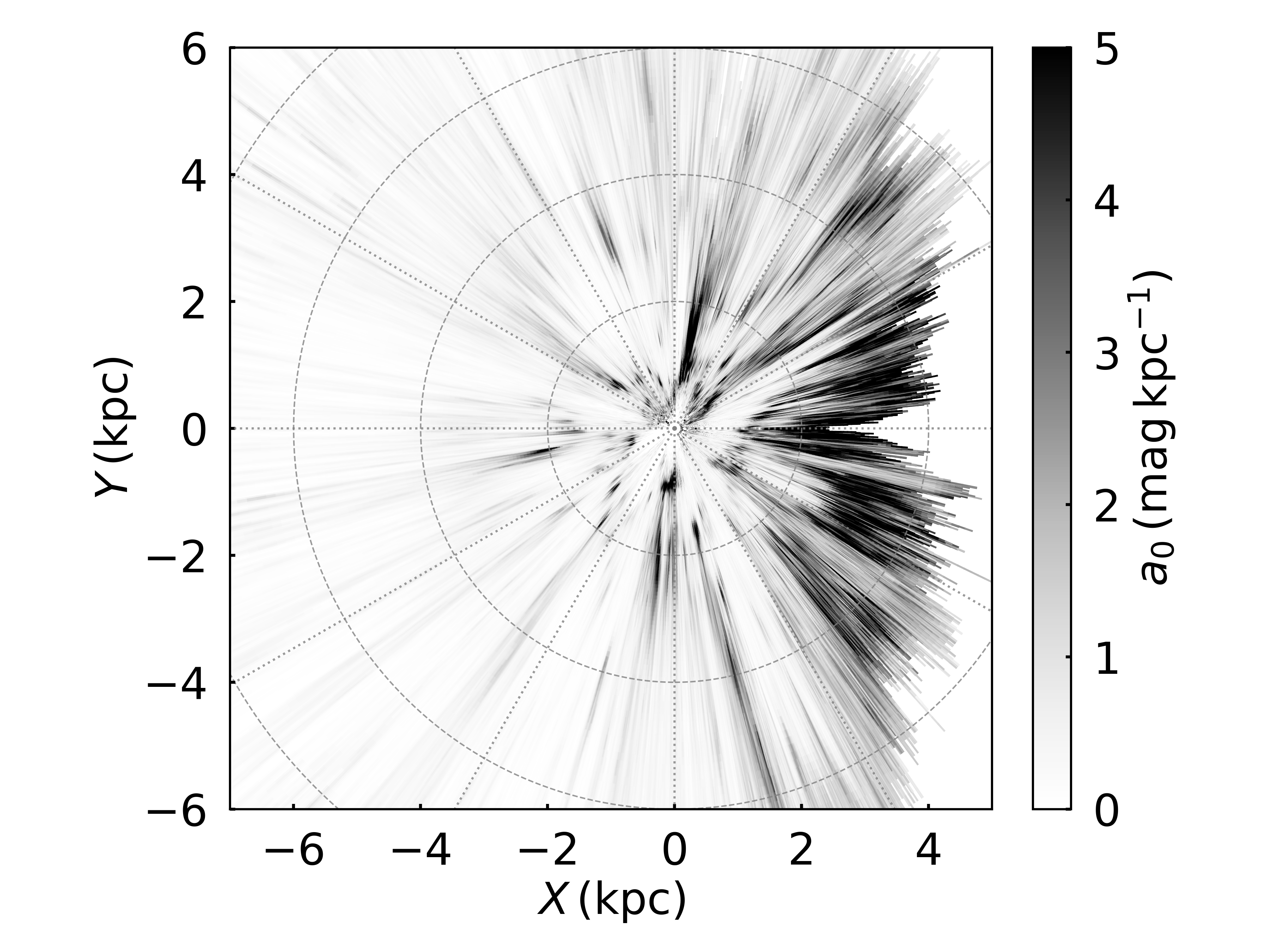}
    \includegraphics[width=0.32\textwidth,keepaspectratio]{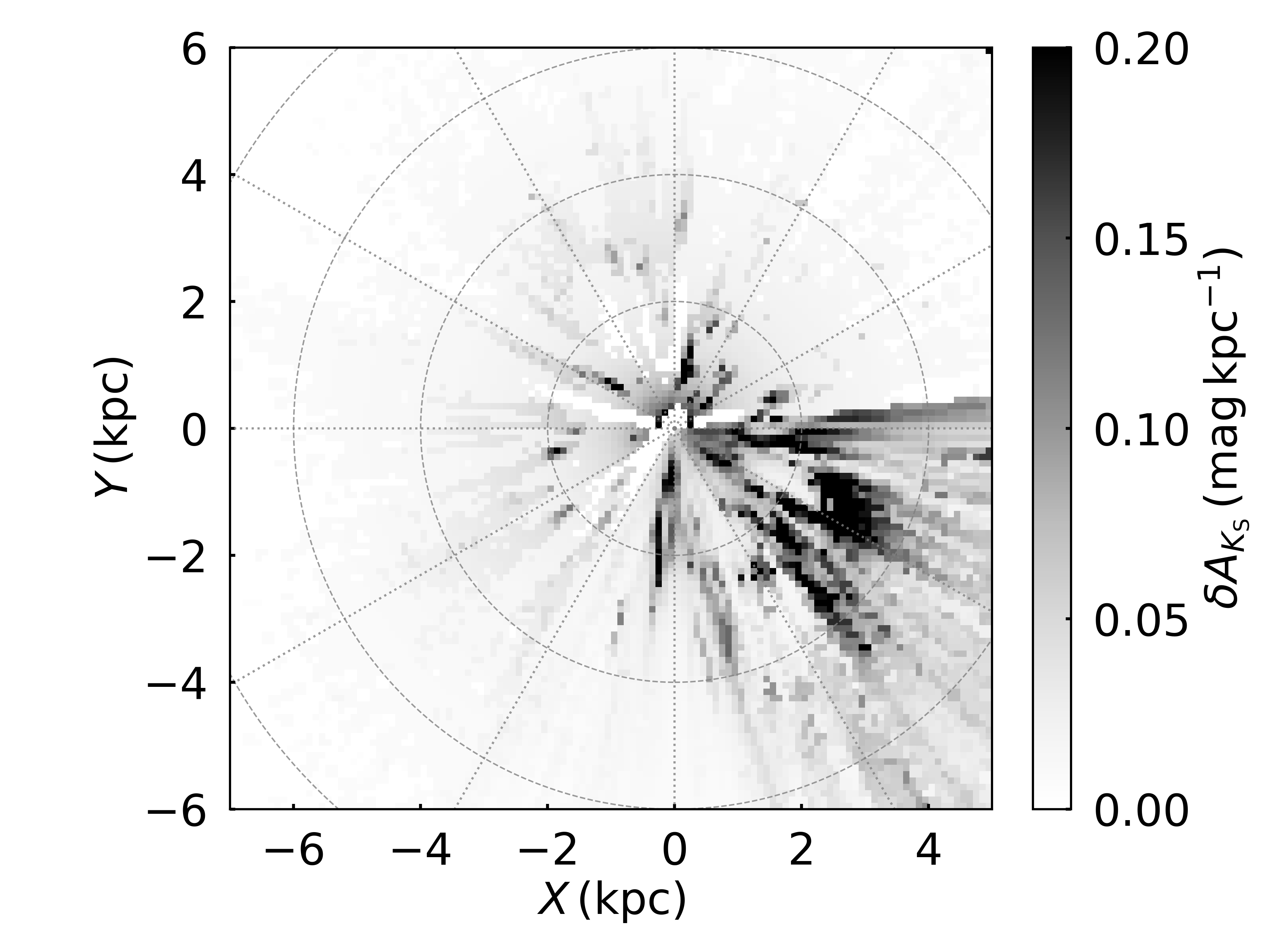}
  \caption{Comparison of our dust density distribution in the LR region (left panel) with the dust maps from \citet{Hottier2020} (middle panel), and \citet{Green2019} and \citet{Zucker2025} (right panel). A polar grid is overlaid on the figure, with radial ticks at 2\,kpc intervals and Galactic longitude lines at $30^\circ$ intervals.}
  \label{fig:100pc}
\end{figure*}

In this subsection, we compare the dust distribution obtained in this work with several dust maps from the literature, including those from \citet{Green2019}, \citet{Hottier2020}, \citet{Vergely2022}, \citet{Edenhofer2024}, and \citet{Zucker2025}, \citet{Marshall2025}.

We compare our results with those in the literature from the smallest to the largest sky regions. In Fig.~\ref{fig:10pc}, we show a comparison between our dust map in the SR region and the dust map from \citet{Edenhofer2024}. \citet{Edenhofer2024} uses distance and extinction estimates for 54 million stars derived from \textit{Gaia} BP/RP spectra from \citet{Zhang2023} and models the logarithmic dust extinction with a Gaussian Process in spherical coordinates via Iterative Charted Refinement and variational inference. Their resulting 3D dust map extends out to 1.25\,kpc from the Sun and achieves parsec-scale resolution. The data from \citet{Zhang2025} used in this work are an updated version of \citet{Zhang2023} and are very similar to those used by \citet{Edenhofer2024}, but we adopt a completely different method. Nonetheless, the overall dust distributions are very consistent. Prominent features, such as the Vela cloud, Split and the Radcliffe wave, are clearly visible in both maps.

In Fig.~\ref{fig:50pc}, we compare the extinction distribution in our MR region with the results from \citet{Vergely2022}. \citet{Vergely2022} calibrates and merges a spectro-photometric reference catalog with a purely photometric catalog. The calibrated merged catalog is then processed with a hierarchical inversion method to produce a high-resolution 3D extinction density map. Their dust density map achieves a resolution of 10\,pc within 3\,kpc of the Sun and 50\,pc beyond. Compared with our results, the overall distribution trends in the two maps agree well within about 2\,kpc of the Sun. Beyond this distance, our extinction map reveals molecular cloud features not detected in \citet{Vergely2022}. For example, the Sagittarius-Carina Complex, which is barely visible in \citet{Vergely2022} as a few discrete clouds, appears in our map as a large kpc-scale complex. The primary reason for this difference is that \citet{Vergely2022} relies primarily on optical photometry combined with \textit{Gaia} parallaxes, whose reliable distance range is limited to within $\sim$2--3\,kpc in the direction of Galactic center. In contrast, our work uses near-infrared photometry and RC stars as distance tracers, which suffer far less from dust extinction and can probe much deeper into the Galactic plane. \citet{Vergely2022} reveals a giant dust-free cavity, the Giant Oval Cavity, with a width of about 2.5\,kpc in the Milky Way. This structure is clearly visible in our results.

Finally, we compare the dust distribution in our LR region with the results from \citet{Hottier2020}, \citet{Green2019}, and \citet{Zucker2025}, as shown in Fig.~\ref{fig:100pc}. \citet{Hottier2020} uses 2MASS near-infrared photometry and \textit{Gaia} astrometric and photometric data, applying a Bayesian deconvolution algorithm based on an empirical Hertzsprung-Russell diagram to map extinction as a function of distance, analyzing over 5.6 million stars. The resulting 3D extinction map of the Galactic disk ($|b| < 0.24^\circ$) extends to 5\,kpc in the Galactic center direction and beyond 7\,kpc in the anti-center direction, as shown in the middle panel of Fig.~\ref{fig:100pc}. The two maps agree well on the large-scale dust structures. The main structures are identifiable in both maps. In particular, the largest structure in their map, the Sagittarius-Carina Complex, which appears as a conical extension between $l=300^\circ$ and $l=30^\circ$, is also clearly visible in our map. \citet{Green2019} constructs a 3D extinction map covering three-quarters of the northern sky using photometry of 800 million stars from Pan-STARRS 1 and 2MASS, dividing the sky into lines of sight and employing a Bayesian probabilistic model to simultaneously infer stellar distances, types, and line-of-sight dust distributions. \citet{Zucker2025} combines multi-band photometry with \textit{Gaia} parallaxes to infer distances, extinctions, and types for over 700 million stars, producing a 3D dust map of the southern Galactic disk ($239^\circ < l < 6^\circ$, $|b| < 10^\circ$). We combine the dust densities from these two works and compare them with our results. We restrict their maps to $|b|<0.25^\circ$ and $|Z|<25$\,pc, as shown in the right panel of Fig.~\ref{fig:100pc}. Both their maps and our map provide a comprehensive view of the dust in the Galactic plane. The overall patterns of the dust maps show a high degree of consistency. \citet{Marshall2025} used 2MASS near-infrared data and the REDLINE method to map dust in the Galactic plane ($|b| \leq 1^{\circ}$; see their Fig.~4). Our LR map shows good agreement with that work, and major features such as the Sagittarius-Carina Complex, Split, and Spur can be seen in both maps.

\subsection{Comparison with the Spatial Distribution of Other Probes}

\begin{figure*}[t!]
    \centering
    \includegraphics[width=0.49\textwidth,keepaspectratio]{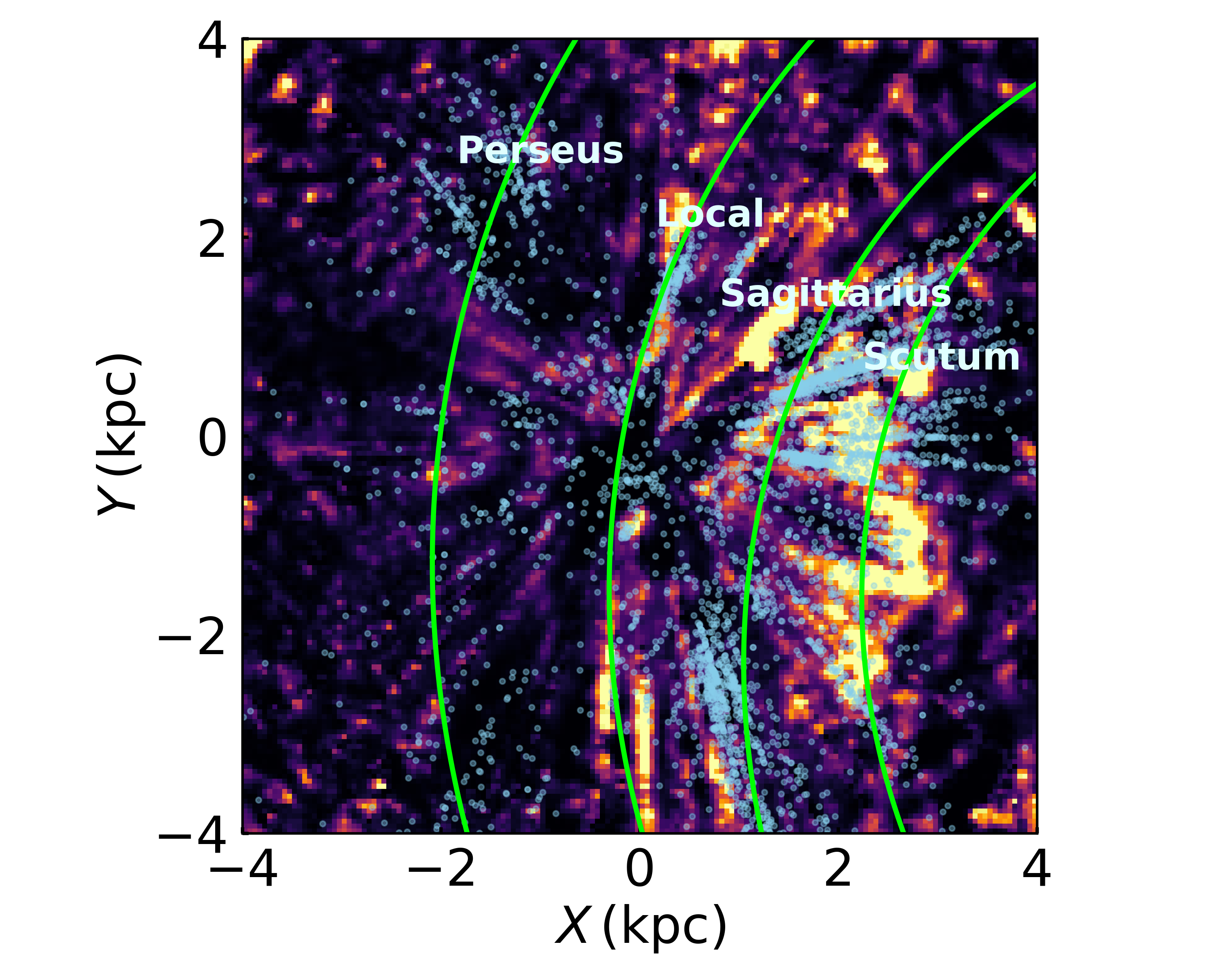}
    \includegraphics[width=0.49\textwidth,keepaspectratio]{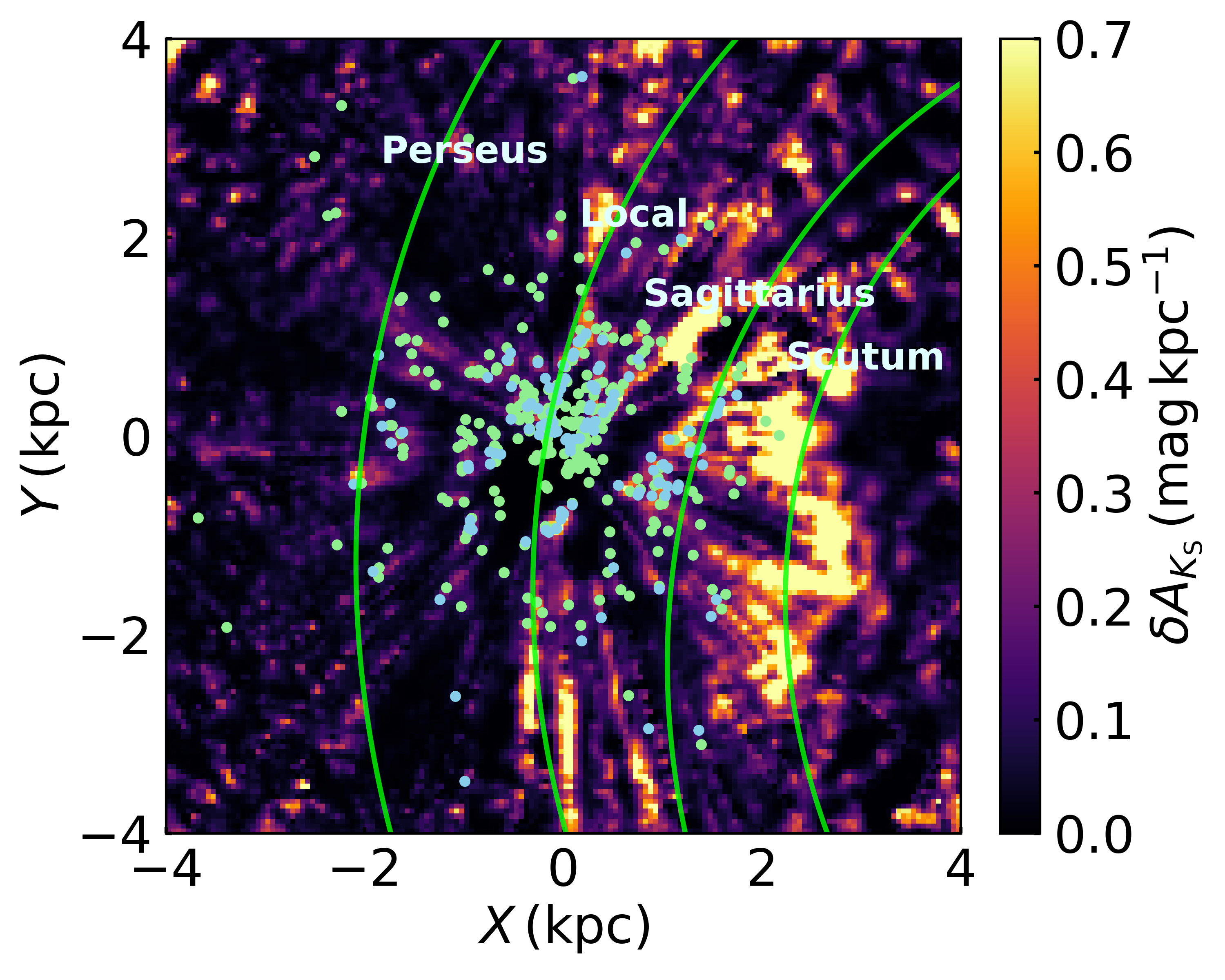}
  \caption{Comparison of the dust distribution in the MR region with the spatial distributions of OB-type stars and molecular clouds. Left panel: OB-type stars from \citet[sky blue points]{Chen2019_OB} used for comparison. The best-fit spiral arms from \citet{Chen2019_OB} are overplotted. From left to right are the Perseus Arm, Local Arm, Sagittarius Arm, and Scutum Arm. The Sun is located at ($X=0$\,kpc, $Y=0$\,kpc). Only OB-type stars with $|b|<2^{\circ}$ and $|Z|<25$\,pc are shown. Right panel: dust distribution in the MR region overlaid with molecular clouds from \citet[sky blue points]{Chen2020} and \citet[light green points]{Wang2025_cloud}. Only molecular clouds with $|b|<2^{\circ}$ and $|Z|<25$\,pc are shown.}
  \label{fig:OB_cloud}
\end{figure*}
In this subsection, we also overlay OB-type stars and molecular clouds from the literature on our dust maps for comparison. In the following discussion, we select only OB-type stars and molecular clouds within $|b|<2^{\circ}$ and $|Z|<25$\,pc for analysis.

In the left panel of Fig.~\ref{fig:OB_cloud}, we overlay the spatial distribution of O and early-B type stars from \citet{Chen2019_OB} on our dust density map of the MR region. In regions of relatively low extinction, where the OB-star sample is fairly complete, an anti-correlation is seen in which OB stars preferentially reside inside or at the edges of large-scale dust cavities, while the dust is concentrated at the cavity boundaries. This configuration points to stellar feedback as the likely mechanism, with ultraviolet radiation, stellar winds, and supernova explosions clearing the ambient material around the stellar birthplaces and piling dust at the cavity rims. In the high-extinction direction toward the Galactic center, by contrast, our dust map reveals dense, continuous dust structures, yet OB stars are largely undetected. This apparent deficit likely results from observational incompleteness rather than from a true physical anti-correlation, since the stars are present but remain hidden behind the heavy foreground dust column.

We also show a comparison between the distribution of molecular clouds from \citet{Chen2020} and \citet{Wang2025_cloud} and the dust distribution from this work, as shown in the right panel of Fig.~\ref{fig:OB_cloud}. Our dust distribution shows excellent agreement with the positions of molecular clouds. Most molecular clouds correspond to high-density regions in our dust map. This spatial consistency observationally validates the reliability and accuracy of our dust density distribution map.

\subsection{Scale Length of the Galactic Dust Disk}

Accurately determining the scale length of the Galactic dust disk is of great astrophysical importance for understanding the coupled evolution of dust and stellar disks in spiral galaxies, the radial distribution of the interstellar medium, and models of galaxy formation. The dust disk scale length directly reflects the extent of interstellar dust in the Milky Way and is closely related to the star formation efficiency, metallicity gradient, and gas disk structure. However, obtaining a reliable dust disk scale length presents significant technical challenges. The core difficulty lies in the need for a complete and high-fidelity 3D mapping of the large-scale dust distribution across the entire Galactic disk, particularly the dust structures in the inner Galaxy. Because interstellar extinction is extremely severe in the optical band and the parallax accuracy for distant objects declines rapidly, traditional methods struggle to simultaneously achieve both large detection depth and high distance accuracy. Beyond a few kiloparsecs, the systematic errors in stellar distance and extinction estimates based on \textit{Gaia} parallaxes become significantly amplified, making it particularly difficult to accurately trace the exponential decay profile of the Galactic disk dust with Galactocentric distance \citep{Guo2021,Zhang2026}. Therefore, constructing a dust distribution map that combines wide coverage with high distance accuracy is a key prerequisite for overcoming this bottleneck. The dust density distribution in our LR region provides an excellent dataset for studying the dust disk scale length.

\begin{figure}
  \centering
    \includegraphics[height=0.35\textwidth, keepaspectratio]{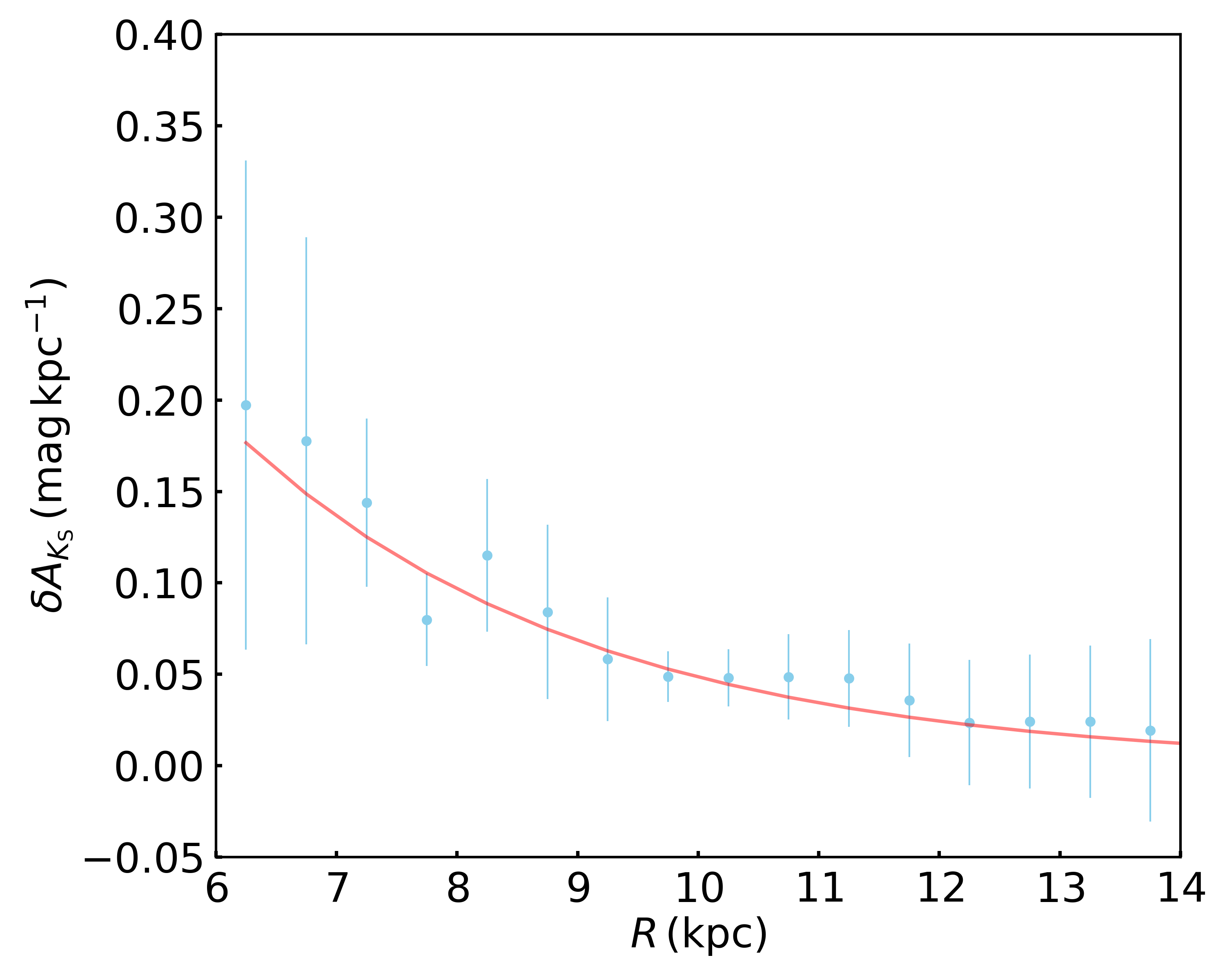}
  \caption{Scale length of the dust disk. The skyblue points show the mean dust density as a function of Galactocentric distance $R$. The red line shows the exponential fit.}
  \label{fig:scale_length}
\end{figure}

From the dust density distribution in the LR region, we derive the dust density of the Galactic disk at different radii to fit the radial profile. The dust disk can be described by an exponential function: $\rho(R)=a\exp\left(-\frac{R}{L}\right)$, where $a$ is the central dust density and $L$ is the scale length. The result is shown in Fig.~\ref{fig:scale_length}. The scale length of the dust disk is 2.90$^{+0.6}_{-0.5}$\,kpc. Since our model does not account for the flaring structure and the vertical distance of the Sun from the Galactic mid-plane, the derived scale length is likely underestimated. Nevertheless, it remains consistent with the scale length reported by \citet{Li2018}. In addition, the dust disk scale length we obtain is slightly larger than the scale lengths of the stellar disk in previous studies \citep{Ivezi2008}, suggesting that the dust disk in a spiral galaxy is generally more extended than its stellar counterpart \citep{Zhang2020}.

\section{Summary} \label{sec:summary}

In this work, we construct a catalog of near 6 million RC candidates using near-infrared and mid-infrared photometric data from 2MASS, VVV, UKIDSS, ALLWISE, and GLIMPSE. The typical selection precision of this RC sample reaches approximately 72.8\%. Using the standard candle properties of RC stars, we derive their distances and extinctions, with a typical distance uncertainty of about 7\%. This RC catalog provides a valuable data foundation for future studies of Galactic structure and interstellar dust. From the RC candidates, we select sources with $|Z|<25$\,pc as the primary extinction sample. To enrich the sample, we add over 6.36 million \textit{Gaia} BP/RP stars from \citet{Zhang2025} and about 52 thousand RCs from \citet{Lucey2020}, merging all sources into a final sample of about 7.7 million stars with distances and extinctions in the Galactic plane.

We train a U-Net model and apply it to the Galactic plane stellar sample to reconstruct the dust density distribution in the Galactic plane ({$|Z|<25$}\,pc). Our dust maps cover the full range of Galactic longitude ($-180^{\circ}<l<180^{\circ}$). The large-scale map (LR) extends to a distance of 7\,kpc with a resolution of 100\,pc, providing a panoramic view of the dust distribution in the Milky Way. The medium-resolution map (MR) reaches 4\,kpc at a resolution of 50\,pc, while the high-resolution map (SR) covers a $3\times3\,\rm kpc$ region around the Sun at a resolution of 10\,pc, enabling detailed studies of local interstellar structures. The dust distribution in the Galactic plane indicates that the Milky Way bears a strong resemblance to the morphology of the Phantom Galaxy (M74) as recently observed with JWST. Comparison with independent dust maps from the literature shows good agreement on the major dust features, and our maps have a greater depth due to the use of infrared data and the RC standard candle method. The spatial distribution of molecular clouds from the literature corresponds well with the high-density dust regions, validating the reliability of our reconstruction. We also find an anti-correlation between young OB-type stars and dense dust regions, with OB-type stars preferentially located inside dust cavities, indicating strong stellar feedback. Finally, we fit the radial profile of the Galactic dust disk and obtain an exponential scale length of $2.90$\,kpc. This value is consistent with previous estimates of the dust disk but slightly larger than the scale length of the stellar thin disk, confirming that the dust disk in a spiral galaxy is more extended than its stellar counterpart. The dust maps presented in this work provide a reliable observational basis for studying the large-scale structure of the Milky Way, the interaction between star formation and the interstellar medium, and for extinction correction in future deep Galactic plane surveys.   

\begin{acknowledgments}

We thank the anonymous referee for the constructive and detailed comments, which have significantly improved the manuscript.
This work is supported by the National Natural Science Foundation of China 12322304 and 12173034, National Natural Science Foundation of Yunnan Province 202301AV070002, the Xingdian talent support programme of Yunnan Province. We acknowledge the science research grants from the China Manned Space Project with no. CMS-CSST-2025-A11. This work is also supported by the Graduate Research Innovation Foundation Project of Yunnan University KC-252512833.

This work has made use of data from the Two Micron All Sky Survey, which is a joint project of the University of Massachusetts and the Infrared Processing and Analysis Center/California Institute of Technology, funded by the National Aeronautics and Space Administration and the National Science Foundation. This work also has made use of data from the ESO Public Survey program ID 179.B-2002 taken with the VISTA telescope, and data products from the Cambridge Astronomical Survey Unit (CASU). This work is based in part on data obtained as part of the UKIRT Infrared Deep Sky Survey. This work is based in part on observations made with the Spitzer Space Telescope, which was operated by the Jet Propulsion Laboratory, California Institute of Technology under a contract with NASA. This publication makes use of data products from AllWISE. AllWISE makes use of data from WISE, which is a joint project of the University of California, Los Angeles, and the Jet Propulsion Laboratory/California Institute of Technology, and NEOWISE, which is a project of the Jet Propulsion Laboratory/California Institute of Technology. WISE and NEOWISE are funded by the National Aeronautics and Space Administration. This work has made use of data from the European Space Agency (ESA) mission {\it Gaia} (\url{https://www.cosmos.esa.int/gaia}), processed by the {\it Gaia}
Data Processing and Analysis Consortium (DPAC,
\url{https://www.cosmos.esa.int/web/gaia/dpac/consortium}). Funding for the DPAC has been provided by national institutions, in particular the institutions participating in the {\it Gaia} Multilateral Agreement.

\end{acknowledgments}




\bibliography{sample701}{}

@MISC{Cutri2013,
       author = {{Cutri}, R.~M. and {Wright}, E.~L. and {Conrow}, T. and {Fowler}, J.~W. and {Eisenhardt}, P.~R.~M. and {Grillmair}, C. and {Kirkpatrick}, J.~D. and {Masci}, F. and {McCallon}, H.~L. and {Wheelock}, S.~L. and {Fajardo-Acosta}, S. and {Yan}, L. and {Benford}, D. and {Harbut}, M. and {Jarrett}, T. and {Lake}, S. and {Leisawitz}, D. and {Ressler}, M.~E. and {Stanford}, S.~A. and {Tsai}, C.~W. and {Liu}, F. and {Helou}, G. and {Mainzer}, A. and {Gettings}, D. and {Gonzalez}, A. and {Hoffman}, D. and {Marsh}, K.~A. and {Padgett}, D. and {Skrutskie}, M.~F. and {Beck}, R.~P. and {Papin}, M. and {Wittman}, M.},
        title = "{Explanatory Supplement to the AllWISE Data Release Products}",
 howpublished = {Explanatory Supplement to the AllWISE Data Release Products, by R. M. Cutri et al.},
         year = 2013,
        month = nov,
        pages = {1},
       adsurl = {https://ui.adsabs.harvard.edu/abs/2013wise.rept....1C}
}

@ARTICLE{Berry2012,
       author = {{Berry}, Michael and {Ivezi{\'c}}, {\v{Z}}eljko and {Sesar}, Branimir and {Juri{\'c}}, Mario and {Schlafly}, Edward F. and {Bellovary}, Jillian and {Finkbeiner}, Douglas and {Vrbanec}, Dijana and {Beers}, Timothy C. and {Brooks}, Keira J. and {Schneider}, Donald P. and {Gibson}, Robert R. and {Kimball}, Amy and {Jones}, Lynne and {Yoachim}, Peter and {Krughoff}, Simon and {Connolly}, Andrew J. and {Loebman}, Sarah and {Bond}, Nicholas A. and {Schlegel}, David and {Dalcanton}, Julianne and {Yanny}, Brian and {Majewski}, Steven R. and {Knapp}, Gillian R. and {Gunn}, James E. and {Allyn Smith}, J. and {Fukugita}, Masataka and {Kent}, Steve and {Barentine}, John and {Krzesinski}, Jurek and {Long}, Dan},
        title = "{The Milky Way Tomography with Sloan Digital Sky Survey. IV. Dissecting Dust}",
      journal = {\apj},
         year = 2012,
        month = oct,
       volume = {757},
       number = {2},
          eid = {166},
        pages = {166},
          doi = {10.1088/0004-637X/757/2/166},
archivePrefix = {arXiv},
       eprint = {1111.4985},
 primaryClass = {astro-ph.GA},
       adsurl = {https://ui.adsabs.harvard.edu/abs/2012ApJ...757..166B}
}

@ARTICLE{Zucker2022,
       author = {{Zucker}, Catherine and {Goodman}, Alyssa A. and {Alves}, Jo{\~a}o and {Bialy}, Shmuel and {Foley}, Michael and {Speagle}, Joshua S. and {Gro{\^I}{\texttwosuperior}schedl}, Josefa and {Finkbeiner}, Douglas P. and {Burkert}, Andreas and {Khimey}, Diana and et al.},
        title = "{Star formation near the Sun is driven by expansion of the Local Bubble}",
      journal = {\nat},
         year = 2022,
        month = jan,
       volume = {601},
       number = {7893},
        pages = {334-337},
          doi = {10.1038/s41586-021-04286-5},
archivePrefix = {arXiv},
       eprint = {2201.05124},
 primaryClass = {astro-ph.GA},
       adsurl = {https://ui.adsabs.harvard.edu/abs/2022Natur.601..334Z}
}

@ARTICLE{Chen2019_dust,
       author = {{Chen}, B.-Q. and {Huang}, Y. and {Yuan}, H.-B. and {Wang}, C. and {Fan}, D.-W. and {Xiang}, M.-S. and {Zhang}, H.-W. and {Tian}, Z.-J. and {Liu}, X.-W.},
        title = "{Three-dimensional interstellar dust reddening maps of the Galactic plane}",
      journal = {\mnras},
         year = 2019,
        month = mar,
       volume = {483},
       number = {4},
        pages = {4277-4289},
          doi = {10.1093/mnras/sty3341},
archivePrefix = {arXiv},
       eprint = {1807.02241},
 primaryClass = {astro-ph.GA},
       adsurl = {https://ui.adsabs.harvard.edu/abs/2019MNRAS.483.4277C}
}

@ARTICLE{Yu2025RAA,
       author = {{Yu}, Zheng and {Chen}, Bing-Qiu and {Liu}, Xiao-Wei},
        title = "{Empirically Predicted Absolute Magnitudes for Red Clump Stars in Mephisto and CSST Filters}",
      journal = {Research in Astronomy and Astrophysics},
         year = 2025,
        month = sep,
       volume = {25},
       number = {9},
          eid = {095004},
        pages = {095004},
          doi = {10.1088/1674-4527/ade7f0},
archivePrefix = {arXiv},
       eprint = {2506.20078},
 primaryClass = {astro-ph.SR},
       adsurl = {https://ui.adsabs.harvard.edu/abs/2025RAA....25i5004Y}
}

@ARTICLE{Benjamin2003,
       author = {{Benjamin}, Robert A. and {Churchwell}, E. and {Babler}, Brian L. and {Bania}, T.~M. and {Clemens}, Dan P. and {Cohen}, Martin and {Dickey}, John M. and {Indebetouw}, R{\'e}my and {Jackson}, James M. and {Kobulnicky}, Henry A. and {Lazarian}, Alex and {Marston}, A.~P. and {Mathis}, John S. and {Meade}, Marilyn R. and {Seager}, Sara and {Stolovy}, S.~R. and {Watson}, C. and {Whitney}, Barbara A. and {Wolff}, Michael J. and {Wolfire}, Mark G.},
        title = "{GLIMPSE. I. An SIRTF Legacy Project to Map the Inner Galaxy}",
      journal = {\pasp},
         year = 2003,
        month = aug,
       volume = {115},
       number = {810},
        pages = {953-964},
          doi = {10.1086/376696},
archivePrefix = {arXiv},
       eprint = {astro-ph/0306274},
 primaryClass = {astro-ph},
       adsurl = {https://ui.adsabs.harvard.edu/abs/2003PASP..115..953B}
}

@ARTICLE{Churchwell2009,
       author = {{Churchwell}, Ed and {Babler}, Brian L. and {Meade}, Marilyn R. and {Whitney}, Barbara A. and {Benjamin}, Robert and {Indebetouw}, Remy and {Cyganowski}, Claudia and {Robitaille}, Thomas P. and {Povich}, Matthew and {Watson}, Christer and {Bracker}, Steve},
        title = "{The Spitzer/GLIMPSE Surveys: A New View of the Milky Way}",
      journal = {\pasp},
         year = 2009,
        month = mar,
       volume = {121},
       number = {877},
        pages = {213},
          doi = {10.1086/597811},
       adsurl = {https://ui.adsabs.harvard.edu/abs/2009PASP..121..213C}
}

@ARTICLE{Werner2004,
       author = {{Werner}, M.~W. and {Roellig}, T.~L. and {Low}, F.~J. and {Rieke}, G.~H. and {Rieke}, M. and {Hoffmann}, W.~F. and {Young}, E. and {Houck}, J.~R. and {Brandl}, B. and {Fazio}, G.~G. and {Hora}, J.~L. and {Gehrz}, R.~D. and {Helou}, G. and {Soifer}, B.~T. and {Stauffer}, J. and {Keene}, J. and {Eisenhardt}, P. and {Gallagher}, D. and {Gautier}, T.~N. and {Irace}, W. and {Lawrence}, C.~R. and {Simmons}, L. and {Van Cleve}, J.~E. and {Jura}, M. and {Wright}, E.~L. and {Cruikshank}, D.~P.},
        title = "{The Spitzer Space Telescope Mission}",
      journal = {\apjs},
         year = 2004,
        month = sep,
       volume = {154},
       number = {1},
        pages = {1-9},
          doi = {10.1086/422992},
archivePrefix = {arXiv},
       eprint = {astro-ph/0406223},
 primaryClass = {astro-ph},
       adsurl = {https://ui.adsabs.harvard.edu/abs/2004ApJS..154....1W}
}

@ARTICLE{Majewski2011,
       author = {{Majewski}, Steven R. and {Zasowski}, Gail and {Nidever}, David L.},
        title = "{Lifting the Dusty Veil with Near- and Mid-infrared Photometry. I. Description and Applications of the Rayleigh-Jeans Color Excess Method}",
      journal = {\apj},
         year = 2011,
        month = sep,
       volume = {739},
       number = {1},
          eid = {25},
        pages = {25},
          doi = {10.1088/0004-637X/739/1/25},
archivePrefix = {arXiv},
       eprint = {1106.2542},
 primaryClass = {astro-ph.GA},
       adsurl = {https://ui.adsabs.harvard.edu/abs/2011ApJ...739...25M}
}

@ARTICLE{Grocholski2002,
       author = {{Grocholski}, Aaron J. and {Sarajedini}, Ata},
        title = "{WIYN Open Cluster Study. X. The K-Band Magnitude of the Red Clump as a Distance Indicator}",
      journal = {\aj},
         year = 2002,
        month = mar,
       volume = {123},
       number = {3},
        pages = {1603-1612},
          doi = {10.1086/339027},
archivePrefix = {arXiv},
       eprint = {astro-ph/0112251},
 primaryClass = {astro-ph},
       adsurl = {https://ui.adsabs.harvard.edu/abs/2002AJ....123.1603G}
}

@ARTICLE{Alves2000,
       author = {{Alves}, David R.},
        title = "{K-Band Calibration of the Red Clump Luminosity}",
      journal = {\apj},
         year = 2000,
        month = aug,
       volume = {539},
       number = {2},
        pages = {732-741},
          doi = {10.1086/309278},
archivePrefix = {arXiv},
       eprint = {astro-ph/0003329},
 primaryClass = {astro-ph},
       adsurl = {https://ui.adsabs.harvard.edu/abs/2000ApJ...539..732A}
}

@ARTICLE{Indebetouw2005,
       author = {{Indebetouw}, R. and {Mathis}, J.~S. and {Babler}, B.~L. and {Meade}, M.~R. and {Watson}, C. and {Whitney}, B.~A. and {Wolff}, M.~J. and {Wolfire}, M.~G. and {Cohen}, M. and {Bania}, T.~M. and {Benjamin}, R.~A. and {Clemens}, D.~P. and {Dickey}, J.~M. and {Jackson}, J.~M. and {Kobulnicky}, H.~A. and {Marston}, A.~P. and {Mercer}, E.~P. and {Stauffer}, J.~R. and {Stolovy}, S.~R. and {Churchwell}, E.},
        title = "{The Wavelength Dependence of Interstellar Extinction from 1.25 to 8.0 {\ensuremath{\mu}}m Using GLIMPSE Data}",
      journal = {\apj},
         year = 2005,
        month = feb,
       volume = {619},
       number = {2},
        pages = {931-938},
          doi = {10.1086/426679},
archivePrefix = {arXiv},
       eprint = {astro-ph/0406403},
 primaryClass = {astro-ph},
       adsurl = {https://ui.adsabs.harvard.edu/abs/2005ApJ...619..931I}
}

@ARTICLE{Zasowski2009,
       author = {{Zasowski}, G. and {Majewski}, S.~R. and {Indebetouw}, R. and {Meade}, M.~R. and {Nidever}, D.~L. and {Patterson}, R.~J. and {Babler}, B. and {Skrutskie}, M.~F. and {Watson}, C. and {Whitney}, B.~A. and {Churchwell}, E.},
        title = "{Lifting the Dusty Veil with Near- and Mid-Infrared Photometry. II. A Large-Scale Study of the Galactic Infrared Extinction Law}",
      journal = {\apj},
         year = 2009,
        month = dec,
       volume = {707},
       number = {1},
        pages = {510-523},
          doi = {10.1088/0004-637X/707/1/510},
archivePrefix = {arXiv},
       eprint = {0910.4403},
 primaryClass = {astro-ph.GA},
       adsurl = {https://ui.adsabs.harvard.edu/abs/2009ApJ...707..510Z}
}

@ARTICLE{Soto2013,
       author = {{Soto}, M. and {Barb{\'a}}, R. and {Gunthardt}, G. and {Minniti}, D. and {Lucas}, P. and {Majaess}, D. and {Irwin}, M. and {Emerson}, J.~P. and {Gonzalez-Solares}, E. and {Hempel}, M. and {Saito}, R.~K. and {Gurovich}, S. and {Roman-Lopes}, A. and {Moni-Bidin}, C. and {Santucho}, M.~V. and {Borissova}, J. and {Kurtev}, R. and {Toledo}, I. and {Geisler}, D. and {Dominguez}, M. and {Beamin}, J.~C.},
        title = "{Milky Way demographics with the VVV survey. II. Color transformations and near-infrared photometry for 136 million stars in the southern Galactic disk}",
      journal = {\aap},
         year = 2013,
        month = apr,
       volume = {552},
          eid = {A101},
        pages = {A101},
          doi = {10.1051/0004-6361/201220046},
archivePrefix = {arXiv},
       eprint = {1305.5902},
 primaryClass = {astro-ph.GA},
       adsurl = {https://ui.adsabs.harvard.edu/abs/2013A&A...552A.101S}
}

@ARTICLE{Wegg2015,
       author = {{Wegg}, Christopher and {Gerhard}, Ortwin and {Portail}, Matthieu},
        title = "{The structure of the Milky Way's bar outside the bulge}",
      journal = {\mnras},
         year = 2015,
        month = jul,
       volume = {450},
       number = {4},
        pages = {4050-4069},
          doi = {10.1093/mnras/stv745},
archivePrefix = {arXiv},
       eprint = {1504.01401},
 primaryClass = {astro-ph.GA},
       adsurl = {https://ui.adsabs.harvard.edu/abs/2015MNRAS.450.4050W}
}

@ARTICLE{Ting2018,
       author = {{Ting}, Yuan-Sen and {Hawkins}, Keith and {Rix}, Hans-Walter},
        title = "{A Large and Pristine Sample of Standard Candles across the Milky Way: {\ensuremath{\sim}}100,000 Red Clump Stars with 3\% Contamination}",
      journal = {\apjl},
         year = 2018,
        month = may,
       volume = {858},
       number = {1},
          eid = {L7},
        pages = {L7},
          doi = {10.3847/2041-8213/aabf8e},
archivePrefix = {arXiv},
       eprint = {1803.06650},
 primaryClass = {astro-ph.SR},
       adsurl = {https://ui.adsabs.harvard.edu/abs/2018ApJ...858L...7T}
}

@ARTICLE{Lucey2020,
       author = {{Lucey}, Madeline and {Ting}, Yuan-Sen and {Ramachandra}, Nesar S. and {Hawkins}, Keith},
        title = "{From the inner to outer Milky Way: a photometric sample of 2.6 million red clump stars}",
      journal = {\mnras},
         year = 2020,
        month = jan,
       volume = {495},
       number = {3},
        pages = {3087-3103},
          doi = {10.1093/mnras/staa1226},
archivePrefix = {arXiv},
       eprint = {2002.02961},
 primaryClass = {astro-ph.SR},
       adsurl = {https://ui.adsabs.harvard.edu/abs/2020MNRAS.495.3087L}
}

@ARTICLE{Zhang2025,
       author = {{Zhang}, Xiangyu and {Green}, Gregory M.},
        title = "{Three-dimensional maps of the interstellar dust extinction curve within the Milky Way galaxy}",
      journal = {Science},
         year = 2025,
        month = mar,
       volume = {387},
       number = {6739},
        pages = {1209-1214},
          doi = {10.1126/science.ado9787},
archivePrefix = {arXiv},
       eprint = {2407.14594},
 primaryClass = {astro-ph.GA},
       adsurl = {https://ui.adsabs.harvard.edu/abs/2025Sci...387.1209Z}
}

@ARTICLE{Zhang2023,
       author = {{Zhang}, Xiangyu and {Green}, Gregory M. and {Rix}, Hans-Walter},
        title = "{Parameters of 220 million stars from Gaia BP/RP spectra}",
      journal = {\mnras},
         year = 2023,
        month = sep,
       volume = {524},
       number = {2},
        pages = {1855-1884},
          doi = {10.1093/mnras/stad1941},
archivePrefix = {arXiv},
       eprint = {2303.03420},
 primaryClass = {astro-ph.SR},
       adsurl = {https://ui.adsabs.harvard.edu/abs/2023MNRAS.524.1855Z}
}

@ARTICLE{Ronneberger2015,
       author = {{Ronneberger}, Olaf and {Fischer}, Philipp and {Brox}, Thomas},
        title = "{U-Net: Convolutional Networks for Biomedical Image Segmentation}",
      journal = {arXiv e-prints},
         year = 2015,
        month = may,
          eid = {arXiv:1505.04597},
        pages = {arXiv:1505.04597},
          doi = {10.48550/arXiv.1505.04597},
archivePrefix = {arXiv},
       eprint = {1505.04597},
 primaryClass = {cs.CV},
       adsurl = {https://ui.adsabs.harvard.edu/abs/2015arXiv150504597R}
}

@ARTICLE{Chen2024,
       author = {{Chen}, Bing-Qiu and {Qin}, Fei and {Li}, Guang-Xing},
        title = "{Constructing the three-dimensional extinction density maps using V-net}",
      journal = {\mnras},
         year = 2024,
        month = mar,
       volume = {528},
       number = {4},
        pages = {7600-7614},
          doi = {10.1093/mnras/stae523},
archivePrefix = {arXiv},
       eprint = {2402.11270},
 primaryClass = {astro-ph.GA},
       adsurl = {https://ui.adsabs.harvard.edu/abs/2024MNRAS.528.7600C}
}

@ARTICLE{Chen2019_OB,
       author = {{Chen}, B.-Q. and {Huang}, Y. and {Hou}, L.-G. and {Tian}, H. and {Li}, G.-X. and {Yuan}, H.-B. and {Wang}, H.-F. and {Wang}, C. and {Tian}, Z.-J. and {Liu}, X.-W.},
        title = "{The Galactic spiral structure as revealed by O- and early B-type stars}",
      journal = {\mnras},
         year = 2019,
        month = jul,
       volume = {487},
       number = {1},
        pages = {1400-1409},
          doi = {10.1093/mnras/stz1357},
archivePrefix = {arXiv},
       eprint = {1905.05542},
 primaryClass = {astro-ph.GA},
       adsurl = {https://ui.adsabs.harvard.edu/abs/2019MNRAS.487.1400C}
}

@ARTICLE{Gontcharov2025,
       author = {{Gontcharov}, G.~A. and {Marchuk}, A.~A. and {Savchenko}, S.~S. and {Mosenkov}, A.~V. and {Il'in}, V.~B. and {Poliakov}, D.~M. and {Smirnov}, A.~A. and {Krayani}, H.},
        title = "{Foreground Extinction to Extended Celestial Objects. I. New Extinction Maps}",
      journal = {Research in Astronomy and Astrophysics},
         year = 2025,
        month = dec,
       volume = {25},
       number = {12},
          eid = {125016},
        pages = {125016},
          doi = {10.1088/1674-4527/ae12a6},
archivePrefix = {arXiv},
       eprint = {2510.02600},
 primaryClass = {astro-ph.GA},
       adsurl = {https://ui.adsabs.harvard.edu/abs/2025RAA....25l5016G}
}

@ARTICLE{Chen2020,
       author = {{Chen}, B.-Q. and {Li}, G.-X. and {Yuan}, H.-B. and {Huang}, Y. and {Tian}, Z.-J. and {Wang}, H.-F. and {Zhang}, H.-W. and {Wang}, C. and {Liu}, X.-W.},
        title = "{A large catalogue of molecular clouds with accurate distances within 4 kpc of the Galactic disc}",
      journal = {\mnras},
         year = 2020,
        month = mar,
       volume = {493},
       number = {1},
        pages = {351-361},
          doi = {10.1093/mnras/staa235},
archivePrefix = {arXiv},
       eprint = {2001.11682},
 primaryClass = {astro-ph.GA},
       adsurl = {https://ui.adsabs.harvard.edu/abs/2020MNRAS.493..351C}
}

@ARTICLE{Wang2025_dust,
       author = {{Wang}, Tao and {Yuan}, Haibo and {Chen}, Bingqiu and {Xiang}, Maosheng and {Zhang}, Ruoyi and {Huang}, Bowen and {Gu}, Hongrui and {Wang}, Shuaicong and {Li}, Jiawei},
        title = "{An All-sky 3D Dust Map Based on Gaia and LAMOST}",
      journal = {\apjs},
         year = 2025,
        month = sep,
       volume = {280},
       number = {1},
          eid = {15},
        pages = {15},
          doi = {10.3847/1538-4365/adea39},
archivePrefix = {arXiv},
       eprint = {2509.07640},
 primaryClass = {astro-ph.GA},
       adsurl = {https://ui.adsabs.harvard.edu/abs/2025ApJS..280...15W}
}

@ARTICLE{Wang2025_cloud,
       author = {{Wang}, Tao and {Yuan}, Haibo and {Chen}, Bingqiu and {Li}, Guangxing and {Huang}, Bowen and {Guo}, Helong and {Zhang}, Ruoyi},
        title = "{A Comprehensive All-sky Catalog of 3345 Molecular Clouds from Three-dimensional Dust Extinction}",
      journal = {\apjs},
         year = 2025,
        month = sep,
       volume = {280},
       number = {1},
          eid = {16},
        pages = {16},
          doi = {10.3847/1538-4365/aded89},
archivePrefix = {arXiv},
       eprint = {2509.07670},
 primaryClass = {astro-ph.GA},
       adsurl = {https://ui.adsabs.harvard.edu/abs/2025ApJS..280...16W}
}

@ARTICLE{Vergely2022,
       author = {{Vergely}, J.~L. and {Lallement}, R. and {Cox}, N.~L.~J.},
        title = "{Three-dimensional extinction maps: Inverting inter-calibrated extinction catalogues}",
      journal = {\aap},
         year = 2022,
        month = aug,
       volume = {664},
          eid = {A174},
        pages = {A174},
          doi = {10.1051/0004-6361/202243319},
archivePrefix = {arXiv},
       eprint = {2205.09087},
 primaryClass = {astro-ph.GA},
       adsurl = {https://ui.adsabs.harvard.edu/abs/2022A&A...664A.174V}
}

@ARTICLE{Marshall2006,
       author = {{Marshall}, D.~J. and {Robin}, A.~C. and {Reyl{\'e}}, C. and {Schultheis}, M. and {Picaud}, S.},
        title = "{Modelling the Galactic interstellar extinction distribution in three dimensions}",
      journal = {\aap},
         year = 2006,
        month = jul,
       volume = {453},
       number = {2},
        pages = {635-651},
          doi = {10.1051/0004-6361:20053842},
archivePrefix = {arXiv},
       eprint = {astro-ph/0604427},
 primaryClass = {astro-ph},
       adsurl = {https://ui.adsabs.harvard.edu/abs/2006A&A...453..635M}
}

@ARTICLE{Zucker2025,
       author = {{Zucker}, Catherine and {Saydjari}, Andrew K. and {Speagle}, Joshua S. and {Schlafly}, Edward F. and {Green}, Gregory M. and {Benjamin}, Robert and {Peek}, Joshua and {Edenhofer}, Gordian and {Goodman}, Alyssa and {Kuhn}, Michael A. and {Finkbeiner}, Douglas P.},
        title = "{A Deep, High-angular-resolution 3D Dust Map of the Southern Galactic Plane}",
      journal = {\apj},
         year = 2025,
        month = oct,
       volume = {992},
       number = {1},
          eid = {39},
        pages = {39},
          doi = {10.3847/1538-4357/adfbe6},
archivePrefix = {arXiv},
       eprint = {2503.02657},
 primaryClass = {astro-ph.GA},
       adsurl = {https://ui.adsabs.harvard.edu/abs/2025ApJ...992...39Z}
}

@ARTICLE{Green2019,
       author = {{Green}, Gregory M. and {Schlafly}, Edward and {Zucker}, Catherine and {Speagle}, Joshua S. and {Finkbeiner}, Douglas},
        title = "{A 3D Dust Map Based on Gaia, Pan-STARRS 1, and 2MASS}",
      journal = {\apj},
         year = 2019,
        month = dec,
       volume = {887},
       number = {1},
          eid = {93},
        pages = {93},
          doi = {10.3847/1538-4357/ab5362},
archivePrefix = {arXiv},
       eprint = {1905.02734},
 primaryClass = {astro-ph.GA},
       adsurl = {https://ui.adsabs.harvard.edu/abs/2019ApJ...887...93G}
}

@ARTICLE{Hottier2020,
       author = {{Hottier}, C. and {Babusiaux}, C. and {Arenou}, F.},
        title = "{FEDReD. II. 3D extinction map with 2MASS and Gaia DR2 data}",
      journal = {\aap},
         year = 2020,
        month = sep,
       volume = {641},
          eid = {A79},
        pages = {A79},
          doi = {10.1051/0004-6361/202037573},
archivePrefix = {arXiv},
       eprint = {2007.03734},
 primaryClass = {astro-ph.GA},
       adsurl = {https://ui.adsabs.harvard.edu/abs/2020A&A...641A..79H}
}

@ARTICLE{Green2015,
       author = {{Green}, Gregory M. and {Schlafly}, Edward F. and {Finkbeiner}, Douglas P. and {Rix}, Hans-Walter and {Martin}, Nicolas and {Burgett}, William and {Draper}, Peter W. and {Flewelling}, Heather and {Hodapp}, Klaus and {Kaiser}, Nicholas and {Kudritzki}, Rolf Peter and {Magnier}, Eugene and {Metcalfe}, Nigel and {Price}, Paul and {Tonry}, John and {Wainscoat}, Richard},
        title = "{A Three-dimensional Map of Milky Way Dust}",
      journal = {\apj},
         year = 2015,
        month = sep,
       volume = {810},
       number = {1},
          eid = {25},
        pages = {25},
          doi = {10.1088/0004-637X/810/1/25},
archivePrefix = {arXiv},
       eprint = {1507.01005},
 primaryClass = {astro-ph.GA},
       adsurl = {https://ui.adsabs.harvard.edu/abs/2015ApJ...810...25G}
}

@ARTICLE{Skrutskie2006,
       author = {{Skrutskie}, M.~F. and {Cutri}, R.~M. and {Stiening}, R. and {Weinberg}, M.~D. and {Schneider}, S. and {Carpenter}, J.~M. and {Beichman}, C. and {Capps}, R. and {Chester}, T. and {Elias}, J. and {Huchra}, J. and {Liebert}, J. and {Lonsdale}, C. and {Monet}, D.~G. and {Price}, S. and {Seitzer}, P. and {Jarrett}, T. and {Kirkpatrick}, J.~D. and {Gizis}, J.~E. and {Howard}, E. and {Evans}, T. and {Fowler}, J. and {Fullmer}, L. and {Hurt}, R. and {Light}, R. and {Kopan}, E.~L. and {Marsh}, K.~A. and {McCallon}, H.~L. and {Tam}, R. and {Van Dyk}, S. and {Wheelock}, S.},
        title = "{The Two Micron All Sky Survey (2MASS)}",
      journal = {\aj},
         year = 2006,
        month = feb,
       volume = {131},
       number = {2},
        pages = {1163-1183},
          doi = {10.1086/498708},
       adsurl = {https://ui.adsabs.harvard.edu/abs/2006AJ....131.1163S}
}

@ARTICLE{Chen2014,
       author = {{Chen}, B.-Q. and {Liu}, X.-W. and {Yuan}, H.-B. and {Zhang}, H.-H. and {Schultheis}, M. and {Jiang}, B.-W. and {Huang}, Y. and {Xiang}, M.-S. and {Zhao}, H.-B. and {Yao}, J.-S. and {Lu}, H.},
        title = "{A three-dimensional extinction map of the Galactic anticentre from multiband photometry}",
      journal = {\mnras},
         year = 2014,
        month = sep,
       volume = {443},
       number = {2},
        pages = {1192-1210},
          doi = {10.1093/mnras/stu1192},
archivePrefix = {arXiv},
       eprint = {1406.3996},
 primaryClass = {astro-ph.SR},
       adsurl = {https://ui.adsabs.harvard.edu/abs/2014MNRAS.443.1192C}
}

@ARTICLE{Wright2010,
       author = {{Wright}, Edward L. and {Eisenhardt}, Peter R.~M. and {Mainzer}, Amy K. and {Ressler}, Michael E. and {Cutri}, Roc M. and {Jarrett}, Thomas and {Kirkpatrick}, J. Davy and {Padgett}, Deborah and {McMillan}, Robert S. and {Skrutskie}, Michael and {Stanford}, S.~A. and {Cohen}, Martin and {Walker}, Russell G. and {Mather}, John C. and {Leisawitz}, David and {Gautier}, III, Thomas N. and {McLean}, Ian and {Benford}, Dominic and {Lonsdale}, Carol J. and {Blain}, Andrew and {Mendez}, Bryan and {Irace}, William R. and {Duval}, Valerie and {Liu}, Fengchuan and {Royer}, Don and {Heinrichsen}, Ingolf and {Howard}, Joan and {Shannon}, Mark and {Kendall}, Martha and {Walsh}, Amy L. and {Larsen}, Mark and {Cardon}, Joel G. and {Schick}, Scott and {Schwalm}, Mark and {Abid}, Mohamed and {Fabinsky}, Beth and {Naes}, Larry and {Tsai}, Chao-Wei},
        title = "{The Wide-field Infrared Survey Explorer (WISE): Mission Description and Initial On-orbit Performance}",
      journal = {\aj},
         year = 2010,
        month = dec,
       volume = {140},
       number = {6},
        pages = {1868-1881},
          doi = {10.1088/0004-6256/140/6/1868},
archivePrefix = {arXiv},
       eprint = {1008.0031},
 primaryClass = {astro-ph.IM},
       adsurl = {https://ui.adsabs.harvard.edu/abs/2010AJ....140.1868W}
}

@ARTICLE{Chen2013,
       author = {{Chen}, B.~Q. and {Schultheis}, M. and {Jiang}, B.~W. and {Gonzalez}, O.~A. and {Robin}, A.~C. and {Rejkuba}, M. and {Minniti}, D.},
        title = "{Three-dimensional interstellar extinction map toward the Galactic bulge}",
      journal = {\aap},
         year = 2013,
        month = feb,
       volume = {550},
          eid = {A42},
        pages = {A42},
          doi = {10.1051/0004-6361/201219682},
archivePrefix = {arXiv},
       eprint = {1211.3092},
 primaryClass = {astro-ph.GA},
       adsurl = {https://ui.adsabs.harvard.edu/abs/2013A&A...550A..42C}
}

@ARTICLE{Schultheis2014,
       author = {{Schultheis}, M. and {Chen}, B.~Q. and {Jiang}, B.~W. and {Gonzalez}, O.~A. and {Enokiya}, R. and {Fukui}, Y. and {Torii}, K. and {Rejkuba}, M. and {Minniti}, D.},
        title = "{Mapping the Milky Way bulge at high resolution: the 3D dust extinction, CO, and X factor maps}",
      journal = {\aap},
         year = 2014,
        month = jun,
       volume = {566},
          eid = {A120},
        pages = {A120},
          doi = {10.1051/0004-6361/201322788},
archivePrefix = {arXiv},
       eprint = {1405.0503},
 primaryClass = {astro-ph.GA},
       adsurl = {https://ui.adsabs.harvard.edu/abs/2014A&A...566A.120S}
}

@ARTICLE{Gaia2016,
       author = {{Gaia Collaboration} and {Prusti}, T. and {de Bruijne}, J.~H.~J. and {Brown}, A.~G.~A. and {Vallenari}, A. and {Babusiaux}, C. and {Bailer-Jones}, C.~A.~L. and {Bastian}, U. and {Biermann}, M. and {Evans}, D.~W. and {Eyer}, L. and {Jansen}, F. and {Jordi}, C. and {Klioner}, S.~A. and {Lammers}, U. and {Lindegren}, L. and {Luri}, X. and {Mignard}, F. and {Milligan}, D.~J. and {Panem}, C. and {Poinsignon}, V. and {Pourbaix}, D. and {Randich}, S. and {Sarri}, G. and {Sartoretti}, P. and {Siddiqui}, H.~I. and {Soubiran}, C. and {Valette}, V. and {van Leeuwen}, F. and {Walton}, N.~A. and {Aerts}, C. and {Arenou}, F. and {Cropper}, M. and {Drimmel}, R. and {H{\o}g}, E. and {Katz}, D. and {Lattanzi}, M.~G. and {O'Mullane}, W. and {Grebel}, E.~K. and {Holland}, A.~D. and {Huc}, C. and {Passot}, X. and {Bramante}, L. and {Cacciari}, C. and {Casta{\~n}eda}, J. and {Chaoul}, L. and {Cheek}, N. and {De Angeli}, F. and {Fabricius}, C. and {Guerra}, R. and {Hern{\'a}ndez}, J. and {Jean-Antoine-Piccolo}, A. and {Masana}, E. and {Messineo}, R. and {Mowlavi}, N. and {Nienartowicz}, K. and {Ord{\'o}{\~n}ez-Blanco}, D. and {Panuzzo}, P. and {Portell}, J. and {Richards}, P.~J. and {Riello}, M. and {Seabroke}, G.~M. and {Tanga}, P. and {Th{\'e}venin}, F. and {Torra}, J. and {Els}, S.~G. and {Gracia-Abril}, G. and {Comoretto}, G. and {Garcia-Reinaldos}, M. and {Lock}, T. and {Mercier}, E. and {Altmann}, M. and {Andrae}, R. and {Astraatmadja}, T.~L. and {Bellas-Velidis}, I. and {Benson}, K. and {Berthier}, J. and {Blomme}, R. and {Busso}, G. and {Carry}, B. and {Cellino}, A. and {Clementini}, G. and {Cowell}, S. and {Creevey}, O. and {Cuypers}, J. and {Davidson}, M. and {De Ridder}, J. and {de Torres}, A. and {Delchambre}, L. and {Dell'Oro}, A. and {Ducourant}, C. and {Fr{\'e}mat}, Y. and {Garc{\'\i}a-Torres}, M. and {Gosset}, E. and {Halbwachs}, J.-L. and {Hambly}, N.~C. and {Harrison}, D.~L. and {Hauser}, M. and {Hestroffer}, D. and {Hodgkin}, S.~T. and {Huckle}, H.~E. and {Hutton}, A. and {Jasniewicz}, G. and {Jordan}, S. and {Kontizas}, M. and {Korn}, A.~J. and {Lanzafame}, A.~C. and {Manteiga}, M. and {Moitinho}, A. and {Muinonen}, K. and {Osinde}, J. and {Pancino}, E. and {Pauwels}, T. and {Petit}, J.-M. and {Recio-Blanco}, A. and {Robin}, A.~C. and {Sarro}, L.~M. and {Siopis}, C. and {Smith}, M. and {Smith}, K.~W. and {Sozzetti}, A. and {Thuillot}, W. and {van Reeven}, W. and {Viala}, Y. and {Abbas}, U. and {Abreu Aramburu}, A. and {Accart}, S. and {Aguado}, J.~J. and {Allan}, P.~M. and {Allasia}, W. and {Altavilla}, G. and {{\'A}lvarez}, M.~A. and {Alves}, J. and {Anderson}, R.~I. and {Andrei}, A.~H. and {Anglada Varela}, E. and {Antiche}, E. and {Antoja}, T. and {Ant{\'o}n}, S. and {Arcay}, B. and {Atzei}, A. and {Ayache}, L. and {Bach}, N. and {Baker}, S.~G. and {Balaguer-N{\'u}{\~n}ez}, L. and {Barache}, C. and {Barata}, C. and {Barbier}, A. and {Barblan}, F. and {Baroni}, M. and {Barrado y Navascu{\'e}s}, D. and {Barros}, M. and {Barstow}, M.~A. and {Becciani}, U. and {Bellazzini}, M. and {Bellei}, G. and {Bello Garc{\'\i}a}, A. and {Belokurov}, V. and {Bendjoya}, P. and {Berihuete}, A. and {Bianchi}, L. and {Bienaym{\'e}}, O. and {Billebaud}, F. and {Blagorodnova}, N. and {Blanco-Cuaresma}, S. and {Boch}, T. and {Bombrun}, A. and {Borrachero}, R. and {Bouquillon}, S. and {Bourda}, G. and {Bouy}, H. and {Bragaglia}, A. and {Breddels}, M.~A. and {Brouillet}, N. and {Br{\"u}semeister}, T. and {Bucciarelli}, B. and {Budnik}, F. and {Burgess}, P. and {Burgon}, R. and {Burlacu}, A. and {Busonero}, D. and {Buzzi}, R. and {Caffau}, E. and {Cambras}, J. and {Campbell}, H. and {Cancelliere}, R. and {Cantat-Gaudin}, T. and {Carlucci}, T. and {Carrasco}, J.~M. and {Castellani}, M. and {Charlot}, P. and {Charnas}, J. and {Charvet}, P. and {Chassat}, F. and {Chiavassa}, A. and {Clotet}, M. and {Cocozza}, G. and {Collins}, R.~S. and {Collins}, P. and {Costigan}, G.},
        title = "{The Gaia mission}",
      journal = {\aap},
         year = 2016,
        month = nov,
       volume = {595},
          eid = {A1},
        pages = {A1},
          doi = {10.1051/0004-6361/201629272},
archivePrefix = {arXiv},
       eprint = {1609.04153},
 primaryClass = {astro-ph.IM},
       adsurl = {https://ui.adsabs.harvard.edu/abs/2016A&A...595A...1G}
}

@ARTICLE{Gaia2018,
       author = {{Gaia Collaboration} and {Brown}, A.~G.~A. and {Vallenari}, A. and {Prusti}, T. and {de Bruijne}, J.~H.~J. and {Babusiaux}, C. and {Bailer-Jones}, C.~A.~L. and {Biermann}, M. and {Evans}, D.~W. and {Eyer}, L. and {Jansen}, F. and {Jordi}, C. and {Klioner}, S.~A. and {Lammers}, U. and {Lindegren}, L. and {Luri}, X. and {Mignard}, F. and {Panem}, C. and {Pourbaix}, D. and {Randich}, S. and {Sartoretti}, P. and {Siddiqui}, H.~I. and {Soubiran}, C. and {van Leeuwen}, F. and {Walton}, N.~A. and {Arenou}, F. and {Bastian}, U. and {Cropper}, M. and {Drimmel}, R. and {Katz}, D. and {Lattanzi}, M.~G. and {Bakker}, J. and {Cacciari}, C. and {Casta{\~n}eda}, J. and {Chaoul}, L. and {Cheek}, N. and {De Angeli}, F. and {Fabricius}, C. and {Guerra}, R. and {Holl}, B. and {Masana}, E. and {Messineo}, R. and {Mowlavi}, N. and {Nienartowicz}, K. and {Panuzzo}, P. and {Portell}, J. and {Riello}, M. and {Seabroke}, G.~M. and {Tanga}, P. and {Th{\'e}venin}, F. and {Gracia-Abril}, G. and {Comoretto}, G. and {Garcia-Reinaldos}, M. and {Teyssier}, D. and {Altmann}, M. and {Andrae}, R. and {Audard}, M. and {Bellas-Velidis}, I. and {Benson}, K. and {Berthier}, J. and {Blomme}, R. and {Burgess}, P. and {Busso}, G. and {Carry}, B. and {Cellino}, A. and {Clementini}, G. and {Clotet}, M. and {Creevey}, O. and {Davidson}, M. and {De Ridder}, J. and {Delchambre}, L. and {Dell'Oro}, A. and {Ducourant}, C. and {Fern{\'a}ndez-Hern{\'a}ndez}, J. and {Fouesneau}, M. and {Fr{\'e}mat}, Y. and {Galluccio}, L. and {Garc{\'\i}a-Torres}, M. and {Gonz{\'a}lez-N{\'u}{\~n}ez}, J. and {Gonz{\'a}lez-Vidal}, J.~J. and {Gosset}, E. and {Guy}, L.~P. and {Halbwachs}, J.-L. and {Hambly}, N.~C. and {Harrison}, D.~L. and {Hern{\'a}ndez}, J. and {Hestroffer}, D. and {Hodgkin}, S.~T. and {Hutton}, A. and {Jasniewicz}, G. and {Jean-Antoine-Piccolo}, A. and {Jordan}, S. and {Korn}, A.~J. and {Krone-Martins}, A. and {Lanzafame}, A.~C. and {Lebzelter}, T. and {L{\"o}ffler}, W. and {Manteiga}, M. and {Marrese}, P.~M. and {Mart{\'\i}n-Fleitas}, J.~M. and {Moitinho}, A. and {Mora}, A. and {Muinonen}, K. and {Osinde}, J. and {Pancino}, E. and {Pauwels}, T. and {Petit}, J.-M. and {Recio-Blanco}, A. and {Richards}, P.~J. and {Rimoldini}, L. and {Robin}, A.~C. and {Sarro}, L.~M. and {Siopis}, C. and {Smith}, M. and {Sozzetti}, A. and {S{\"u}veges}, M. and {Torra}, J. and {van Reeven}, W. and {Abbas}, U. and {Abreu Aramburu}, A. and {Accart}, S. and {Aerts}, C. and {Altavilla}, G. and {{\'A}lvarez}, M.~A. and {Alvarez}, R. and {Alves}, J. and {Anderson}, R.~I. and {Andrei}, A.~H. and {Anglada Varela}, E. and {Antiche}, E. and {Antoja}, T. and {Arcay}, B. and {Astraatmadja}, T.~L. and {Bach}, N. and {Baker}, S.~G. and {Balaguer-N{\'u}{\~n}ez}, L. and {Balm}, P. and {Barache}, C. and {Barata}, C. and {Barbato}, D. and {Barblan}, F. and {Barklem}, P.~S. and {Barrado}, D. and {Barros}, M. and {Barstow}, M.~A. and {Bartholom{\'e} Mu{\~n}oz}, S. and {Bassilana}, J.-L. and {Becciani}, U. and {Bellazzini}, M. and {Berihuete}, A. and {Bertone}, S. and {Bianchi}, L. and {Bienaym{\'e}}, O. and {Blanco-Cuaresma}, S. and {Boch}, T. and {Boeche}, C. and {Bombrun}, A. and {Borrachero}, R. and {Bossini}, D. and {Bouquillon}, S. and {Bourda}, G. and {Bragaglia}, A. and {Bramante}, L. and {Breddels}, M.~A. and {Bressan}, A. and {Brouillet}, N. and {Br{\"u}semeister}, T. and {Brugaletta}, E. and {Bucciarelli}, B. and {Burlacu}, A. and {Busonero}, D. and {Butkevich}, A.~G. and {Buzzi}, R. and {Caffau}, E. and {Cancelliere}, R. and {Cannizzaro}, G. and {Cantat-Gaudin}, T. and {Carballo}, R. and {Carlucci}, T. and {Carrasco}, J.~M. and {Casamiquela}, L. and {Castellani}, M. and {Castro-Ginard}, A. and {Charlot}, P. and {Chemin}, L. and {Chiavassa}, A. and {Cocozza}, G. and {Costigan}, G. and {Cowell}, S. and {Crifo}, F. and {Crosta}, M. and {Crowley}, C. and {Cuypers}, J. and {Dafonte}, C. and {Damerdji}, Y. and {Dapergolas}, A. and {David}, P. and {David}, M. and {de Laverny}, P. and {De Luise}, F.},
        title = "{Gaia Data Release 2. Summary of the contents and survey properties}",
      journal = {\aap},
         year = 2018,
        month = aug,
       volume = {616},
          eid = {A1},
        pages = {A1},
          doi = {10.1051/0004-6361/201833051},
archivePrefix = {arXiv},
       eprint = {1804.09365},
 primaryClass = {astro-ph.GA},
       adsurl = {https://ui.adsabs.harvard.edu/abs/2018A&A...616A...1G}
}

@ARTICLE{Minniti2010,
       author = {{Minniti}, D. and {Lucas}, P.~W. and {Emerson}, J.~P. and {Saito}, R.~K. and {Hempel}, M. and {Pietrukowicz}, P. and {Ahumada}, A.~V. and {Alonso}, M.~V. and {Alonso-Garcia}, J. and {Arias}, J.~I. and {Bandyopadhyay}, R.~M. and {Barb{\'a}}, R.~H. and {Barbuy}, B. and {Bedin}, L.~R. and {Bica}, E. and {Borissova}, J. and {Bronfman}, L. and {Carraro}, G. and {Catelan}, M. and {Clari{\'a}}, J.~J. and {Cross}, N. and {de Grijs}, R. and {D{\'e}k{\'a}ny}, I. and {Drew}, J.~E. and {Fari{\~n}a}, C. and {Feinstein}, C. and {Fern{\'a}ndez Laj{\'u}s}, E. and {Gamen}, R.~C. and {Geisler}, D. and {Gieren}, W. and {Goldman}, B. and {Gonzalez}, O.~A. and {Gunthardt}, G. and {Gurovich}, S. and {Hambly}, N.~C. and {Irwin}, M.~J. and {Ivanov}, V.~D. and {Jord{\'a}n}, A. and {Kerins}, E. and {Kinemuchi}, K. and {Kurtev}, R. and {L{\'o}pez-Corredoira}, M. and {Maccarone}, T. and {Masetti}, N. and {Merlo}, D. and {Messineo}, M. and {Mirabel}, I.~F. and {Monaco}, L. and {Morelli}, L. and {Padilla}, N. and {Palma}, T. and {Parisi}, M.~C. and {Pignata}, G. and {Rejkuba}, M. and {Roman-Lopes}, A. and {Sale}, S.~E. and {Schreiber}, M.~R. and {Schr{\"o}der}, A.~C. and {Smith}, M. and {Sodr{\'e}}, Jr., L. and {Soto}, M. and {Tamura}, M. and {Tappert}, C. and {Thompson}, M.~A. and {Toledo}, I. and {Zoccali}, M. and {Pietrzynski}, G.},
        title = "{VISTA Variables in the Via Lactea (VVV): The public ESO near-IR variability survey of the Milky Way}",
      journal = {\na},
         year = 2010,
        month = jul,
       volume = {15},
       number = {5},
        pages = {433-443},
          doi = {10.1016/j.newast.2009.12.002},
archivePrefix = {arXiv},
       eprint = {0912.1056},
 primaryClass = {astro-ph.GA},
       adsurl = {https://ui.adsabs.harvard.edu/abs/2010NewA...15..433M}
}

@ARTICLE{Lawrence2007,
       author = {{Lawrence}, A. and {Warren}, S.~J. and {Almaini}, O. and {Edge}, A.~C. and {Hambly}, N.~C. and {Jameson}, R.~F. and {Lucas}, P. and {Casali}, M. and {Adamson}, A. and {Dye}, S. and {Emerson}, J.~P. and {Foucaud}, S. and {Hewett}, P. and {Hirst}, P. and {Hodgkin}, S.~T. and {Irwin}, M.~J. and {Lodieu}, N. and {McMahon}, R.~G. and {Simpson}, C. and {Smail}, I. and {Mortlock}, D. and {Folger}, M.},
        title = "{The UKIRT Infrared Deep Sky Survey (UKIDSS)}",
      journal = {\mnras},
         year = 2007,
        month = aug,
       volume = {379},
       number = {4},
        pages = {1599-1617},
          doi = {10.1111/j.1365-2966.2007.12040.x},
archivePrefix = {arXiv},
       eprint = {astro-ph/0604426},
 primaryClass = {astro-ph},
       adsurl = {https://ui.adsabs.harvard.edu/abs/2007MNRAS.379.1599L}
}

@ARTICLE{Lucas2008,
       author = {{Lucas}, P.~W. and {Hoare}, M.~G. and {Longmore}, A. and {Schr{\"o}der}, A.~C. and {Davis}, C.~J. and {Adamson}, A. and {Bandyopadhyay}, R.~M. and {de Grijs}, R. and {Smith}, M. and {Gosling}, A. and {Mitchison}, S. and {G{\'a}sp{\'a}r}, A. and {Coe}, M. and {Tamura}, M. and {Parker}, Q. and {Irwin}, M. and {Hambly}, N. and {Bryant}, J. and {Collins}, R.~S. and {Cross}, N. and {Evans}, D.~W. and {Gonzalez-Solares}, E. and {Hodgkin}, S. and {Lewis}, J. and {Read}, M. and {Riello}, M. and {Sutorius}, E.~T.~W. and {Lawrence}, A. and {Drew}, J.~E. and {Dye}, S. and {Thompson}, M.~A.},
        title = "{The UKIDSS Galactic Plane Survey}",
      journal = {\mnras},
         year = 2008,
        month = nov,
       volume = {391},
       number = {1},
        pages = {136-163},
          doi = {10.1111/j.1365-2966.2008.13924.x},
archivePrefix = {arXiv},
       eprint = {0712.0100},
 primaryClass = {astro-ph},
       adsurl = {https://ui.adsabs.harvard.edu/abs/2008MNRAS.391..136L}
}

@ARTICLE{Mainzer2011,
       author = {{Mainzer}, A. and {Bauer}, J. and {Grav}, T. and {Masiero}, J. and {Cutri}, R.~M. and {Dailey}, J. and {Eisenhardt}, P. and {McMillan}, R.~S. and {Wright}, E. and {Walker}, R. and {Jedicke}, R. and {Spahr}, T. and {Tholen}, D. and {Alles}, R. and {Beck}, R. and {Brandenburg}, H. and {Conrow}, T. and {Evans}, T. and {Fowler}, J. and {Jarrett}, T. and {Marsh}, K. and {Masci}, F. and {McCallon}, H. and {Wheelock}, S. and {Wittman}, M. and {Wyatt}, P. and {DeBaun}, E. and {Elliott}, G. and {Elsbury}, D. and {Gautier}, IV, T. and {Gomillion}, S. and {Leisawitz}, D. and {Maleszewski}, C. and {Micheli}, M. and {Wilkins}, A.},
        title = "{Preliminary Results from NEOWISE: An Enhancement to the Wide-field Infrared Survey Explorer for Solar System Science}",
      journal = {\apj},
         year = 2011,
        month = apr,
       volume = {731},
       number = {1},
          eid = {53},
        pages = {53},
          doi = {10.1088/0004-637X/731/1/53},
archivePrefix = {arXiv},
       eprint = {1102.1996},
 primaryClass = {astro-ph.EP},
       adsurl = {https://ui.adsabs.harvard.edu/abs/2011ApJ...731...53M}
}

@ARTICLE{Barnes2023,
       author = {{Barnes}, Ashley. T. and {Watkins}, Elizabeth J. and {Meidt}, Sharon E. and {Kreckel}, Kathryn and {Sormani}, Mattia C. and {Tre{\ss}}, Robin G. and {Glover}, Simon C.~O. and {Bigiel}, Frank and {Chandar}, Rupali and {Emsellem}, Eric and {Lee}, Janice C. and {Leroy}, Adam K. and {Sandstrom}, Karin M. and {Schinnerer}, Eva and {Rosolowsky}, Erik and {Belfiore}, Francesco and {Blanc}, Guillermo A. and {Boquien}, M{\'e}d{\'e}ric and {Brok}, Jakob den and {Cao}, Yixian and {Chevance}, M{\'e}lanie and {Dale}, Daniel A. and {Egorov}, Oleg V. and {Eibensteiner}, Cosima and {Grasha}, Kathryn and {Groves}, Brent and {Hassani}, Hamid and {Henshaw}, Jonathan D. and {Jeffreson}, Sarah and {Jim{\'e}nez-Donaire}, Mar{\'\i}a J. and {Keller}, Benjamin W. and {Klessen}, Ralf S. and {Koch}, Eric W. and {Kruijssen}, J.~M. Diederik and {Larson}, Kirsten L. and {Li}, Jing and {Liu}, Daizhong and {Lopez}, Laura A. and {Murphy}, Eric J. and {Neumann}, Lukas and {Pety}, J{\'e}r{\^o}me and {Pinna}, Francesca and {Querejeta}, Miguel and {Renaud}, Florent and {Saito}, Toshiki and {Sarbadhicary}, Sumit K. and {Sardone}, Amy and {Smith}, Rowan J. and {Stuber}, Sophia K. and {Sun}, Jiayi and {Thilker}, David A. and {Usero}, Antonio and {Whitmore}, Bradley C. and {Williams}, Thomas G.},
        title = "{PHANGS-JWST First Results: Multiwavelength View of Feedback-driven Bubbles (the Phantom Voids) across NGC 628}",
      journal = {\apjl},
         year = 2023,
        month = feb,
       volume = {944},
       number = {2},
          eid = {L22},
        pages = {L22},
          doi = {10.3847/2041-8213/aca7b9},
archivePrefix = {arXiv},
       eprint = {2212.00812},
 primaryClass = {astro-ph.GA},
       adsurl = {https://ui.adsabs.harvard.edu/abs/2023ApJ...944L..22B}
}

@ARTICLE{Chen2017,
       author = {{Chen}, Y.~Q. and {Casagrande}, L. and {Zhao}, G. and {Bovy}, J. and {Silva Aguirre}, V. and {Zhao}, J.~K. and {Jia}, Y.~P.},
        title = "{Absolute Magnitudes of Seismic Red Clumps in the Kepler  Field and SAGA: The Age Dependency of the Distance Scale}",
      journal = {\apj},
         year = 2017,
        month = may,
       volume = {840},
       number = {2},
          eid = {77},
        pages = {77},
          doi = {10.3847/1538-4357/aa6d0f},
archivePrefix = {arXiv},
       eprint = {1704.03903},
 primaryClass = {astro-ph.SR},
       adsurl = {https://ui.adsabs.harvard.edu/abs/2017ApJ...840...77C}
}

@ARTICLE{Salaris2002,
       author = {{Salaris}, Maurizio and {Girardi}, L{\'e}o},
        title = "{Population effects on the red giant clump absolute magnitude: the K band}",
      journal = {\mnras},
         year = 2002,
        month = nov,
       volume = {337},
       number = {1},
        pages = {332-340},
          doi = {10.1046/j.1365-8711.2002.05917.x},
archivePrefix = {arXiv},
       eprint = {astro-ph/0208057},
 primaryClass = {astro-ph},
       adsurl = {https://ui.adsabs.harvard.edu/abs/2002MNRAS.337..332S}
}

@ARTICLE{Plevne2020,
       author = {{Plevne}, Olcay and {{\"O}nal Ta{\c{s}}}, {\"O}zgecan and {Bilir}, Sel{\c{c}}uk and {Seabroke}, George M.},
        title = "{Multiwavelength Absolute Magnitudes and Colors of Red Clump Stars in the Gaia Era}",
      journal = {\apj},
         year = 2020,
        month = apr,
       volume = {893},
       number = {2},
          eid = {108},
        pages = {108},
          doi = {10.3847/1538-4357/ab80bb},
archivePrefix = {arXiv},
       eprint = {2003.07887},
 primaryClass = {astro-ph.GA},
       adsurl = {https://ui.adsabs.harvard.edu/abs/2020ApJ...893..108P}
}

@ARTICLE{Alonso2017,
       author = {{Alonso-Garc{\'\i}a}, Javier and {Minniti}, Dante and {Catelan}, M{\'a}rcio and {Contreras Ramos}, Rodrigo and {Gonzalez}, Oscar A. and {Hempel}, Maren and {Lucas}, Philip W. and {Saito}, Roberto K. and {Valenti}, Elena and {Zoccali}, Manuela},
        title = "{Extinction Ratios in the Inner Galaxy as Revealed by the VVV Survey}",
      journal = {\apjl},
         year = 2017,
        month = nov,
       volume = {849},
       number = {1},
          eid = {L13},
        pages = {L13},
          doi = {10.3847/2041-8213/aa92c3},
archivePrefix = {arXiv},
       eprint = {1710.04854},
 primaryClass = {astro-ph.GA},
       adsurl = {https://ui.adsabs.harvard.edu/abs/2017ApJ...849L..13A}
}

@ARTICLE{Edenhofer2024,
       author = {{Edenhofer}, Gordian and {Zucker}, Catherine and {Frank}, Philipp and {Saydjari}, Andrew K. and {Speagle}, Joshua S. and {Finkbeiner}, Douglas and {En{\ss}lin}, Torsten A.},
        title = "{A parsec-scale Galactic 3D dust map out to 1.25 kpc from the Sun}",
      journal = {\aap},
         year = 2024,
        month = may,
       volume = {685},
          eid = {A82},
        pages = {A82},
          doi = {10.1051/0004-6361/202347628},
archivePrefix = {arXiv},
       eprint = {2308.01295},
 primaryClass = {astro-ph.GA},
       adsurl = {https://ui.adsabs.harvard.edu/abs/2024A&A...685A..82E}
}

@ARTICLE{Li2018,
       author = {{Li}, Linlin and {Shen}, Shiyin and {Hou}, Jinliang and {Yuan}, Haibo and {Xiang}, Maosheng and {Chen}, Bingqiu and {Huang}, Yang and {Liu}, Xiaowei},
        title = "{Three-dimensional Structure of the Milky Way Dust: Modeling of LAMOST Data}",
      journal = {\apj},
         year = 2018,
        month = may,
       volume = {858},
       number = {2},
          eid = {75},
        pages = {75},
          doi = {10.3847/1538-4357/aabaef},
archivePrefix = {arXiv},
       eprint = {1803.10540},
 primaryClass = {astro-ph.GA},
       adsurl = {https://ui.adsabs.harvard.edu/abs/2018ApJ...858...75L}
}

@ARTICLE{Ivezi2008,
       author = {{Ivezi{\'c}}, {\v{Z}}eljko and {Sesar}, Branimir and {Juri{\'c}}, Mario and {Bond}, Nicholas and {Dalcanton}, Julianne and {Rockosi}, Constance M. and {Yanny}, Brian and {Newberg}, Heidi J. and {Beers}, Timothy C. and {Allende Prieto}, Carlos and {Wilhelm}, Ron and {Lee}, Young Sun and {Sivarani}, Thirupathi and {Norris}, John E. and {Bailer-Jones}, Coryn A.~L. and {Re Fiorentin}, Paola and {Schlegel}, David and {Uomoto}, Alan and {Lupton}, Robert H. and {Knapp}, Gillian R. and {Gunn}, James E. and {Covey}, Kevin R. and {Allyn Smith}, J. and {Miknaitis}, Gajus and {Doi}, Mamoru and {Tanaka}, Masayuki and {Fukugita}, Masataka and {Kent}, Steve and {Finkbeiner}, Douglas and {Munn}, Jeffrey A. and {Pier}, Jeffrey R. and {Quinn}, Tom and {Hawley}, Suzanne and {Anderson}, Scott and {Kiuchi}, Furea and {Chen}, Alex and {Bushong}, James and {Sohi}, Harkirat and {Haggard}, Daryl and {Kimball}, Amy and {Barentine}, John and {Brewington}, Howard and {Harvanek}, Mike and {Kleinman}, Scott and {Krzesinski}, Jurek and {Long}, Dan and {Nitta}, Atsuko and {Snedden}, Stephanie and {Lee}, Brian and {Harris}, Hugh and {Brinkmann}, Jonathan and {Schneider}, Donald P. and {York}, Donald G.},
        title = "{The Milky Way Tomography with SDSS. II. Stellar Metallicity}",
      journal = {\apj},
         year = 2008,
        month = sep,
       volume = {684},
       number = {1},
        pages = {287-325},
          doi = {10.1086/589678},
archivePrefix = {arXiv},
       eprint = {0804.3850},
 primaryClass = {astro-ph},
       adsurl = {https://ui.adsabs.harvard.edu/abs/2008ApJ...684..287I}
}

@ARTICLE{Zhang2026,
       author = {{Zhang}, Ruoyi and {Yuan}, Haibo and {Chen}, Bingqiu and {Xiang}, Maosheng and {Huang}, Yang and {Liu}, Xiaowei and {Liu}, Jifeng},
        title = "{A Tale of Two Dust Disks in Our Milky Way}",
      journal = {\apj},
         year = 2026,
        month = feb,
       volume = {998},
       number = {2},
          eid = {344},
        pages = {344},
          doi = {10.3847/1538-4357/ae381d},
archivePrefix = {arXiv},
       eprint = {2601.02724},
 primaryClass = {astro-ph.GA},
       adsurl = {https://ui.adsabs.harvard.edu/abs/2026ApJ...998..344Z}
}

@ARTICLE{Zhang2020,
       author = {{Ruoyi}, Zhang and {Haibo}, Yuan},
        title = "{Detections of Dust in the Outskirts of M31 and M33}",
      journal = {\apjl},
         year = 2020,
        month = dec,
       volume = {905},
       number = {2},
          eid = {L20},
        pages = {L20},
          doi = {10.3847/2041-8213/abccc4},
archivePrefix = {arXiv},
       eprint = {2011.12658},
 primaryClass = {astro-ph.GA},
       adsurl = {https://ui.adsabs.harvard.edu/abs/2020ApJ...905L..20R}
}

@ARTICLE{Guo2021,
       author = {{Guo}, H.-L. and {Chen}, B.-Q. and {Yuan}, H.-B. and {Huang}, Y. and {Liu}, D.-Z. and {Yang}, Y. and {Li}, X.-Y. and {Sun}, W.-X. and {Liu}, X.-W.},
        title = "{Three-dimensional Distribution of the Interstellar Dust in the Milky Way}",
      journal = {\apj},
         year = 2021,
        month = jan,
       volume = {906},
       number = {1},
          eid = {47},
        pages = {47},
          doi = {10.3847/1538-4357/abc68a},
archivePrefix = {arXiv},
       eprint = {2010.14092},
 primaryClass = {astro-ph.GA},
       adsurl = {https://ui.adsabs.harvard.edu/abs/2021ApJ...906...47G}
}

@ARTICLE{Drimmel2001,
       author = {{Drimmel}, Ronald and {Spergel}, David N.},
        title = "{Three-dimensional Structure of the Milky Way Disk: The Distribution of Stars and Dust beyond 0.35 R$_{solar}$}",
      journal = {\apj},
         year = 2001,
        month = jul,
       volume = {556},
       number = {1},
        pages = {181-202},
          doi = {10.1086/321556},
archivePrefix = {arXiv},
       eprint = {astro-ph/0101259},
 primaryClass = {astro-ph},
       adsurl = {https://ui.adsabs.harvard.edu/abs/2001ApJ...556..181D}
}

@ARTICLE{Lallement2019,
       author = {{Lallement}, R. and {Babusiaux}, C. and {Vergely}, J.~L. and {Katz}, D. and {Arenou}, F. and {Valette}, B. and {Hottier}, C. and {Capitanio}, L.},
        title = "{Gaia-2MASS 3D maps of Galactic interstellar dust within 3 kpc}",
      journal = {\aap},
         year = 2019,
        month = may,
       volume = {625},
          eid = {A135},
        pages = {A135},
          doi = {10.1051/0004-6361/201834695},
archivePrefix = {arXiv},
       eprint = {1902.04116},
 primaryClass = {astro-ph.GA},
       adsurl = {https://ui.adsabs.harvard.edu/abs/2019A&A...625A.135L}
}

@ARTICLE{Dobbs2006,
       author = {{Dobbs}, C.~L. and {Bonnell}, I.~A.},
        title = "{Spurs and feathering in spiral galaxies}",
      journal = {\mnras},
         year = 2006,
        month = apr,
       volume = {367},
       number = {3},
        pages = {873-878},
          doi = {10.1111/j.1365-2966.2006.10146.x},
archivePrefix = {arXiv},
       eprint = {astro-ph/0602100},
 primaryClass = {astro-ph},
       adsurl = {https://ui.adsabs.harvard.edu/abs/2006MNRAS.367..873D}
}

@ARTICLE{Chen2025,
       author = {{Chen}, Bingqiu and {Li}, Guangxing and {Yuan}, Haibo and {Xiang}, Maosheng and {Zhou}, Jixuan and {Chen}, Pinjian and {Krause}, Martin and {Coombs}, Ashley},
        title = "{A large, long-lived, slowly-expanding superbubble across the Perseus arm}",
      journal = {Nature Communications},
         year = 2025,
        month = nov,
       volume = {16},
       number = {1},
          eid = {10558},
        pages = {10558},
          doi = {10.1038/s41467-025-65591-5},
archivePrefix = {arXiv},
       eprint = {2512.21927},
 primaryClass = {astro-ph.GA},
       adsurl = {https://ui.adsabs.harvard.edu/abs/2025NatCo..1610558C}
}

@ARTICLE{Carpenter1995,
       author = {{Carpenter}, John M. and {Snell}, Ronald L. and {Schloerb}, F. Peter},
        title = "{Anatomy of the Gemini OB1 Molecular Cloud Complex}",
      journal = {\apj},
         year = 1995,
        month = may,
       volume = {445},
        pages = {246},
          doi = {10.1086/175692},
       adsurl = {https://ui.adsabs.harvard.edu/abs/1995ApJ...445..246C}
}

@ARTICLE{Reid2019,
       author = {{Reid}, M.~J. and {Menten}, K.~M. and {Brunthaler}, A. and {Zheng}, X.~W. and {Dame}, T.~M. and {Xu}, Y. and {Li}, J. and {Sakai}, N. and {Wu}, Y. and {Immer}, K. and {Zhang}, B. and {Sanna}, A. and {Moscadelli}, L. and {Rygl}, K.~L.~J. and {Bartkiewicz}, A. and {Hu}, B. and {Quiroga-Nu{\~n}ez}, L.~H. and {van Langevelde}, H.~J.},
        title = "{Trigonometric Parallaxes of High-mass Star-forming Regions: Our View of the Milky Way}",
      journal = {\apj},
         year = 2019,
        month = nov,
       volume = {885},
       number = {2},
          eid = {131},
        pages = {131},
          doi = {10.3847/1538-4357/ab4a11},
archivePrefix = {arXiv},
       eprint = {1910.03357},
 primaryClass = {astro-ph.GA},
       adsurl = {https://ui.adsabs.harvard.edu/abs/2019ApJ...885..131R}
}

@MISC{Majewski2007,
       author = {{Majewski}, Steven and {Babler}, Brian and {Churchwell}, Edward and {Indebetouw}, Remy and {Meade}, Marilyn and {Nidever}, David and {Patterson}, Richard and {Rocha-Pinto}, Helio and {Skrutskie}, Michael and {Watson}, Christer},
        title = "{Galactic Structure and Star Formation in Vela-Carina}",
 howpublished = {Spitzer Proposal ID 40791},
         year = 2007,
        month = may,
        pages = {40791},
       adsurl = {https://ui.adsabs.harvard.edu/abs/2007sptz.prop40791M}
}

@ARTICLE{An2024,
       author = {{An}, Deokkeun and {Beers}, Timothy C. and {Chiti}, Anirudh},
        title = "{A Blueprint for the Milky Way's Stellar Populations. V. 3D Local Dust Extinction}",
      journal = {\apjs},
         year = 2024,
        month = may,
       volume = {272},
       number = {1},
          eid = {20},
        pages = {20},
          doi = {10.3847/1538-4365/ad3641},
archivePrefix = {arXiv},
       eprint = {2404.14626},
 primaryClass = {astro-ph.GA},
       adsurl = {https://ui.adsabs.harvard.edu/abs/2024ApJS..272...20A}
}

@ARTICLE{Dharmawardena2024,
       author = {{Dharmawardena}, T.~E. and {Bailer-Jones}, C.~A.~L. and {Fouesneau}, M. and {Foreman-Mackey}, D. and {Coronica}, P. and {Colnaghi}, T. and {M{\"u}ller}, T. and {Wilson}, A.~G.},
        title = "{All-sky three-dimensional dust density and extinction Maps of the Milky Way out to 2.8 kpc}",
      journal = {\mnras},
         year = 2024,
        month = aug,
       volume = {532},
       number = {3},
        pages = {3480-3498},
          doi = {10.1093/mnras/stae1474},
archivePrefix = {arXiv},
       eprint = {2406.06740},
 primaryClass = {astro-ph.GA},
       adsurl = {https://ui.adsabs.harvard.edu/abs/2024MNRAS.532.3480D}
}

@ARTICLE{Hawkins2017,
       author = {{Hawkins}, Keith and {Leistedt}, Boris and {Bovy}, Jo and {Hogg}, David W.},
        title = "{Red clump stars and Gaia: calibration of the standard candle using a hierarchical probabilistic model}",
      journal = {\mnras},
         year = 2017,
        month = oct,
       volume = {471},
       number = {1},
        pages = {722-729},
          doi = {10.1093/mnras/stx1655},
archivePrefix = {arXiv},
       eprint = {1705.08988},
 primaryClass = {astro-ph.GA},
       adsurl = {https://ui.adsabs.harvard.edu/abs/2017MNRAS.471..722H}
}

@ARTICLE{Ruiz-Dern2018,
       author = {{Ruiz-Dern}, L. and {Babusiaux}, C. and {Arenou}, F. and {Turon}, C. and {Lallement}, R.},
        title = "{Empirical photometric calibration of the Gaia red clump: Colours, effective temperature, and absolute magnitude}",
      journal = {\aap},
         year = 2018,
        month = jan,
       volume = {609},
          eid = {A116},
        pages = {A116},
          doi = {10.1051/0004-6361/201731572},
archivePrefix = {arXiv},
       eprint = {1710.05803},
 primaryClass = {astro-ph.SR},
       adsurl = {https://ui.adsabs.harvard.edu/abs/2018A&A...609A.116R}
}

@ARTICLE{Qin2023,
       author = {{Qin}, Fei and {Parkinson}, David and {Hong}, Sungwook E. and {Sabiu}, Cristiano G.},
        title = "{Reconstructing the cosmological density and velocity fields from redshifted galaxy distributions using V-net}",
      journal = {\jcap},
         year = 2023,
        month = jun,
       volume = {2023},
       number = {6},
          eid = {062},
        pages = {062},
          doi = {10.1088/1475-7516/2023/06/062},
archivePrefix = {arXiv},
       eprint = {2302.02087},
 primaryClass = {astro-ph.CO},
       adsurl = {https://ui.adsabs.harvard.edu/abs/2023JCAP...06..062Q}
}

@ARTICLE{Marshall2025,
       author = {{Marshall}, D.~J. and {Montillaud}, J. and {Cambr{\'e}sy}, L. and {Cornu}, D.},
        title = "{A new dust map of the Milky Way: I. Principal features}",
      journal = {\aap},
         year = 2025,
        month = dec,
       volume = {704},
          eid = {A118},
        pages = {A118},
          doi = {10.1051/0004-6361/202554622},
       adsurl = {https://ui.adsabs.harvard.edu/abs/2025A&A...704A.118M}
}

@ARTICLE{Lindegren2021,
       author = {{Lindegren}, L. and {Bastian}, U. and {Biermann}, M. and {Bombrun}, A. and {de Torres}, A. and {Gerlach}, E. and {Geyer}, R. and {Hern{\'a}ndez}, J. and {Hilger}, T. and {Hobbs}, D. and {Klioner}, S.~A. and {Lammers}, U. and {McMillan}, P.~J. and {Ramos-Lerate}, M. and {Steidelm{\"u}ller}, H. and {Stephenson}, C.~A. and {van Leeuwen}, F.},
        title = "{Gaia Early Data Release 3. Parallax bias versus magnitude, colour, and position}",
      journal = {\aap},
         year = 2021,
        month = may,
       volume = {649},
          eid = {A4},
        pages = {A4},
          doi = {10.1051/0004-6361/202039653},
archivePrefix = {arXiv},
       eprint = {2012.01742},
 primaryClass = {astro-ph.IM},
       adsurl = {https://ui.adsabs.harvard.edu/abs/2021A&A...649A...4L}
}

@ARTICLE{Gordon2023,
       author = {{Gordon}, Karl D. and {Clayton}, Geoffrey C. and {Decleir}, Marjorie and {Fitzpatrick}, E.~L. and {Massa}, Derck and {Misselt}, Karl A. and {Tollerud}, Erik J.},
        title = "{One Relation for All Wavelengths: The Far-ultraviolet to Mid-infrared Milky Way Spectroscopic R(V)-dependent Dust Extinction Relationship}",
      journal = {\apj},
         year = 2023,
        month = jun,
       volume = {950},
       number = {2},
          eid = {86},
        pages = {86},
          doi = {10.3847/1538-4357/accb59},
archivePrefix = {arXiv},
       eprint = {2304.01991},
 primaryClass = {astro-ph.GA},
       adsurl = {https://ui.adsabs.harvard.edu/abs/2023ApJ...950...86G}
}
\bibliographystyle{aasjournalv7}


\end{CJK*}
\end{document}